\documentclass[twocolumn, showkeys, superscriptaddress, amsmath, amssymb,
 aps, pra, floatfix]{revtex4-1}

\usepackage{graphicx}% Include figure files
\usepackage{dcolumn}% Align table columns on decimal point
\usepackage{bm}% bold math
\usepackage{hyperref}% add hypertext capabilities
\usepackage{xcolor,tikz, circuitikz}
\usetikzlibrary{shapes, positioning}
\usepackage{algpseudocode}
\usepackage{multirow}
\usepackage{etoolbox}

\usepackage{newfloat}

\DeclareFloatingEnvironment[
    fileext=loa,
    listname=List of Algorithms,
    name=ALGORITHM,
]{algorithm}

\AtEndEnvironment{algorithm*}{\noindent\hrulefill\par\nobreak\vskip-5pt}

\hypersetup{
colorlinks,
linkcolor={blue},%{blue!50!black},
citecolor={blue},
urlcolor={blue!80!black}
}

\usepackage{amsthm}

\newcommand{\ket}[1]{\vert{#1}\rangle}
\newcommand{\bra}[1]{\langle{#1}\vert}
\newcommand{\braket}[2]{\langle{#1}\vert{#2}\rangle}
\newcommand{\kb}[2]{\ket{#1}\bra{#2}}

\DeclareMathOperator{\Real}{Re}
\DeclareMathOperator{\tr}{Tr}

\newtheorem{defination}{Definition}
\newtheorem{theorem}{Theorem}
\newtheorem{lemma}[theorem]{Lemma}
\newtheorem{corollary}{Corollary}

\algnewcommand\algorithmicswitch{\textbf{switch}}
\algnewcommand\algorithmiccase{\textbf{case}}
\algnewcommand\algorithmicassert{\texttt{assert}}
\algnewcommand\Assert[1]{\State \algorithmicassert(#1)}%
\algdef{SE}[SWITCH]{Switch}{EndSwitch}[1]{\algorithmicswitch\ #1\ \algorithmicdo}{\algorithmicend\ \algorithmicswitch}%
\algdef{SE}[CASE]{Case}{EndCase}[1]{\algorithmiccase\ #1}{\algorithmicend\ \algorithmiccase}%
\begin{document}

\preprint{APS/123-QED}

\title{Improved Routing of Multiparty Entanglement over Quantum Networks}% Force line breaks with \\

\author{Nirupam Basak}
\email{nirupambasak2020@iitkalumni.org}
\affiliation{Cryptology and Security Research Unit,
Indian Statistical Institute,
Kolkata 700108, India} 
\author{Goutam Paul}
\email{goutam.paul@isical.ac.in}
\affiliation{Cryptology and Security Research Unit,
Indian Statistical Institute,
Kolkata 700108, India} 

\begin{abstract}
Effective routing of entanglements over a quantum network is a fundamental problem in quantum communication. Due to the fragility of quantum states, it is difficult to route entanglements at long distances. Graph states can be utilized for this purpose, reducing the need for long-distance entanglement routing by leveraging local operations. In this paper, we propose two graph state-based routing protocols for sharing GHZ states, achieving larger sizes than the existing works, for given network topologies. For this improvement, we consider tree structures connecting the users participating in the final GHZ states, as opposed to the linear configurations used in the earlier ones. For arbitrary network topologies, we show that if such a tree is balanced, it achieves a larger size than unbalanced trees. In particular, for grid networks, we show special constructions of the above-mentioned tree that achieve optimal results. Moreover, if the user nodes among whom the entanglement is to be routed are pre-specified, we propose a strategy to accomplish the required routing.
\end{abstract}

\maketitle

%\tableofcontents

\section{Introduction}

In today’s technological landscape, secure communication has become one of the most critical requirements. Traditional classical communication security depends on hardness assumptions on  mathematical problems such as discrete logarithms and integer factorization~\cite{sharma2023post}. However, with the advent of quantum computing, these classical security measures are under significant threat. Shor's Algorithm~\cite{365700} and Grover's Algorithm~\cite{10.1145/237814.237866} present formidable challenges to classical encryption methods. To counteract these threats, researchers have proposed several quantum communication protocols~\cite{BENNETT20147,PhysRevLett.67.661,PhysRevLett.94.230504,liao2017satellite,RevModPhys.92.025002,grunenfelder2023fast,doi:10.1142/S0219749920500380,PhysRevA.68.042317,das2021quantum,PhysRevApplied.19.014036,PhysRevApplied.16.024012,PhysRevA.59.1829,PhysRevA.69.052307,PhysRevA.71.044301,GUO2003247,PhysRevA.61.042311,PhysRevA.63.042301,PhysRevLett.121.150502,PhysRevA.103.032410,CHONG20101192,liu2013multiparty,das2020comment,yang2022detector}. While these protocols have been experimentally validated~\cite{chen2021twin,avesani2021full} for point-to-point secure quantum communication, issues such as errors and signal losses still limit the effectiveness of these systems~\cite{pirandola2017fundamental}. To address these challenges, entanglement swapping using Bell states has been proposed as a solution~\cite{dahlberg2020transforming}. Recent experimental advancements have successfully demonstrated increased distances and success rates in implementing these protocols, including groundbreaking efforts in intercontinental quantum communication via satellites~\cite{PhysRevLett.120.030501}.

Having achieved these successes, the next ambitious goal is to progress towards multiparty scenarios and eventually establish a fully operational quantum internet~\cite{simon2017towards}. The quantum internet represents a global network that interconnects quantum devices worldwide and facilitates advanced quantum communication. This network underpins a variety of quantum communication protocols, including quantum key distribution~\cite{BENNETT20147,PhysRevLett.67.661,PhysRevLett.94.230504,liao2017satellite,RevModPhys.92.025002,grunenfelder2023fast}, quantum secret sharing~\cite{PhysRevA.59.1829,PhysRevA.69.052307,PhysRevA.71.044301,GUO2003247,PhysRevA.61.042311,PhysRevA.63.042301,PhysRevLett.121.150502,PhysRevA.103.032410}, quantum secure direct communication~\cite{doi:10.1142/S0219749920500380,PhysRevA.68.042317,das2021quantum,PhysRevApplied.19.014036,PhysRevApplied.16.024012}, quantum voting~\cite{HILLERY200675,doi:10.1142/S0219749917500071,mishra2022quantum}, and quantum key agreement~\cite{CHONG20101192,liu2013multiparty,das2020comment,yang2022detector}, all of which leverage quantum entanglement as a crucial resource. One notable form of entanglement is the GHZ state (Greenberger-Horne-Zeilinger state)\cite{10.1119/1.16243}, which is widely used in numerous quantum protocols~\cite{PhysRevA.59.1829,PhysRevA.65.012308,christandl2005quantum,HILLERY200675,PhysRevLett.112.080801,komar2014quantum,PhysRevLett.123.070504}. To effectively utilize this entanglement, it must be shared among the users who are going to participate in the communication. In practice, users may spread over a large region. The delicate nature of quantum states makes long-distance entanglement routing particularly challenging.

To overcome these challenges, Hahn et al.~\cite{hahn2019quantum} used the notion of the graph states~\cite{PhysRevA.69.062311} in a quantum network to propose a protocol using only local operations for routing of $3$ and $4$-party GHZ states over a network. A graph state can be constructed by performing local operations in the nodes of the networks when they share a maximally entangled state with their neighbors. This only needs to share entanglement within a short distance. This method significantly reduces the need for long-distance entanglement routing, allowing for the distribution of large entangled states across extensive networks.

The number of users participating in the communication depends on the size of the shared GHZ state. To facilitate a large number of participants in a group chat, group video conferencing, or broadcasting, the size of the shared GHZ state must be larger. Mannalath and Pathak~\cite{PhysRevA.108.062614} enhanced the protocol described in Ref.~\cite{hahn2019quantum} to increase the number of users in the final shared state. However, for grid networks, the increased size mentioned in Ref.\cite{PhysRevA.108.062614} is not always compatible with the proposed protocol.

Both previous works considered configurations where the users of the final GHZ state are initially connected in a linear arrangement within the graph state. This makes us curious to think about \emph{whether we can further improve the size of the GHZ state if the users are initially connected in a tree structure.}

Our investigation into this question reveals that using a tree structure, as opposed to a linear configuration, can indeed enhance the size of the final GHZ state. Our analysis, which includes both balanced and unbalanced trees, demonstrates that these structures can support a larger number of users in the final GHZ state. Additionally, we have explored scenarios where nodes have multiple quantum memories and showed how these memories can be integrated into the GHZ state.

For grid networks, which are among the most practical, employing a tree structure significantly improves performance and efficiency. We proposed a method to extract GHZ states from a grid network that performed better than existing results. We have also shown that our method provides {\em optimal} results for any $n\times n$ grid. Finally, we suggest a method for selecting the repeater tree when entanglement needs to be shared among specific users.

\subsection{\label{sec:motiv}Motivation and Contribution}

Quantum communication using entangled quantum states requires entanglement sharing over a quantum network. Quantum states are extremely delicate, making it challenging to transmit or share them over long distances. This restricts us from performing long-distance communication over quantum networks. Considering the limitations of the quantum states, to perform long-distance quantum communication, we need some process for routing of entanglement that works with the sharing of quantum states within short-distance and some local operations. Entanglement swapping is one such process~\cite{PhysRevLett.71.4287}. In this process, a locally produced entangled state is shared by swapping protocol, which is very similar to the teleportation protocol~\cite{PhysRevLett.70.1895}. Although it works for point-to-point quantum communication, it may not be suitable for large and complex networks. Hahn et al.~\cite{hahn2019quantum} proposed a protocol using the structure of the network to share GHZ states. They used the notion of graph states, connecting a graph and a quantum state shared over a network, to produce the shared entanglement. Later Mannalath and Pathak~\cite{PhysRevA.108.062614} improved the above protocol to share a larger GHZ state over the network.

\begin{figure*}
\begin{tikzpicture}
\node[rectangle, draw, very thick, text width = 1cm, align = center] at (0, 0) (ms) {Main Server};
\node[rectangle, draw, thick, text width = 1cm, align = center] at (-7, -2) (rs1) {Server $R_1$};
\node[rectangle, draw, thick, text width = 1cm, align = center] at (-2, -2) (rs2) {Server $R_2$};
\node[rectangle, text width = 1cm, align = center] at (2, -2) {$\cdots$};
\node[rectangle, draw, thick, text width = 1cm, align = center] at (7, -2) (rsn) {Server $R_n$};
\node[rectangle, draw, text width = 1cm, align = center] at (-8, -4) (ls11) {Server $L_{11}$};
\node[rectangle, text width = 1cm, align = center] at (-6.5, -4) {$\cdots$};
\node[rectangle, draw, text width = 1cm, align = center] at (-5, -4) (ls1n) {Server $L_{1n}$};
\node[rectangle, draw, text width = 1cm, align = center] at (-2.5, -4) (ls21) {Server $L_{21}$};
\node[rectangle, text width = 1cm, align = center] at (-1, -4) {$\cdots$};
\node[rectangle, draw, text width = 1cm, align = center] at (0.5, -4) (ls2n) {Server $L_{2n}$};
\node[rectangle, text width = 1cm, align = center] at (2.25, -4) {$\cdots$};
\node[rectangle, draw, text width = 1cm, align = center] at (5, -4) (lsn1) {Server $L_{n1}$};
\node[rectangle, text width = 1cm, align = center] at (6.5, -4) {$\cdots$};
\node[rectangle, draw, text width = 1cm, align = center] at (8, -4) (lsnn) {Server $L_{nn}$};
\draw[thick] (ms.250) -- (rs1.north) (ms.south) -- (rs2.north) (ms.290) -- (rsn.north) (rs1.east) -- (rs2.west) (rs1.260) -- (ls11.north) (rs1.280) -- (ls1n.north) (rs2.260) -- (ls21.north) (rs2.280) -- (ls2n.north) (rsn.260) -- (lsn1.north) (rsn.280) -- (lsnn.north);
\node[circle, draw] at (-8.75, -7) (111) {};
\node[circle, draw] at (-8.5, -6) (112) {};
\node[circle, draw] at (-7.5, -7) (113) {};
\node[circle, draw] at (-7.25, -6) (114) {};
\node[circle, draw] at (-8, -6.5) (115) {};
\node[circle, draw] at (-5.75, -7) (1n1) {};
\node[circle, draw] at (-5.5, -6) (1n2) {};
\node[circle, draw] at (-4.5, -7) (1n3) {};
\node[circle, draw] at (-4.25, -6) (1n4) {};
\node[circle, draw] at (-5, -6.5) (1n5) {};
\node[circle, draw] at (-3.25, -7) (211) {};
\node[circle, draw] at (-3, -6) (212) {};
\node[circle, draw] at (-2, -7) (213) {};
\node[circle, draw] at (-1.75, -6) (214) {};
\node[circle, draw] at (-2.5, -6.5) (215) {};
\node[circle, draw] at (-0.25, -7) (2n1) {};
\node[circle, draw] at (0, -6) (2n2) {};
\node[circle, draw] at (1, -7) (2n3) {};
\node[circle, draw] at (1.25, -6) (2n4) {};
\node[circle, draw] at (0.5, -6.5) (2n5) {};
\node[circle, draw] at (4.25, -7) (n11) {};
\node[circle, draw] at (4.5, -6) (n12) {};
\node[circle, draw] at (5.5, -7) (n13) {};
\node[circle, draw] at (5.75, -6) (n14) {};
\node[circle, draw] at (5, -6.5) (n15) {};
\node[circle, draw] at (6.75, -7) (nn1) {};
\node[circle, draw] at (7, -6) (nn2) {};
\node[circle, draw] at (8, -7) (nn3) {};
\node[circle, draw] at (8.25, -6) (nn4) {};
\node[circle, draw] at (7.5, -6.5) (nn5) {};
\foreach \i in {1, 2, n}:
\foreach \j in {1, n}:
\draw (\i\j3) -- (\i\j4) -- (\i\j2);
\foreach \i in {1, 2, n}:
\foreach \j in {1, n}:
\foreach \k in {1, ..., 4}
\draw (\i\j5) -- (\i\j\k);
\foreach \i in {1, 2, n}:
\foreach \j in {1, n}:
\draw (\i\j2) -- ++(-0.3, 0) (\i\j2) -- ++(0.06, 0.3) (\i\j3) -- ++(0.3, 0) (\i\j3) -- ++(-0.06, -0.3) (\i\j4) -- ++(0.06, 0.3) (\i\j4) -- ++(0.3, 0) (ls\i\j.south) -- (\i\j2) (ls\i\j.south) -- (\i\j4) (ls\i\j.south) -- (\i\j5) (ls\i\j.south) -- (\i\j3);
\end{tikzpicture}
\caption{A graphical representation of a network. In general, a network can be thought of as a combination of different layers. The top layer consists of at least one main server connecting the whole network. The next layer has a few regional servers $R_i$ providing service to a big region that may be covering one or multiple states. The following layer may have some local servers $L_j$ providing service within a city. Finally, the last layer consists of users. A server may be connected with another nearby server. This graphical representation has a hidden tree structure with the main server as root. Therefore, this tree structure can be used to route entanglement.}
\label{fig:network}
\end{figure*}
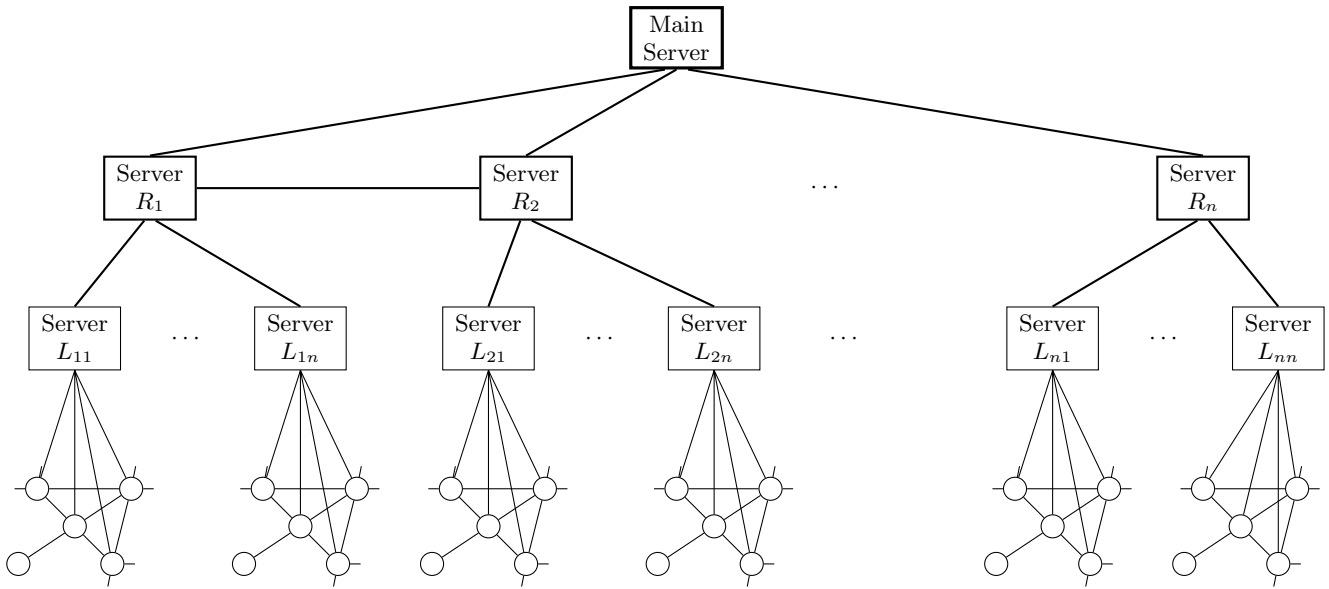

Any network can be represented using a graph, where the vertices represent nodes in the network and the edges denote the connections between the corresponding nodes. Similarly, in a quantum network, users and servers can be defined as vertices, while the connections between users, or between users and servers, are represented as edges. This creates a graph that models the network's structure. One example of such a graph is given in Fig.~\ref{fig:network}. Considering the main server as the root, a tree structure is hidden in this graph, where the users are the leaves and the regional and local servers are the intermediate nodes. This motivates us to think about a tree that connects the users to route the shared GHZ state.

As the definition of $X$ measurement consists of a sequence of 4 operations, it may not be clear to everyone the actual effect of it on a graph. Therefore, we start with a simple result in Theorem~\ref{th:merge} to show how $X$ measurement affects a graph. Then as our main contribution, we introduce the concept of \emph{repeater tree} to extract a GHZ state from it as Theorem~\ref{th:extract}. We also show that a tree structure is necessary to generate GHZ state in Theorem~\ref{th:opt}. We also considered multimemory nodes when constructing GHZ states. For a grid network, we propose an algorithm to produce a repeater tree which along with Theorem~\ref{th:extract} would construct a GHZ state. The state that our proposed algorithm produces is a graph state corresponding to a star graph. The application of a local complement on the center would transform this graph into a complete graph. This complete graph may contain some of the servers that should not be part of the final GHZ state. However, if any of the servers mentioned above are in the final complete graph, can be removed by performing $Z$ measurements on the server nodes. Note that, after such $Z$ measurements, the remaining graph is still complete. From this graph state, the required GHZ state can be generated by some local operation. Our proposed algorithm using tree structures provides larger GHZ states than both of the proposals mentioned above. Also, our algorithm for the grid networks provides optimal GHZ states.

Preliminary discussions on the tools that we use throughout the article are mentioned in Section~\ref{sec:prelim}. We revisit the work of Mannalath and Pathak~\cite{PhysRevA.108.062614} and produce the corrected size of the GHZ state for the grid network according to their proposed protocol in Section~\ref{sec:prev}. In the next two sections, we discuss our proposal of using the tree structure connecting the users. The discussion in Section~\ref{sec:GHZ} consists of the general graph network and GHZ state routing over that graph. This section also contains a discussion about GHZ state routing over multi-memory nodes in the network. In Section~\ref{sec:grid}, We consider the grid network to share GHZ state. Finally, in Section~\ref{sec:concl}, we conclude our work.

\section{\label{sec:prelim}Preliminaries}

An undirected simple finite graph $G=(V,E)$ is defined by a set $V\subset\mathbb N$ of vertices and a set of edges $E\subseteq V\times V$ such that $(i,i)\notin E$ for all $i\in V$. For any vertex $a$, the set $N^G_a = \{v\in V:(a,v)\in E\}$ is called the neighbourhood of $a$ in graph $G$.

\begin{defination}[Vertex Deletion]
Deletion of a vertex $v$ from a graph $G=(V,E)$ is a graph operation that removes the vertex $v$ and all the edges adjacent to $v$. The resulting graph becomes,
\[
G-v=(V\backslash v,\{(i,j)\in E:i\neq v\neq j\}).
\]
\end{defination}

\begin{defination}[Local Complementation]
A local complementation $LC_v$, at vertex $v$, is a graph operation that takes a graph $G=(V,E)$ to $LC_v(G)=(V,E\Delta K_{N^G_v})$, where $K_{N^G_v}$ is the set of edges of the complete graph on the vertex set $N^G_v$, and $E\Delta K_{N^G_v}$ is the symmetric difference of two sets $E$ and $K_{N^G_v}$.
\end{defination}

\begin{defination}[Vertex-minor]
A vertex-minor of a graph $G$ is another graph $H$ which can be found by a sequence of vertex deletions and local complementation.
\end{defination}

\begin{defination}[Graph state]A graph state~\cite{PhysRevA.69.062311} $\ket{G}$ is a pure quantum state associated with a graph $G=(V,E)$ and defined as
\begin{equation}
\label{graph_state}
\ket{G}:=\prod_{(i,j)\in E}CZ_{i,j}\ket{+}^{\otimes V}.
\end{equation}
\end{defination}

The local complementations on graph $G$, defined above, are local Clifford operations on graph state $\ket{G}$~\cite{PhysRevA.69.022316}. Local Pauli measurements on the graph states can be realized by vertex deletions and local complementations~\cite{PhysRevA.69.062311,PhysRevA.108.062614}. Table~\ref{tab:Pauli_meas} shows the correspondence between Pauli measurement and the graph operations.

\begin{table}
\centering
\begin{tabular}{|c|c|}
\hline
\textbf{Pauli measurement}&\textbf{Realization in Graph}\\
\hline
\hline
$Z$ measurement&$Z_v(G)=G-v$\\
$Y$ measurement&$Y_v(G)=Z_vLC_v(G)$\\
$X$ measurement&$X_v(G)=LC_uZ_vLC_vLC_u(G)$\\
\hline
\end{tabular}
\caption{Pauli measurement on the graph state $\ket{G}$ and the corresponding graph operation on vertex $v$ in graph $G$. Vertex $v$ corresponds to the qubit to measure. $u$ is one vertex in the neighborhood of $v$.}
\label{tab:Pauli_meas}
\end{table}

A graph $G=(V,E)$ is called a star graph if there is a vertex $a\in V$ such that $(a,v)\in E$ for all $v\in V\backslash a$ and $(v_1,v_2)\notin E$ if $v_1\neq a\neq v_2$. $a$ is called the center of the graph. Let $K_V$ be a complete graph with vertex set $V$. Then local complementation on any vertex $v\in V$ results in a star graph with center $v$. Also, local complementation on the center of the star graph results in a complete graph. Application of Hadamard gates on the qubits corresponding to the vertex other than the center of a star graph makes the graph state a GHZ state~\cite{PhysRevA.69.022316,PhysRevA.100.052333}.

\section{\label{sec:prev} A Flaw in Previous Work and Our Correction}

Mannalath and Pathak~\cite{PhysRevA.108.062614} improved Hahn et al.'s $X$ protocol~\cite{hahn2019quantum} to extract a larger GHZ state from a given graph state. They showed that if the underlying graph has a path connecting $2n-3$ nodes, labeled as $1,2,\dots,2n-3$, an $n$-party GHZ state can be extracted from this which contains all nodes with an even label including two nodes at each end. Such a path is called \emph{repeater line}. In this protocol, after isolating the path from the graph by $Z$ measurement, $X$ measurements have been performed on all the odd-labeled nodes except the first and the last nodes. This protocol is called \emph{generalized $X$ protocol}. They also showed that if the isolation happens after the $X$ measurements, the same GHZ state can be produced with at most the same number of Pauli measurements. They applied their protocol in $n\times n$ grid network and conjectured a bound for the largest GHZ state as $L_M=\lfloor\frac{n+1}{4}\rfloor(3\lfloor\frac{n-1}{2}\rfloor+4)-2$. This bound improves the previous bound $L_B=\lceil n/2\rceil^2$ conjectured by Briegel and Raussendorf~\cite{PhysRevLett.86.910}.

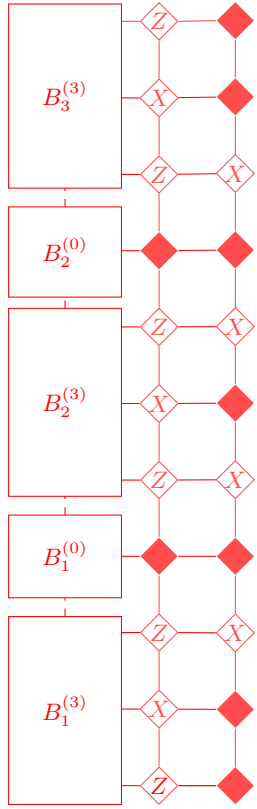
\begin{figure}
\begin{tikzpicture}
\node[red, diamond, draw = red!70, inner sep = 0pt, minimum size = 5mm] at (0, 0) (00) {$Z$};
\draw[red!70] (00.east) -- ++(0.5, 0) node[diamond, anchor = west, fill = red!70, minimum size = 5mm] (10) {} (00.north) -- ++(0, 0.5) node[diamond, anchor = south, draw = red!70, inner sep = 0pt, minimum size = 5mm] (01) {$X$} (10.north) -- ++(0, 0.5) node[diamond, anchor = south, fill = red!70, minimum size = 5mm] (11) {} (01.north) -- ++(0, 0.5) node[diamond, anchor = south, draw = red!70, inner sep = 0pt, minimum size = 5mm] (02) {$Z$} (11.north) -- ++(0, 0.5) node[diamond, anchor = south, draw = red!70, inner sep = 0pt, minimum size = 5mm] (12) {$X$} (02.north) -- ++(0, 0.5) node[diamond, anchor = south, fill = red!70, minimum size = 5mm] (03) {} (12.north) -- ++(0, 0.5) node[diamond, anchor = south, fill = red!70, minimum size = 5mm] (13) {} (03.north) -- ++(0, 0.5) node[diamond, anchor = south, draw = red!70, inner sep = 0pt, minimum size = 5mm] (04) {$Z$} (13.north) -- ++(0, 0.5) node[diamond, anchor = south, draw = red!70, inner sep = 0pt, minimum size = 5mm] (14) {$X$} (04.north) -- ++(0, 0.5) node[diamond, anchor = south, draw = red!70, inner sep = 0pt, minimum size = 5mm] (05) {$X$} (14.north) -- ++(0, 0.5) node[diamond, anchor = south, fill = red!70, minimum size = 5mm] (15) {} (05.north) -- ++(0, 0.5) node[diamond, anchor = south, draw = red!70, inner sep = 0pt, minimum size = 5mm] (06) {$Z$} (15.north) -- ++(0, 0.5) node[diamond, anchor = south, draw = red!70, inner sep = 0pt, minimum size = 5mm] (16) {$X$} (06.north) -- ++(0, 0.5) node[diamond, anchor = south, fill = red!70, minimum size = 5mm] (07) {} (16.north) -- ++(0, 0.5) node[diamond, anchor = south, fill = red!70, minimum size = 5mm] (17) {} (07.north) -- ++(0, 0.5) node[diamond, anchor = south, draw = red!70, inner sep = 0pt, minimum size = 5mm] (08) {$Z$} (17.north) -- ++(0, 0.5) node[diamond, anchor = south, draw = red!70, inner sep = 0pt, minimum size = 5mm] (18) {$X$} (08.north) -- ++(0, 0.5) node[diamond, anchor = south, draw = red!70, inner sep = 0pt, minimum size = 5mm] (09) {$X$} (18.north) -- ++(0, 0.5) node[diamond, anchor = south, fill = red!70, minimum size = 5mm] (19) {} (09.north) -- ++(0, 0.5) node[diamond, anchor = south, draw = red!70, inner sep = 0pt, minimum size = 5mm] (88) {$Z$} (19.north) -- ++(0, 0.5) node[diamond, anchor = south, fill = red!70, minimum size = 5mm] (99) {} (88.east) -- (99.west) (88) -- ++(-0.5, 0);
\foreach \y in {0,...,9}
\draw[red] (0\y) -- (1\y) (0\y) -- ++(-0.5, 0);
\draw[red] (-2, -0.25) rectangle (-0.5, 2.25) node[pos = 0.5] {$B^{(3)}_1$} (-2, 2.5) rectangle (-0.5, 3.6) node[pos = 0.5] {$B^{(0)}_1$} (-2, 3.85) rectangle (-0.5, 6.35) node[pos = 0.5] {$B^{(3)}_2$} (-2, 6.5) rectangle (-0.5, 7.7) node[pos = 0.5] {$B^{(0)}_2$} (-2, 7.95) rectangle (-0.5, 10.4) node[pos = 0.5] {$B^{(3)}_3$};
\draw[dashed, red] (-1.25, 2.25) -- (-1.25, 2.5) (-1.25, 3.6) -- (-1.25, 3.85) (-1.25, 6.35) -- (-1.25, 6.5) (-1.25, 7.7) -- (-1.25, 7.95);
\end{tikzpicture}
\caption{Reconstructing GHZ state using generalized $X$ protocol proposed by Mannalath and Pathak~\cite{PhysRevA.108.062614}. Only the last two columns from a $11\times 11$ grid are shown here. The blocks $B_i^{(3)}$ and $B_j^{(0)}$ contains the remaining $9$ columns constructed as in Ref.~\cite[Fig.9(b)]{PhysRevA.108.062614}. Dotted lines have been used to denote the edges between vertices from two different blocks. There are a total of $45$ vertices from the final GHZ state inside the blocks. The last two columns cannot have more than $9$ vertices in the final GHZ state as shown in this figure. This creates a $54$-party GHZ state. However, the conjecture in Ref.~\cite{PhysRevA.108.062614} claims it as $55$-party GHZ.}
\label{fig:reconst}
\end{figure}

However, the bound given by Mannalath et al. is not compatible with their construction~\cite[Fig.9(b)]{PhysRevA.108.062614} of a repeater line to perform generalized $X$ protocol when $n=4k+3$ with $k\geq2$. For $k=2$, we try to reconstruct the repeater line in Fig~\ref{fig:reconst} with their construction. Here we have drawn only the last two columns of the grid, and the blocks $B^{(3)}_i$ and $B^{(0)}_j$ denote the remaining parts of the grid constructed with the construction given in Ref.~\cite[Fig.~9(b)]{PhysRevA.108.062614}. There are total $y=\lfloor\frac{n+1}{4}\rfloor=3$ number of $B^{(3)}_i$ blocks each containing $3x=3\lfloor\frac{n-1}{2}\rfloor=15$ nodes from final GHZ states. On the other hand, the numbers of $B^{(0)}_j$ blocks is $y-1=2$, and they contain no node from the final GHZ state. Therefore these five blocks contain a total of $45$ nodes from the final GHZ state. Now consider the last two columns from Fig.~\ref{fig:reconst} constructed using the instruction given in Ref.~\cite[Fig.~9(b)]{PhysRevA.108.062614}. It contains $9$ nodes (filled nodes) from the final GHZ states indicating the size of the GHZ state as $54$. However, as mentioned in Ref.~\cite{PhysRevA.108.062614}, this size should be $55$. If any node, labeled as $Z$ or $X$, is considered as a part of the final GHZ state, it violates the condition ``the underlying graph has a repeater line as vertex-minor, connecting all $n$ nodes of the final GHZ state with an extra node in between every pair of $n-2$ intermediate nodes" for \emph{Theorem III.1} of Ref.~\cite{PhysRevA.108.062614}. Therefore, the size of the final GHZ state cannot be $55$ with this construction. The reason behind this is that only one node from the last column on the left of the block $B^{(3)}_2$ comes inside the final GHZ state, which was considered as $2$ in Ref.~\cite{PhysRevA.108.062614}. A similar argument holds for $k>2$ also. Thus, we get one node in the final GHZ state from the three nodes in the last column on the left of all intermediate $B_i^{(3)}$ blocks. Since, there are $y-2$ such blocks, the corrected bound is given by $3xy+(2y-(y-2))+2(y-1)=3y(x+1)=3\lfloor\frac{n+1}{4}\rfloor\lfloor\frac{n+1}{2}\rfloor$.

\section{\label{sec:GHZ}GHZ State routing over Arbitrary Network}

Let us consider a quantum network where a graph state is shared between multiple users. This graph state corresponds to some graph where each node (vertex of the graph) corresponds to one user in the network. Our task is to build an entangled GHZ state shared between multiple users so that they can use that state to perform their quantum protocol. The number of quantum memory required by a user may vary. To build a GHZ state from a graph state, the $X$ protocol conducts a sequence of $Z$ and $X$ measurements on the graph state. From Table~\ref{tab:Pauli_meas}, it is clear that a $Z$ measurement on a graph is nothing but deleting the corresponding node. However, the impact of the $X$ measurement is not straightforward from the definition. Considering each node has single-qubit memory, we first see the effect of $X$ measurement in the following theorem.

\begin{theorem}[$X$ measurement]\label{th:merge}
Suppose $G=(V,E)$ be a graph, and $G_1=(V_1,E_1),G_2=(V_2,E_2)$ be two non-intersecting subgraphs of $G$ such that $\{u\}=V-(V_1\cup V_2)$ and $u$ have only two vertices $v_i\in V_i,i=0,1$ in its neighborhood. Also consider that, there is no edge between $G_1$ and $G_2$, that is, $E=E_1\cup E_2\cup\{(v_1,u),(u,v_2)\}$. Then, without loss of generality, $X_u(G)=(V\backslash u,E_1\cup E')$, where $E'=\{(v_1,v):v\in \{v_2\}\cup N^{G_2}_{v_2}\}\cup E_2-\{(v_2,v):v\in N^{G_2}_{v_2}\}$.
\end{theorem}

\begin{proof}
By definition, without loss of generality, we can write,
\begin{equation}
\label{eq:x_meas}
X_{u}(G)=LC_{v_2}Z_uLC_uLC_{v_2}(G)
\end{equation}
Let $E^G_A$ be the set of all edges between the vertices $A$ in graph G. Also, let $G'$ be the subgraph of $G_2$, containing $N^{G_2}_{v_2}$ as vertices and the corresponding edges, that is, $G'=(N^{G_2}_{v_2},E^{G_2}_{N^{G_2}_{v_2}})$. Then we can write $E$ as a union of disjoint sets,
\begin{eqnarray*}
E&=&E_1\cup E^{G_2}_{N^{G_2}_{V_2}}\cup E^{G_2}_{V_2-N^{G_2}_{V_2}}\\
&&\cup\{(u,v_1),(u,v_2)\}\cup\{(v_2,v):v\in N^{G_2}_{v_2}\}\\
&&\cup\left(E_2\cap\{(u_1,u_2):u_1\in N^{G_2}_{v_2},u_2\in V_2\backslash v_2-N^{G_2}_{v_2}\}\right).
\end{eqnarray*}

Since, $N^G_{v_2}=N^{G_2}_{v_2}\cup\{u\}$, after the operation $LC_{v_2}$, neighbourhood of $u$ becomes $N_u=N^{G_2}_{v_2}\cup\{v_1, v_2\}$, and the second set in the above disjoint union becomes $E^{G_2}_{N^{G_2}_{v_2}}\Delta K_{N^{G_2}_{v_2}}$ keeping other sets unchanged. Therefore the current set of edges is given by
\begin{eqnarray}
&&E_1\cup E^{G_2}_{N^{G_2}_{v_2}}\Delta K_{N^{G_2}_{v_2}}\cup E^{G_2}_{V_2-N^{G_2}_{V_2}}\nonumber\\
&&\cup\{(u,v_1),(u,v_2)\}\cup\{(v_2,v),(u,v):v\in N^{G_2}_{v_2}\}\nonumber\\
&&\cup\left(E_2\cap\{(u_1,u_2):u_1\in N^{G_2}_{v_2},u_2\in V_2\backslash v_2-N^{G_2}_{v_2}\}\right).\label{eq:edge_1}
\end{eqnarray}

Now, the operation $LC_u$ will again change the second set in~\eqref{eq:edge_1} to $E^{G_2}_{N^{G_2}_{v_2}}$, and establish edges between $v_1$ and $N^{G_2}_{v_2}\cup\{v_2\}$ removing edges between $v_2$ and $N^{G_2}_{v_2}$. Thus, after the operation $Z_u$, the new set of edges becomes
\begin{eqnarray}
&&E_1\cup E^{G_2}_{N^{G_2}_{v_2}}\cup E^{G_2}_{V_2-N^{G_2}_{V_2}}\cup\{(v_1,v):v\in N^{G_2}_{v_2}\cup\{v_2\}\}\nonumber\\
&&\cup\left(E_2\cap\{(u_1,u_2):u_1\in N^{G_2}_{v_2},u_2\in V_2\backslash v_2-N^{G_2}_{v_2}\}\right).\label{eq:edge_2}
\end{eqnarray}

At this point, $v_2$ has only one neighbor $v_1$. Therefore, the operation $LC_{v_2}$ has no effect. Therefore,~\eqref {eq:edge_2} gives the new set of edges and the new set of vertices is $V\backslash u$. Note that, the union of second, third and fourth sets in~\eqref{eq:edge_2} is the same as $E_2-\{(v_2,v):v\in N^{G_2}_{v_2}\}$.\\
This completes the proof.
\end{proof}

\begin{figure*}
\begin{tikzpicture}
\node[circle, fill = red!50, minimum size = 7mm] at (0,0) (v1) {$v_1$};
\draw[dashed, thick, black!50] (v1.north) --++(0, .5);
\draw[thick] (v1.south) --++(0, -0.5) node[circle, fill = green!50, anchor = north, minimum size = 7mm] (u) {$u$} (u.south) --++(0, -0.5) node[circle, fill = red!50, anchor = north, minimum size = 7mm] (v2) {$v_2$} (v2.south east) --++(1, -1) node[circle, fill = red!50, anchor = north west, minimum size = 7mm] (v3) {} (v2.south west) --++(-1, -1) node[circle, fill = red!50, anchor = north east, minimum size = 7mm] (v4) {} (v2.south) --++(0.5, -2) node[circle, fill = red!50, anchor = north, minimum size = 7mm] (v5) {} (v2.south) --++(-0.5, -2) node[circle, fill = red!50, anchor = north, minimum size = 7mm] (v6) {};
\draw[dashed, thick, black!50] (v3.south east) --++(0.35, -0.35) (v4.south west) --++(-0.35, -0.35) (v5.south) --++(0, -0.5) (v6.south) --++(0, -0.5);
\draw[thick] (v3.west) -- (v4.east) (v4.south east) -- (v6.north west) (v6.east) -- (v5.west);
\node at (0, -6.5) {(a)};
\end{tikzpicture}
\hspace{.5cm}
\begin{tikzpicture}
\node[circle, fill = red!50, minimum size = 7mm] at (0,0) (v1) {$v_1$};
\draw[dashed, thick, black!50] (v1.north) --++(0, .5);
\draw[thick] (v1.south) --++(0, -0.5) node[circle, fill = green!50, anchor = north, minimum size = 7mm] (u) {$u$} (u.south) --++(0, -0.5) node[circle, fill = red!50, anchor = north, minimum size = 7mm] (v2) {$v_2$} (v2.south east) --++(1, -1) node[circle, fill = red!50, anchor = north west, minimum size = 7mm] (v3) {} (v2.south west) --++(-1, -1) node[circle, fill = red!50, anchor = north east, minimum size = 7mm] (v4) {} (v2.south) --++(0.5, -2) node[circle, fill = red!50, anchor = north, minimum size = 7mm] (v5) {} (v2.south) --++(-0.5, -2) node[circle, fill = red!50, anchor = north, minimum size = 7mm] (v6) {};
\draw[dashed, thick, black!50] (v3.south east) --++(0.35, -0.35) (v4.south west) --++(-0.35, -0.35) (v5.south) --++(0, -0.5) (v6.south) --++(0, -0.5);
\draw[dashed, thick, red] (v3.west) -- (v4.east) (v4.south east) -- (v6.north west) (v6.east) -- (v5.west);
\draw[thick, green] (v3.south west) -- (v5.north east) (v3.west) -- (v6.north east) (v4.east) -- (v5.north west) (u.south east) -- (v3.north) (u.south west) -- (v4.north) (u.south) to[bend left=25] (v5.north) (u.south) to[bend right=25] (v6.north);
\node at (0, -6.5) {(b)};
\end{tikzpicture}
\hspace{.5cm}
\begin{tikzpicture}
\node[circle, fill = red!50, minimum size = 7mm] at (0,0) (v1) {$v_1$};
\draw[dashed, thick, black!50] (v1.east) --++(0.5, 0);
\draw[thick] (v1.west) --++(-1, 0) node[circle, fill = green!50, anchor = east, minimum size = 7mm] (u) {$u$} (u.south) --++(0, -1.5) node[circle, fill = red!50, anchor = north, minimum size = 7mm] (v2) {$v_2$};
\draw[dashed, thick, red] (v2.south east) --++(1.5, -1.5) node[circle, fill = red!50, anchor = north west, minimum size = 7mm] (v3) {} (v2.south west) --++(-1.5, -1.5) node[circle, fill = red!50, anchor = north east, minimum size = 7mm] (v4) {} (v2.south) --++(1, -3) node[circle, fill = red!50, anchor = north, minimum size = 7mm] (v5) {} (v2.south) --++(-1, -3) node[circle, fill = red!50, anchor = north, minimum size = 7mm] (v6) {};
\draw[dashed, thick, black!50] (v3.south east) --++(0.35, -0.35) (v4.south west) --++(-0.35, -0.35) (v5.south) --++(0, -0.5) (v6.south) --++(0, -0.5);
\draw[thick, green] (v3.west) -- (v4.east) (v4.south east) -- (v6.north west) (v6.east) -- (v5.west);
\draw[dashed, thick, red] (v3.south west) -- (v5.north east) (v3.west) -- (v6.north east) (v4.east) -- (v5.north west);
\draw[thick] (u.south east) -- (v3.north) (u.south west) -- (v4.north) (u.south) to[bend left=25] (v5.north) (u.south) to[bend right=25] (v6.north);
\draw[thick, green] (v1.south west) -- (v2.north east) (v1.south west) to[bend right=25] (v4.north) (v1.south) -- (v3.north) (v1.south) -- (v5.north) (v1.south) -- (v6.north);
\node[at = (v2), below = 4.5cm] {(c)};
\end{tikzpicture}\\
\vspace{.2cm}
\begin{tikzpicture}
\node[circle, fill = red!50, minimum size = 7mm] at (0,0) (v1) {$v_1$};
\draw[dashed, thick, black!50] (v1.north east) --++(0.35, 0.35);
\draw[thick] (v1.north) --++(0, 0.5) node[circle, fill = red!50, anchor = south, minimum size = 7mm] (v2) {$v_2$};
\draw[thick] (v1.south east) --++(0.75, -0.75) node[circle, fill = red!50, anchor = north west, minimum size = 7mm] (v3) {} (v1.south west) --++(-0.75, -0.75) node[circle, fill = red!50, anchor = north east, minimum size = 7mm] (v4) {} (v1.south) --++(0.5, -1.5) node[circle, fill = red!50, anchor = north, minimum size = 7mm] (v5) {} (v1.south) --++(-0.5, -1.5) node[circle, fill = red!50, anchor = north, minimum size = 7mm] (v6) {};
\draw[dashed, thick, black!50] (v3.south east) --++(0.35, -0.35) (v4.south west) --++(-0.35, -0.35) (v5.south) --++(0, -0.5) (v6.south) --++(0, -0.5);
\draw[thick] (v3.west) -- (v4.east) (v4.south east) -- (v6.north west) (v6.east) -- (v5.west);
\node[at = (v1), below = 3.2cm] {(d)};
\end{tikzpicture}
\hspace{.5cm}
\begin{tikzpicture}
\node[circle, fill = green!50, minimum size = 7mm] at (0,0) (u) {$u$};
\draw[thick] (u.150) --++ (-0.43, 0.25) node[circle, fill = red!50, anchor = 330, minimum size = 7mm] (v1) {$v_1$} (u.30) --++ (0.43, 0.25) node[circle, fill = red!50, anchor = 210, minimum size = 7mm] (v2) {$v_2$} (u.south) --++ (0, -0.5) node[circle, fill = red!50, anchor = north, minimum size = 7mm] (v3) {$v_3$};
\draw[dashed, thick] (v1.150) --++ (-0.43, 0.25) node[rectangle, fill = red!50, anchor = 330, minimum size = 7mm] {$G_1$} (v2.30) --++ (0.43, 0.25) node[rectangle, fill = red!50, anchor = 210, minimum size = 7mm] {$G_2$} (v3.south) --++ (0, -0.5) node[rectangle, fill = red!50, anchor = north, minimum size = 7mm] {$G_3$};
\draw[->, thick] (1, -0.5) --++ (1.5, 0);
\node[circle, fill = red!50, minimum size = 7mm] at (3.5, -0.5) (v2') {$v_2$};
\node[rectangle, fill = red!50, minimum size = 7mm] at (5.5, -0.5) (G2) {$G_2$};
\draw[thick] (v2'.north east) --++ (0.35, 0.35) node[circle, fill = red!50, anchor = south west, minimum size = 7mm] (v1') {$v_1$} (v2'.south east) --++ (0.35, -0.35) node[circle, fill = red!50, anchor = north west, minimum size = 7mm] (v3') {$v_3$};
\draw[dashed, thick] (v1'.north) --++ (0, 0.5) node[rectangle, fill = red!50, anchor = south, minimum size = 7mm] {$G_1$} (v1'.south east) -- (G2.north west) (v3'.south) --++ (0, -0.5) node[rectangle, fill = red!50, anchor = north, minimum size = 7mm] {$G_3$} (v3'.north east) -- (G2.south west);
\node at (1.75, -3) {(e)};
\end{tikzpicture}
\caption{\label{fig:merge} Application of $X$ measurement on vertex $u$. (a) A part of graph $G$, that will change due to $X$ measurement. The remaining part (omitted with the dashed line) of the graph will remain unchanged. (b) Local complement on $v_2$. Red edges (dashed lines) have been deleted and green edges have been added. (c) Local complement on $u$. (d) $Z$ measurement on $u$. Since $v_2$ has only one neighbor $v_1$, this is the final state. (e) Effect of $X$ measurement on vertex $u$ (degree $3$) of an arbitrary graph. $G_i(i=1,2,3)$ in squares denotes the rest of the graph. The neighbors of $v_3$ in $G_2$ are the same as the neighbors of $v_1$ in $G_2$.}
\end{figure*}
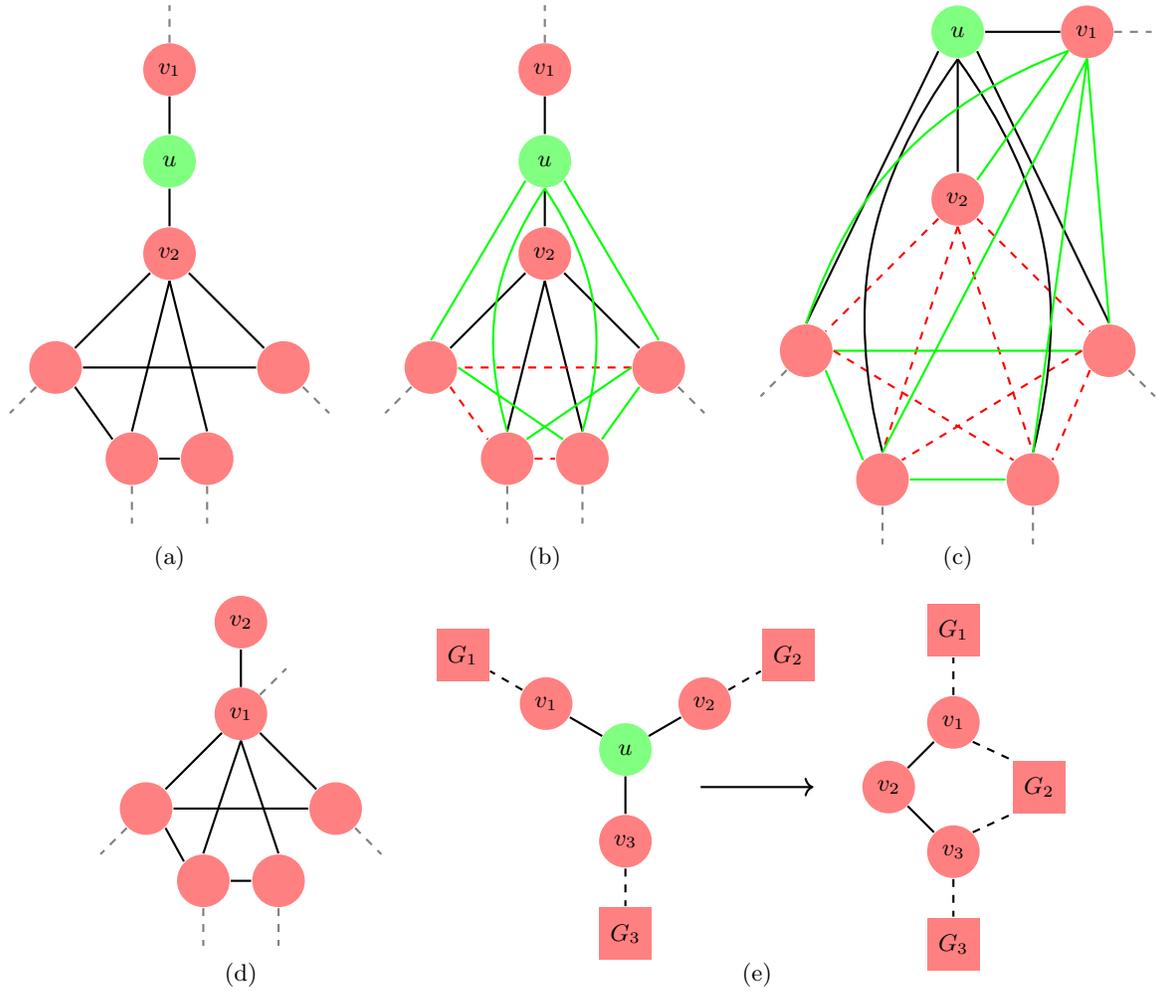

A visual representation of Theorem~\ref{th:merge} is given in Fig.~\ref{fig:merge}(a-d). This theorem can be generalized where the vertex $u$ has more than two neighbors. Let $v_3$ be the third neighbor and $G_3$ be the remaining subgraph connected through the vertex $v_3$. This generalized version is shown in Fig.~\ref{fig:merge}(e). This shows that performing a $Z$ measurement on $v_3$ followed by an $X$ measurement on $u$ gives the same graph found by an $X$ measurement on $u$ followed by a $Z$ measurement on $v_3$. Mannalath and Pathak~\cite[Lemma III.4]{PhysRevA.108.062614} showed that the latter one requires at most many measurements as the first one.

\begin{corollary}\label{cr:merge}
If $G_2=(V_2, E_2)$ is a star graph with $v_2$ as its center, then the final graph will be given by, $G_1\cup G'$, where $G'$ is a star graph with center $v_1$ and $V_2$ as other vertices.
\end{corollary}

\begin{proof}
Note that, in Theorem~\ref{th:merge}, $X$ measurement makes no change in $G_1$. And also, since $G_2$ is a star graph, $X$ measurement establishes edges between $v_1$ and $N^{G_2}_{v_2}\cup\{v_2\}$, removes edges between $v_2$ and $N^{G_2}_{v_2}$. Hence the result.
\end{proof}

Now, in Corollary~\ref{cr:merge}, if $G_1$ is also a star graph, then $G_1\cup G'$ is clearly a star graph with $v_1$ as its center. Also, the total number of vertices is $m+n$. Thus we can state our next corollary.

\begin{corollary}\label{cr:merge_star}
If $G_1$ and $G_2$ are both star graphs, with $m$ and $n$ vertices, then the final graph will be a new star graph with $m+n$ vertices.
\end{corollary}

Since, the graph state of a star graph corresponds to a GHZ state, using the above corollary, two or more GHZ states can be merged to get a larger GHZ state, when vertices like $u$, mentioned in theorem~\ref{th:merge}, exist.

A vertex that corresponds to a qubit in the final GHZ state will be called \textit{$g$-type} vertex, and other vertices (that may help to extract the GHZ state) will be called \textit{$h$-type} vertex.
\begin{defination}[Repeater Tree]
A tree is a repeater tree if it satisfies
\begin{enumerate}
\item the root and all the leaves are $g$-type,
\item all root-to-leaf path has the form $g$-$h$-$\cdots$-$h$-$g$-$g$, where $g$ and $h$ denotes $g$-type and $h$-type vertices respectively,
\item $h$-type vertices have exactly one child, and
\item subtree rooting at any $g$-type vertex, except leaf, is also a repeater tree.
\end{enumerate}
\end{defination}

Note that, any leaf-to-leaf path in a repeater tree is a repeater line mentioned in Section~\ref{sec:prev}.

\begin{theorem}[Extraction of GHZ state with $n$-qubits]\label{th:extract}
An $n$-partite GHZ state can be extracted from a graph state $\ket{G}$ when the underlying graph has a repeater tree as vertex-minor.

Given such a repeater tree, by performing $X$ measurement on $h$-type vertices, the GHZ state can be extracted.
\end{theorem}

\begin{proof}
Since, the repeater tree, say $R$, is a vertex-minor of the graph $G$, we can assume that the repeater tree is isolated with $g_0$ as its root. Suppose, $g_0$-$h_1$-$g_1$-$\cdots$-$h_{k-1}$-$g_{k-1}$-$g_k$ be a root-to-leaf path in the repeater tree with $k+1$ $g$-type and $k-1$ $h$-type vertices. Let us consider, $G_1=(V_1=\{g_0,h_1,g_1,\cdots,h_{k-2},g_{k-2}\},E_2=E^R_{V_1}), G_2=(V_2=\{g_{k-1},g_k\},E_2=\{(g_{k-1},g_k)\})$ and $u=h_{k-1}$. Here $G_2$ is a star graph with $2$ vertices. Then, using Corollary~\ref{cr:merge}, the resulting graph after $X$ measurement on $h_{k-1}$ can be written as $G_1\cup G'$, where $G'$ is a star graph with center $g_{k-2}$. Repeating $X$ measurement on $h_{k-2},\cdots,h_1$, in this order, the root-to-leaf path will be transformed into a star graph with center $g_0$ and $k$ vertices. Repeating this for all root-to-leaf paths, finally, we will have a star graph with $n$ vertices centered at $g_0$ as stated in Corollary~\ref{cr:merge_star}. Then applying the Hadamard gate as required, we can get the GHZ state.
\end{proof}

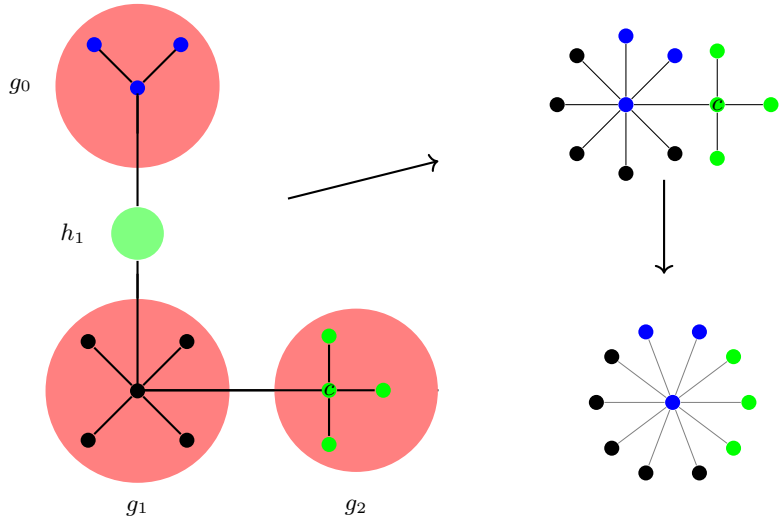
\begin{figure*}
\begin{tikzpicture}
\node[circle, fill = red!50, minimum size = 7mm] at (0,0) (root) {
\begin{tikzpicture}
\node[circle, fill = blue, minimum size = 2mm, inner sep = 0pt] at (0, 0) (c) {};
\draw[thick] (c.north east) --++(0.5, 0.5) node[circle, fill = blue, minimum size = 2mm, inner sep = 0pt] {} (c.north west) --++(-0.5, 0.5) node[circle, fill = blue, minimum size = 2mm, inner sep = 0pt] {} (c.south) --++(0, -0.5);
\end{tikzpicture}};
\node[at = (root), left = 1.2cm, minimum size = 7mm] {$g_0$};
\draw[thick] (0,-0.1) --++(0, -1.5) node[circle, fill = green!50, minimum size = 7mm, anchor = north] (h) {} (h.south) --++(0, -0.5) node[circle, fill = red!50, minimum size = 7mm, anchor =  north] (g) {
\begin{tikzpicture}
\node[circle, fill = black, minimum size = 2mm, inner sep = 0pt] at (0, 0) (c) {};
\draw[thick] (c.south west) --++(-0.5, -0.5) node[circle, fill = black, minimum size = 2mm, inner sep = 0pt, anchor = north east] {} (c.north west) --++(-0.5, 0.5) node[circle, fill = black, minimum size = 2mm, inner sep = 0pt, anchor = south east] {} (c.north east) --++(0.5, 0.5) node[circle, fill = black, minimum size = 2mm, inner sep = 0pt, anchor = south west] () {} (c.south east) --++(0.5, -0.5) node[circle, fill = black, minimum size = 2mm, inner sep = 0pt, anchor = north west] () {};
\end{tikzpicture}};
\draw[thick] (0, -4.05) --++(4, 0) node[circle, fill = red!50, minimum size = 7mm, anchor =  east] (leaf) {
\begin{tikzpicture}
\node[circle, fill = green, minimum size = 2mm, inner sep = 0pt] at (0, 0) (c) {$c$};
\draw[thick] (c.south) --++(0, -0.5) node[circle, fill = green, minimum size = 2mm, inner sep = 0pt, anchor = north] {} (c.east) --++(0.5, 0) node[circle, fill = green, minimum size = 2mm, inner sep = 0pt, anchor = west] {} (c.north) --++(0, 0.5) node[circle, fill = green, minimum size = 2mm, inner sep = 0pt, anchor = south] () {};
\end{tikzpicture}};
\node[at = (h), left = 0.5cm, minimum size = 7mm] {$h_1$};
\node[at = (g), below = 1.2cm, minimum size = 7mm] {$g_1$};
\node[at = (leaf), below = 1.2cm, minimum size = 7mm] {$g_2$};
\draw[thick] (0, -2.5) -- (0,-4.05) --++(2.45, 0);
\draw[->, thick] (2, -1.5) -- (4, -1);
\node[circle, minimum size = 7mm] at (7, -.25) {
\begin{tikzpicture}
\node[circle, fill = blue, minimum size = 2mm, inner sep = 0pt] at (0, 0) (c) {};
\draw(c.north) --++(0, 0.7) node[circle, fill = blue, minimum size = 2mm, inner sep = 0pt, anchor = south] () {} (c.west) --++(-0.7, 0) node[circle, fill = black, minimum size = 2mm, inner sep = 0pt, anchor = east] () {} (c.south) --++(0, -0.7) node[circle, fill = black, minimum size = 2mm, inner sep = 0pt, anchor = north] () {} (c.north east) --++(0.5, 0.5) node[circle, fill = blue, minimum size = 2mm, inner sep = 0pt, anchor = south west] () {} (c.north west) --++(-0.5, 0.5) node[circle, fill = black, minimum size = 2mm, inner sep = 0pt, anchor = south east] () {} (c.south east) --++(0.5, -0.5) node[circle, fill = black, minimum size = 2mm, inner sep = 0pt, anchor = north west] () {} (c.south west) --++(-0.5, -0.5) node[circle, fill = black, minimum size = 2mm, inner sep = 0pt, anchor = north east] () {} (c.east) --++(1, 0) node[circle, fill = green, minimum size = 2mm, inner sep = 0pt, anchor = west] (leaf) {$c$} (leaf.north) --++(0, 0.5) node[circle, fill = green, minimum size = 2mm, inner sep = 0pt, anchor = south] () {} (leaf.east) --++(0.5, 0) node[circle, fill = green, minimum size = 2mm, inner sep = 0pt, anchor = west] () {} (leaf.south) --++(0, -0.5) node[circle, fill = green, minimum size = 2mm, inner sep = 0pt, anchor = north] () {};
\end{tikzpicture}};
\draw[->, thick] (7, -1.25) --++(0, -1.25);
\node[circle, minimum size = 7mm] at (6, -5.25) {
\tikz[pin distance = 0.8cm]
% \tikzstyle{every pin}=[fill = black, minimum size = 2mm]
\begin{tikzpicture}
\node[circle, fill = blue, minimum size = 2mm, inner sep = 0pt, pin = {[circle, fill=green, minimum size = 2mm, inner sep = 0pt]0:}, pin = {[circle, fill=green, minimum size = 2mm, inner sep = 0pt]36:}, , pin = {[circle, fill=blue, minimum size = 2mm, inner sep = 0pt]72:}, , pin = {[circle, fill=blue, minimum size = 2mm, inner sep = 0pt]108:}, , pin = {[circle, fill=black, minimum size = 2mm, inner sep = 0pt]144:}, , pin = {[circle, fill=black, minimum size = 2mm, inner sep = 0pt]180:}, , pin = {[circle, fill=black, minimum size = 2mm, inner sep = 0pt]216:}, , pin = {[circle, fill=black, minimum size = 2mm, inner sep = 0pt]252:}, , pin = {[circle, fill=black, minimum size = 2mm, inner sep = 0pt]288:}, , pin = {[circle, fill=green, minimum size = 2mm, inner sep = 0pt]324:}] {};
\end{tikzpicture}};
\end{tikzpicture}
\caption{\label{fig:mult_mem}All memories from the multi-memory leaf cannot be part of the final GHZ state. Here we consider a simple tree with one $h$-type node and one single root-to-leaf path. Note that, After $X$ measurement on $h_1$, only one memory, $c$, from the leaf vertex becomes part of the GHZ state (star graph). After applying X measurement on $c$, the remaining memories are in the final GHZ state.}
\end{figure*}

In this theorem, we have considered that the $X$ measurements have been performed after isolating the repeater tree from the graph. However, as mentioned above, one can perform $x$ measurements followed by $Z$ measurements to get the same GHZ state and that can be done by at most many measurements required by the above theorem.

Theorem~\ref{th:extract} shows that the repeater tree is a sufficient condition for extracting a GHZ state from a given graph state. We can also show that a tree structure is necessary for a GHZ state of maximum size.

\begin{theorem}
\label{th:opt}
Let $G$ be a graph and $H$ be a vertex-minor of $G$ containing a cycle such that if $Z$ measurement is applied on vertex $u$, the resulting graph $H'=H\backslash\{u\}$ would be a repeater tree. Then the size of any GHZ state extracted from the graph $H$ is at most the maximum size of the GHZ state extracted from the graph $H'$.
\end{theorem}

\begin{proof} Assume subgraphs $G_1, G_2$ from Fig.~\ref{fig:merge}(e) along with vertices $v_1, u$ and $v_2$ form a cycle of size $l$. After $X$ measurement at $u$, it transforms into another cycle with $G_1, G_2$ and $v_1$ of size $l-2$. Therefore, an $X$ measurement cannot remove any cyclic structure in a graph, only reducing the cycle size by 2. Also, an $Y$ measurement at vertex $u$ would connect the vertices $v_1$ and $v_2$ followed by deleting the vertex $u$. Thus, an $Y$ measurement also cannot remove any cyclic structure in a graph, it only reduces the cycle size by 1.

Now, note that a local complement is self-invertible. A local complement on any vertex $v$ of a complete graph gives a star graph centered at $v$ and a local complement on any leaves of a star graph is equivalent to a $Z$ measurement on that vertex. Therefore, a cycle cannot be transformed into a star graph with only local complementation.

Therefore, only local operation is a $Z$ measurement that can remove a cycle from a graph reducing the graph size by 1. Since a single $Z$ measurement transforms $H$ to $H'$, the size of any GHZ state extracted from the graph $H$ is at most the maximum size of the GHZ state extracted from the graph $H'$. 
\end{proof}

Till now we have considered that each node has a single-qubit memory. Suppose, the vertex $g_i$ (except leaf nodes) has $m_i$-qubit memories. Then by local single-qubit and two-qubit quantum operations, one can construct $m_i$-partite GHZ state, or equivalently, a star graph with $m_i$ vertices at $g_i$. Then Corollary~\ref{cr:merge_star} tells that if the center (say, $c_i$) of the star graph at $g_i$ is part of a repeater tree, then Theorem~\ref{th:extract} will give $M$-partite ($M=\sum m_i$) GHZ state with $m_i$ qubits at $g_i$. But if the leaves also have multiple-qubit memory then they will not be part of the final GHZ state (see Fig.~\ref{fig:mult_mem}). But this restriction can be overcome if there is an $h$-type vertex between the leaf and its parent vertex. That is, consider one $h$-type vertex between $g_{k-1}$ and $g_k$ in each root-to-leaf path in the repeater tree. Also, if $c$ is the center of a star graph in a leaf $g_k$ having $m_k$-qubit memory, After performing the $X$ protocol, one extra $X$ measurement at $c$ would produce a GHZ state including remaining $m_k-1$ memories from the leaf. This does not require the existence of $h$-type vertex between $g_{k-1}$ and $g_k$.

In~\cite{PhysRevLett.86.910}, authors conjectured a bound of $n/2$ for the size of GHZ state from a graph state with $n$ vertices. Later, Mannalath and Pathak~\cite{PhysRevA.108.062614}, show that this bound can be $\lfloor\frac{n+3}{2}\rfloor$, which is optimal for $n$-vertex linear cluster network~\cite{PhysRevResearch.6.013330}. But we can extract GHZ state of size $1+\frac{nm}{m+1}$ if the repeater tree is balanced, where all $g$-type vertices except leaves have $m$ children, and if it is not balanced, then the bound would be at least $\lfloor\frac{n+m+1}{2}\rfloor$, where $m$ is the number of children of the root.

\begin{lemma}
GHZ state of size $1+\frac{nm}{m+1}$ can be extracted from an $n$-vertex repeater tree with all $g$-type vertices except leaves having $m$ children and all root-to-leaf paths have the same length.
\end{lemma}

\begin{proof}
Let $2k$ be the length of all root-to-leaf paths $g_0$-$h_1$-$g_1$-$\cdots$-$h_{k-1}$-$g_{k-1}$-$g_k$. The size of the GHZ state is given by the number of $g$-type vertices in the tree. Since all $g$-type vertices except leaves have $m$ children, and all $h$-type vertices have one child, the number of $g_j$ vertices, as well as $h_j$ vertices, is given by $m^j$. Therefore,
\[
n=1+m^k+2\sum_{j=1}^{k-1}m^j=\frac{(m^k-1)(m+1)}{m-1}.
\]
The size of the GHZ state is given by,
\[
\sum_{j=0}^km^j=\frac{m^{k+1}-1}{m-1}=1+\frac{nm}{m+1}.
\]
\end{proof}

\begin{figure}
\centering
\includegraphics[width = \columnwidth]{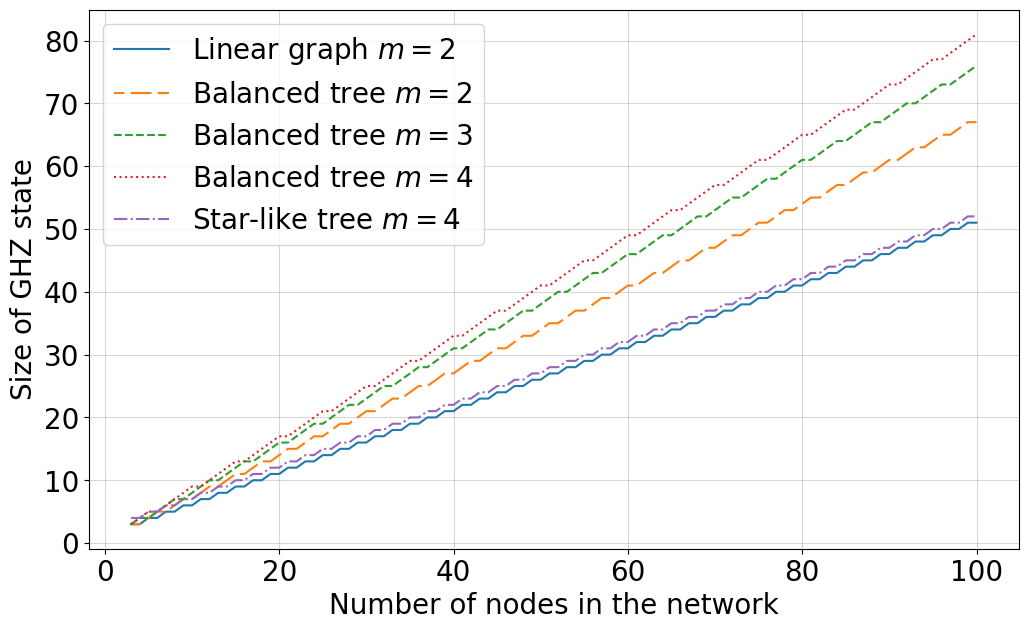}
\caption{Size of GHZ states that can be extracted from a network starting from different initial trees connecting the participants in the GHZ state. $m$ is the number of children of the $g$-type nodes. The linear graph corresponds to the result given in Ref.~\cite{PhysRevA.108.062614}.}
\label{fig:GHZ_size}
\end{figure}

Note that, if the tree is itself a star graph, then $n=1+m$, and the size of the GHZ state will be $1+\frac{(m+1)m}{m+1}=1+m=n$. Now, if the tree is not balanced, then all the vertices except the leaves have at least $1$ child. Suppose, the root has $m$ children, and all other non-leaf vertices have $1$ children, and the lengths of the root-to-leaf paths are $2k_1,2k_2,\cdots,2k_m$. Then $n=1+\sum_{j=1}^m(2k_j-1)\implies\sum_jk_j=\frac{n+m-1}{2}$. The size of the GHZ state will be $1+\sum_jk_j=\frac{n+m+1}{2}$. For linear graph, $m=2$, and so, the size of the GHZ state is given by $\frac{n+3}{2}$~\cite{PhysRevA.108.062614}. Figure~\ref{fig:GHZ_size} shows, according to our proposal, the maximum sizes of different graphs are required to get an $n$-partite GHZ state.

\section{\label{sec:grid}GHZ State routing over Grid Network}

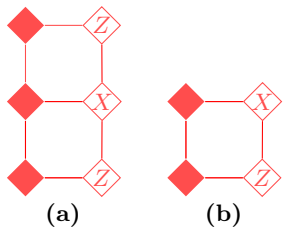
\begin{figure}
\begin{tikzpicture}
\node[diamond, fill = red!70, minimum size = 5mm] at (0, 0) (00) {};
\draw[red!70] (00.east) -- ++(0.5, 0) node[diamond, anchor = west, draw = red!70, inner sep = 0pt, minimum size = 5mm] (10) {$Z$} (00.north) -- ++(0, 0.5) node[diamond, anchor = south, fill = red!70, minimum size = 5mm] (01) {} (01.east) -- ++(0.5, 0) node[diamond, anchor = west, draw = red!70, inner sep = 0pt, minimum size = 5mm] (11) {$X$} (01.north) -- ++(0, 0.5) node[diamond, anchor = south, fill = red!70, inner sep = 0pt, minimum size = 5mm] (02) {} (02.east) -- ++(0.5, 0) node[diamond, anchor = west, draw = red!70, inner sep = 0pt, minimum size = 5mm] (12) {$Z$};
\foreach \x in {0, 1} 
\draw[red] (\x0) -- (\x1) (0\x) -- (1\x);
\draw[red] (11) -- (12);
\node at (0.5, -0.5) {\textbf{(a)}};
\end{tikzpicture}
\hspace{0.5cm}\begin{tikzpicture}
\node[diamond, fill = red!70, minimum size = 5mm] at (0, 0) (00) {};
\draw[red!70] (00.east) -- ++(0.5, 0) node[diamond, anchor = west, draw = red!70, inner sep = 0pt, minimum size = 5mm] (10) {$Z$} (00.north) -- ++(0, 0.5) node[diamond, anchor = south, fill = red!70, minimum size = 5mm] (01) {} (01.east) -- ++(0.5, 0) node[diamond, anchor = west, draw = red!70, inner sep = 0pt, minimum size = 5mm] (11) {$X$};
\foreach \x in {0, 1}
\draw[red] (\x0) -- (\x1) (0\x) -- (1\x);
\node at (0.5, -0.5) {\textbf{(b)}};
\end{tikzpicture}
\caption{\label{fig:blocks}Key building blocks to construct repeater tree in grid network. The nodes with labels $Z$ and $X$ indicate that Pauli $Z$ and Pauli $X$ measurements are to be applied on those nodes in the repeater tree to get the GHZ state. The filled nodes are labeled $G$ and would be in the final GHZ state.}
\end{figure}

\begin{algorithm*}
\vskip-15pt\hrulefill
\vskip-10pt\caption{\label{alg:const}Repeater tree construction for grid network}
\vskip-6pt\hrulefill
\begin{algorithmic}[1]
\Require $n\times n$ grid graph with $n\geq1$.
\State\label{S:init} Label all nodes as $Z$.
\State Label rows and columns as $1, 2, \dots, n$ from bottom to top and left to right respectively.% \Comment{The node in the intersection of $i$-th row and $j$-th column is denoted as $(i,j)$}
\If{$n$ is odd}
\State $k\gets\lfloor\frac{n}{4}\rfloor$.
\State $r\gets n\bmod4$.
\For{$0\leq i\leq k-1$}\label{S:l1}
\For{$0\leq j\leq \frac{n-3}{2}$}
\State\label{S:oij} Replace the block of nodes consisting of columns $2j+1, 2j+2$ and rows $4i+1, 4i+2, 4i+3$ as the block in Fig.~\ref{fig:blocks}(a).
\EndFor
\State Change the labels of the nodes at $(4i+1,n)$ and $(4i+3, n)$ to $X$.\Comment{In tuple we write (row number, column number)}
\State\label{S:oi} Change the labels of the nodes at $(4i+2,n), (4i+4,n-1)$ and $(4i+4, n)$ to $G$.
\EndFor
\If{$r=3$}
\For{$0\leq j\leq \frac{n-3}{2}$}
\State\label{S:o3j} Replace the block of nodes consisting of columns $2j+1, 2j+2$ and rows $n-2, n-1, n$ as the block in Fig.~\ref{fig:blocks}(a).
\EndFor
\State\label{S:o3} Change the labels of the nodes at $(1,n), (n-1, n)$ and $(n, n)$ to $G$.
\State Change the labels of the nodes at $(n-2,n)$ to $X$.
\Else\Comment{$r=1$}
\For{$1\leq j\leq \frac{n-3}{2}$}
\State\label{S:o1j} Replace the block of nodes consisting of columns $2j, 2j+1$ and rows $n-1, n$ as the block in Fig.~\ref{fig:blocks}(b).
\EndFor
\State Change the label of the node at $(n-1,n-1)$ to $X$.
\State\label{S:o1} Change the labels of the nodes at $(n,1),(n,n-1)$ and $(1,n)$ to $G$.
\EndIf
\Else\Comment{$n$ is even}
\State $k\gets\lfloor\frac{n}{6}\rfloor$.
\State $r\gets n\bmod6$.
\For{$0\leq i\leq k-1$}\label{S:l2}
\For{$0\leq j\leq\frac{n-4}{2}$}
\State\label{S:eij1} Replace the block of nodes consisting of columns $2j+1, 2j+2$ and rows $6i+1, 6i+2, 6i+3$ as the block in Fig.~\ref{fig:blocks}(a).
\State\label{S:eij2} Replace the block of nodes consisting of columns $2j+2, 2j+3$ and rows $6i+4, 6i+5, 6i+6$ as the block in Fig.~\ref{fig:blocks}(a).
\EndFor
\State\label{S:eix} Change the label of the node at $(6i+2,n), (6i+4,n)$ and $(6i+6,n)$ to $X$.
\State\label{S:ei} Change the labels of the nodes at $(6i+2,n-1),(6i+3,n),(6i+5,n)$ and $(6i+5,1)$ to $G$.
\EndFor
\If{$r=0$}
\State\label{S:e0} Change the label of the node at $(n,n)$ to $G$.
\Else
\If{$r=2$}
\For{$0\leq j\leq\frac{n-4}{2}$}
\State\label{S:e2j} Replace the block of nodes consisting of columns $2j+1, 2j+2$ and rows $n-1, n$ as the block in Fig.~\ref{fig:blocks}(b).
\EndFor
\State\label{S:e2} Change the label of the node at $(n,n-1)$ and $(n-1,n)$ to $G$.
\Else\Comment{$r=4$}
\For{$0\leq j\leq\frac{n-4}{2}$}
\State\label{S:e4j} Replace the block of nodes consisting of columns $2j+1, 2j+2$ and rows $n-3, n-2, n-1$ as the block in Fig.~\ref{fig:blocks}(a).
\EndFor
\State\label{S:e4} Change the labels of the nodes at $(n-2,n-1),(n-1,n),(n,n-2)$ and $(n,n-1)$ to $G$.
\EndIf
\State\label{S:e4x} Change the label of the node at $(n,n)$ to $X$.
\EndIf
\State\label{S:e} Change the label of the node at $(1,n-1)$ to $G$.
\EndIf
\end{algorithmic}
\end{algorithm*}

\begin{table*}
\centering
\begin{tabular}{|c|c|c|c|l|}
\hline
&\textbf{Steps in Algorithm~\ref{alg:const}}&\textbf{Number of repetitions}&\textbf{Number of $G$ nodes}&\multicolumn{1}{c|}{\textbf{Size of GHZ state}}\\
\hline
\multirow{3}{*}{$n=4k+3$}&Step~\ref{S:oij}&$k\times\frac{n-1}{2}$&$3\times k\times \frac{n-1}{2}$&\multirow{2}{*}{$3k\frac{n-1}{2}+3k+3\frac{n-1}{2}+3$}\\
\cline{2-4}
\multirow{3}{*}{$k\geq0$}&Step~\ref{S:oi}&$k$&$3\times k$&\multirow{2}{*}{$=3(k+1)\frac{n+1}{2}$}\\
\cline{2-4}
&Step~\ref{S:o3j}&$\frac{n-1}{2}$&$\frac{3\times (n-1)}{2}$&\multirow{2}{*}{$=\frac{3}{8}(n+1)^2$}\\
\cline{2-4}
&Step~\ref{S:o3}&$1$&$3$&\\
\hline
\multirow{3}{*}{$n=4k+1$}&Step~\ref{S:oij}&$k\times \frac{n-1}{2}$&$3\times k\times \frac{n-1}{2}$&\multirow{2}{*}{$3k\frac{n-1}{2}+3k+2\frac{n-3}{2}+3$}\\
\cline{2-4}
\multirow{3}{*}{$k\geq0$}&Step~\ref{S:oi}&$k$&$3\times k$&\multirow{2}{*}{$=\frac{3k}{2}(n+1)+n$}\\
\cline{2-4}
&Step~\ref{S:o1j}&$\frac{n-3}{2}$&$\frac{2\times (n-3)}{2}$&\multirow{2}{*}{$=\frac{3}{8}(n^2-1)+n$}\\
\cline{2-4}
&Step~\ref{S:o1}&$1$&$3$&\\
\hline
\multirow{4}{*}{$n=6k$}&Step~\ref{S:eij1}&$k\times \frac{n-2}{2}$&$3\times k\times \frac{n-2}{2}$&\multirow{3}{*}{$3k\frac{n-2}{2}+3k\frac{n-2}{2}+4k+2$}\\
\cline{2-4}
\multirow{4}{*}{$k\geq1$}&Step~\ref{S:eij2}&$k\times \frac{n-2}{2}$&$3\times k\times \frac{n-2}{2}$&\multirow{3}{*}{$=k(3n-2)+2$}\\
\cline{2-4}
&Step~\ref{S:ei}&$k$&$4\times k$&\multirow{3}{*}{$=\frac{3n^2-2n+12}{6}$}\\
\cline{2-4}
&Step~\ref{S:e0}&$1$&$1$&\\
\cline{2-4}
&Step~\ref{S:e}&$1$&$1$&\\
\hline
\multirow{5}{*}{$n=6k+2$}&Step~\ref{S:eij1}&$k\times \frac{n-2}{2}$&$3\times k\times \frac{n-2}{2}$&\multirow{4}{*}{$2\times 3k\frac{n-2}{2}+4k+2\frac{n-2}{2}+3$}\\
\cline{2-4}
\multirow{5}{*}{$k\geq0$}&Step~\ref{S:eij2}&$k\times \frac{n-2}{2}$&$3\times k\times \frac{n-2}{2}$&\multirow{4}{*}{$=k(3n-2)+n+1$}\\
\cline{2-4}
&Step~\ref{S:ei}&$k$&$4\times k$&\multirow{4}{*}{$=\frac{n-2}{6}(3n-2)+n+1$}\\
\cline{2-4}
&Step~\ref{S:e2j}&$\frac{n-2}{2}$&$2\times \frac{n-2}{2}$&\multirow{4}{*}{$=\frac{3n^2-2n+10}{6}$}\\
\cline{2-4}
&Step~\ref{S:e2}&$1$&$2$&\\
\cline{2-4}
&Step~\ref{S:e}&$1$&$1$&\\
\hline
\multirow{5}{*}{$n=6k+4$}&Step~\ref{S:eij1}&$k\times \frac{n-2}{2}$&$3\times k\times \frac{n-2}{2}$&\multirow{4}{*}{$2\times 3k\frac{n-2}{2}+4k+3\frac{n-2}{2}+5$}\\
\cline{2-4}
\multirow{5}{*}{$k\geq0$}&Step~\ref{S:eij2}&$k\times \frac{n-2}{2}$&$3\times k\times \frac{n-2}{2}$&\multirow{4}{*}{$=k(3n-2)+3\frac{n-2}{2}+5$}\\
\cline{2-4}
&Step~\ref{S:ei}&$k$&$4\times k$&\multirow{4}{*}{$=\frac{n-4}{6}(3n-2)+3\frac{n-2}{2}+5$}\\
\cline{2-4}
&Step~\ref{S:e4j}&$\frac{n-2}{2}$&$3\times \frac{n-2}{2}$&\multirow{4}{*}{$=\frac{3n^2-5n+20}{6}$}\\
\cline{2-4}
&Step~\ref{S:e4}&$1$&$4$&\\
\cline{2-4}
&Step~\ref{S:e}&$1$&$1$&\\
\hline
\end{tabular}
\caption{Size of final GHZ state achieved from an $n\times n$ grid network using Algorithm~\ref{alg:const} and repetition tree. The first column indicates the values of $n$, the steps in the algorithm that provide the nodes labeled as $G$ are given in the second column, the third column contains the number of repetitions of the steps in the second column, and the fourth column is the number of nodes labeled as $G$ in that step including repetitions. Finally, in the last column, we provide the size of the final GHZ state for corresponding $n$. In all of the five cases, the size is larger than the previous conjectures $L_B$ and $L_M^c$.}
\label{tab:GHZ_size}
\end{table*}

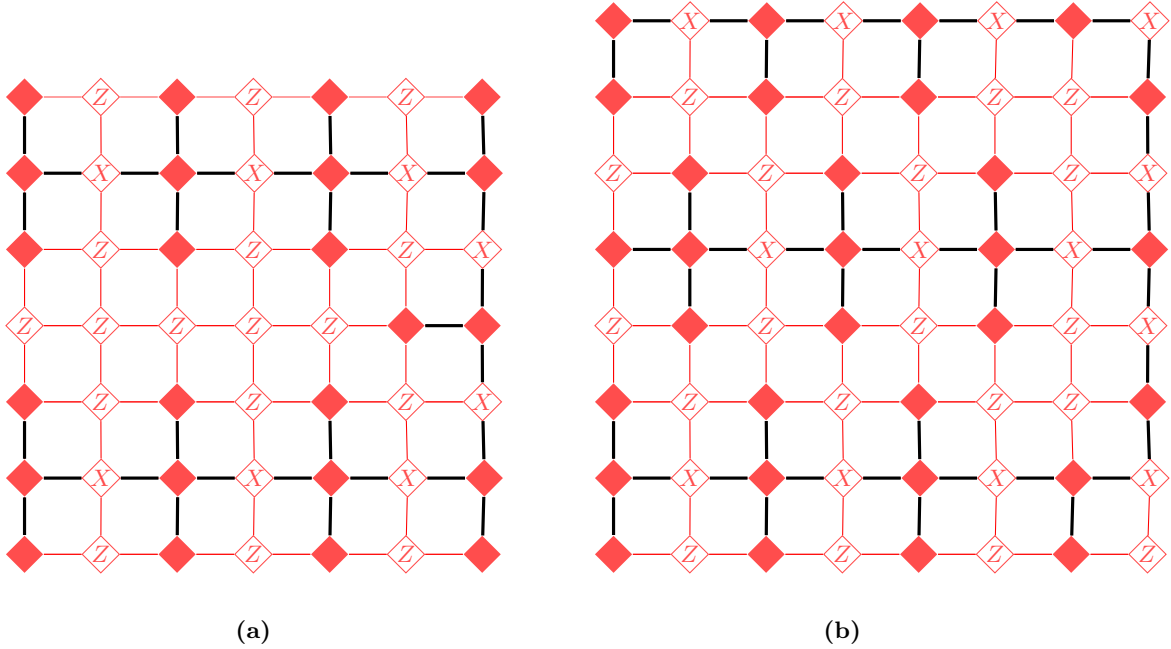
\begin{figure*}
\begin{tikzpicture}
\node[diamond, fill = red!70, minimum size = 5mm] at (0, 0) (00) {};
\draw[red!70] (00.east) -- ++(0.5, 0) node[diamond, anchor = west, draw = red!70, inner sep = 0pt, minimum size = 5mm] (10) {$Z$} (10.east) -- ++(0.5, 0) node[diamond, anchor = west, fill = red!70, minimum size = 5mm] (20) {} (20.east) -- ++(0.5, 0) node[diamond, anchor = west, draw = red!70, inner sep = 0pt, minimum size = 5mm] (30) {$Z$} (30.east) -- ++(0.5, 0) node[diamond, anchor = west, fill = red!70, minimum size = 5mm] (40) {} (40.east) -- ++(0.5, 0) node[diamond, anchor = west, draw = red!70, inner sep = 0pt, minimum size = 5mm] (50) {$Z$} (50.east) -- ++(0.5, 0) node[diamond, anchor = west, fill = red!70, minimum size = 5mm] (60) {} (00.north) -- ++(0, 0.5) node[diamond, anchor = south, fill = red!70, minimum size = 5mm] (01) {} (01.east) -- ++(0.5, 0) node[diamond, anchor = west, draw = red!70, inner sep = 0pt, minimum size = 5mm] (11) {$X$} (11.east) -- ++(0.5, 0) node[diamond, anchor = west, fill = red!70, minimum size = 5mm] (21) {} (21.east) -- ++(0.5, 0) node[diamond, anchor = west, draw = red!70, inner sep = 0pt, minimum size = 5mm] (31) {$X$} (31.east) -- ++(0.5, 0) node[diamond, anchor = west, fill = red!70, minimum size = 5mm] (41) {} (41.east) -- ++(0.5, 0) node[diamond, anchor = west, draw = red!70, inner sep = 0pt, minimum size = 5mm] (51) {$X$} (51.east) -- ++(0.5, 0) node[diamond, anchor = west, fill = red!70, inner sep=0pt, minimum size = 5mm] (61) {} (01.north) -- ++(0, 0.5) node[diamond, anchor = south, fill = red!70, inner sep = 0pt, minimum size = 5mm] (02) {} (02.east) -- ++(0.5, 0) node[diamond, anchor = west, draw = red!70, inner sep = 0pt, minimum size = 5mm] (12) {$Z$} (12.east) -- ++(0.5, 0) node[diamond, anchor = west, fill = red!70, minimum size = 5mm] (22) {} (22.east) -- ++(0.5, 0) node[diamond, anchor = west, draw = red!70, inner sep = 0pt, minimum size = 5mm] (32) {$Z$} (32.east) -- ++(0.5, 0) node[diamond, anchor = west, fill = red!70, minimum size = 5mm] (42) {} (42.east) -- ++(0.5, 0) node[diamond, anchor = west, draw = red!70, inner sep = 0pt, minimum size = 5mm] (52) {$Z$} (52.east) -- ++(0.5, 0) node[diamond, anchor = west, draw = red!70, inner sep = 0pt, minimum size = 5mm] (62) {$X$} (02.north) -- ++(0, 0.5) node[diamond, anchor = south, draw = red!70, inner sep = 0pt, minimum size = 5mm] (03) {$Z$} (03.east) -- ++(0.5, 0) node[diamond, anchor = west, draw = red!70, inner sep = 0pt, minimum size = 5mm] (13) {$Z$} (13.east) -- ++(0.5, 0) node[diamond, anchor = west, draw = red!70, inner sep = 0pt, minimum size = 5mm] (23) {$Z$} (23.east) -- ++(0.5, 0) node[diamond, anchor = west, draw = red!70, inner sep = 0pt, minimum size = 5mm] (33) {$Z$} (33.east) -- ++(0.5, 0) node[diamond, anchor = west, draw = red!70, inner sep = 0pt, minimum size = 5mm] (43) {$Z$} (43.east) -- ++(0.5, 0) node[diamond, anchor = west, fill = red!70, inner sep = 0pt, minimum size = 5mm] (53) {} (53.east) -- ++(0.5, 0) node[diamond, anchor = west, fill = red!70, inner sep = 0pt, minimum size = 5mm] (63) {} (03.north) -- ++(0, 0.5) node[diamond, anchor = south, fill = red!70, inner sep = 0pt, minimum size = 5mm] (04) {} (04.east) -- ++(0.5, 0) node[diamond, anchor = west, draw = red!70, inner sep = 0pt, minimum size = 5mm] (14) {$Z$} (14.east) -- ++(0.5, 0) node[diamond, anchor = west, fill = red!70, minimum size = 5mm] (24) {} (24.east) -- ++(0.5, 0) node[diamond, anchor = west, draw = red!70, inner sep = 0pt, minimum size = 5mm] (34) {$Z$} (34.east) -- ++(0.5, 0) node[diamond, anchor = west, fill = red!70, minimum size = 5mm] (44) {} (44.east) -- ++(0.5, 0) node[diamond, anchor = west, draw = red!70, inner sep = 0pt, minimum size = 5mm] (54) {$Z$} (54.east) -- ++(0.5, 0) node[diamond, anchor = west, draw = red!70, inner sep = 0pt, minimum size = 5mm] (64) {$X$} (04.north) -- ++(0, 0.5) node[diamond, anchor = south, fill = red!70, minimum size = 5mm] (05) {} (05.east) -- ++(0.5, 0) node[diamond, anchor = west, draw = red!70, inner sep = 0pt, minimum size = 5mm] (15) {$X$} (15.east) -- ++(0.5, 0) node[diamond, anchor = west, fill = red!70, minimum size = 5mm] (25) {} (25.east) -- ++(0.5, 0) node[diamond, anchor = west, draw = red!70, inner sep = 0pt, minimum size = 5mm] (35) {$X$} (35.east) -- ++(0.5, 0) node[diamond, anchor = west, fill = red!70, minimum size = 5mm] (45) {} (45.east) -- ++(0.5, 0) node[diamond, anchor = west, draw = red!70, inner sep = 0pt, minimum size = 5mm] (55) {$X$} (55.east) -- ++(0.5, 0) node[diamond, anchor = west, fill = red!70, inner sep=0pt, minimum size = 5mm] (65) {} (05.north) -- ++(0, 0.5) node[diamond, fill = red!70, minimum size = 5mm, anchor = south] (06) {} (06.east) -- ++(0.5, 0) node[diamond, anchor = west, draw = red!70, inner sep = 0pt, minimum size = 5mm] (16) {$Z$} (16.east) -- ++(0.5, 0) node[diamond, anchor = west, fill = red!70, minimum size = 5mm] (26) {} (26.east) -- ++(0.5, 0) node[diamond, anchor = west, draw = red!70, inner sep = 0pt, minimum size = 5mm] (36) {$Z$} (36.east) -- ++(0.5, 0) node[diamond, anchor = west, fill = red!70, minimum size = 5mm] (46) {} (46.east) -- ++(0.5, 0) node[diamond, anchor = west, draw = red!70, inner sep = 0pt, minimum size = 5mm] (56) {$Z$} (56.east) -- ++(0.5, 0) node[diamond, anchor = west, fill = red!70, minimum size = 5mm] (66) {};
\foreach \x in {0,...,5}
\foreach \y [count=\yi] in {0,...,5}  
\draw[red] (\x\y) -- (\x\yi) (\y\x) -- (\yi\x);
\foreach \y [count=\yi] in {0,...,5}  
\draw[very thick] (6\y) -- (6\yi);
\foreach \x in {1, 5}
\foreach \y [count=\yi] in {0,...,5}  
\draw[very thick] (\y\x) -- (\yi\x);
\foreach \x in {0, 2, 4}
\foreach \y in {1, 5}
\pgfmathtruncatemacro{\yi}{\y+1}
\pgfmathtruncatemacro{\yj}{\y-1}
\draw[very thick] (\x\y) -- (\x\yi) (\x\y) -- (\x\yj);
\draw[very thick] (53) -- (63);
\node[below = 0.5cm of 30] {\textbf{(a)}};
\end{tikzpicture}
\hspace{1cm}
\begin{tikzpicture}
\node[diamond, fill = red!70, minimum size = 5mm] at (0, 0) (00) {};
\draw[red!70] (00.east) -- ++(0.5, 0) node[diamond, anchor = west, draw = red!70, inner sep = 0pt, minimum size = 5mm] (10) {$Z$} (10.east) -- ++(0.5, 0) node[diamond, anchor = west, fill = red!70, minimum size = 5mm] (20) {} (20.east) -- ++(0.5, 0) node[diamond, anchor = west, draw = red!70, inner sep = 0pt, minimum size = 5mm] (30) {$Z$} (30.east) -- ++(0.5, 0) node[diamond, anchor = west, fill = red!70, minimum size = 5mm] (40) {} (40.east) -- ++(0.5, 0) node[diamond, anchor = west, draw = red!70, inner sep = 0pt, minimum size = 5mm] (50) {$Z$} (50.east) -- ++(0.5, 0) node[diamond, anchor = west, fill = red!70, minimum size = 5mm] (60) {} (60.east) -- ++(0.5, 0) node[diamond, anchor = west, draw = red!70, inner sep = 0pt, minimum size = 5mm] (70) {$Z$} (00.north) -- ++(0, 0.5) node[diamond, anchor = south, fill = red!70, minimum size = 5mm] (01) {} (01.east) -- ++(0.5, 0) node[diamond, anchor = west, draw = red!70, inner sep = 0pt, minimum size = 5mm] (11) {$X$} (11.east) -- ++(0.5, 0) node[diamond, anchor = west, fill = red!70, minimum size = 5mm] (21) {} (21.east) -- ++(0.5, 0) node[diamond, anchor = west, draw = red!70, inner sep = 0pt, minimum size = 5mm] (31) {$X$} (31.east) -- ++(0.5, 0) node[diamond, anchor = west, fill = red!70, minimum size = 5mm] (41) {} (41.east) -- ++(0.5, 0) node[diamond, anchor = west, draw = red!70, inner sep = 0pt, minimum size = 5mm] (51) {$X$} (51.east) -- ++(0.5, 0) node[diamond, anchor = west, fill = red!70, inner sep=0pt, minimum size = 5mm] (61) {} (61.east) -- ++(0.5, 0) node[diamond, anchor = west, draw = red!70, inner sep=0pt, minimum size = 5mm] (71) {$X$} (01.north) -- ++(0, 0.5) node[diamond, anchor = south, fill = red!70, inner sep = 0pt, minimum size = 5mm] (02) {} (02.east) -- ++(0.5, 0) node[diamond, anchor = west, draw = red!70, inner sep = 0pt, minimum size = 5mm] (12) {$Z$} (12.east) -- ++(0.5, 0) node[diamond, anchor = west, fill = red!70, minimum size = 5mm] (22) {} (22.east) -- ++(0.5, 0) node[diamond, anchor = west, draw = red!70, inner sep = 0pt, minimum size = 5mm] (32) {$Z$} (32.east) -- ++(0.5, 0) node[diamond, anchor = west, fill = red!70, minimum size = 5mm] (42) {} (42.east) -- ++(0.5, 0) node[diamond, anchor = west, draw = red!70, inner sep = 0pt, minimum size = 5mm] (52) {$Z$} (52.east) -- ++(0.5, 0) node[diamond, anchor = west, draw = red!70, inner sep = 0pt, minimum size = 5mm] (62) {$Z$} (62.east) -- ++(0.5, 0) node[diamond, anchor = west, fill = red!70, inner sep = 0pt, minimum size = 5mm] (72) {} (02.north) -- ++(0, 0.5) node[diamond, anchor = south, draw = red!70, inner sep = 0pt, minimum size = 5mm] (03) {$Z$} (03.east) -- ++(0.5, 0) node[diamond, anchor = west, fill = red!70, inner sep = 0pt, minimum size = 5mm] (13) {} (13.east) -- ++(0.5, 0) node[diamond, anchor = west, draw = red!70, inner sep = 0pt, minimum size = 5mm] (23) {$Z$} (23.east) -- ++(0.5, 0) node[diamond, anchor = west, fill = red!70, inner sep = 0pt, minimum size = 5mm] (33) {} (33.east) -- ++(0.5, 0) node[diamond, anchor = west, draw = red!70, inner sep = 0pt, minimum size = 5mm] (43) {$Z$} (43.east) -- ++(0.5, 0) node[diamond, anchor = west, fill = red!70, inner sep = 0pt, minimum size = 5mm] (53) {} (53.east) -- ++(0.5, 0) node[diamond, anchor = west, draw = red!70, inner sep = 0pt, minimum size = 5mm] (63) {$Z$} (63.east) -- ++(0.5, 0) node[diamond, anchor = west, draw = red!70, inner sep = 0pt, minimum size = 5mm] (73) {$X$} (03.north) -- ++(0, 0.5) node[diamond, anchor = south, fill = red!70, inner sep = 0pt, minimum size = 5mm] (04) {} (04.east) -- ++(0.5, 0) node[diamond, anchor = west, fill = red!70, inner sep = 0pt, minimum size = 5mm] (14) {} (14.east) -- ++(0.5, 0) node[diamond, anchor = west, draw = red!70, inner sep = 0pt, minimum size = 5mm] (24) {$X$} (24.east) -- ++(0.5, 0) node[diamond, anchor = west, fill = red!70, inner sep = 0pt, minimum size = 5mm] (34) {} (34.east) -- ++(0.5, 0) node[diamond, anchor = west, draw = red!70, inner sep = 0pt, minimum size = 5mm] (44) {$X$} (44.east) -- ++(0.5, 0) node[diamond, anchor = west, fill = red!70, inner sep = 0pt, minimum size = 5mm] (54) {} (54.east) -- ++(0.5, 0) node[diamond, anchor = west, draw = red!70, inner sep = 0pt, minimum size = 5mm] (64) {$X$} (64.east) -- ++(0.5, 0) node[diamond, anchor = west, fill = red!70, inner sep = 0pt, minimum size = 5mm] (74) {} (04.north) -- ++(0, 0.5) node[diamond, anchor = south, draw = red!70, inner sep = 0pt, minimum size = 5mm] (05) {$Z$} (05.east) -- ++(0.5, 0) node[diamond, anchor = west, fill = red!70, inner sep = 0pt, minimum size = 5mm] (15) {} (15.east) -- ++(0.5, 0) node[diamond, anchor = west, draw = red!70, inner sep = 0pt, minimum size = 5mm] (25) {$Z$} (25.east) -- ++(0.5, 0) node[diamond, anchor = west, fill = red!70, inner sep = 0pt, minimum size = 5mm] (35) {} (35.east) -- ++(0.5, 0) node[diamond, anchor = west, draw = red!70, inner sep = 0pt, minimum size = 5mm] (45) {$Z$} (45.east) -- ++(0.5, 0) node[diamond, anchor = west, fill = red!70, inner sep = 0pt, minimum size = 5mm] (55) {} (55.east) -- ++(0.5, 0) node[diamond, anchor = west, draw = red!70, inner sep=0pt, minimum size = 5mm] (65) {$Z$} (65.east) -- ++(0.5, 0) node[diamond, anchor = west, draw = red!70, inner sep = 0pt, minimum size = 5mm] (75) {$X$} (05.north) -- ++(0, 0.5) node[diamond, fill = red!70, minimum size = 5mm, anchor = south] (06) {} (06.east) -- ++(0.5, 0) node[diamond, anchor = west, draw = red!70, inner sep = 0pt, minimum size = 5mm] (16) {$Z$} (16.east) -- ++(0.5, 0) node[diamond, anchor = west, fill = red!70, minimum size = 5mm] (26) {} (26.east) -- ++(0.5, 0) node[diamond, anchor = west, draw = red!70, inner sep = 0pt, minimum size = 5mm] (36) {$Z$} (36.east) -- ++(0.5, 0) node[diamond, anchor = west, fill = red!70, minimum size = 5mm] (46) {} (46.east) -- ++(0.5, 0) node[diamond, anchor = west, draw = red!70, inner sep = 0pt, minimum size = 5mm] (56) {$Z$} (56.east) -- ++(0.5, 0) node[diamond, anchor = west, draw = red!70, inner sep = 0pt, minimum size = 5mm] (66) {$Z$} (66.east) -- ++(0.5, 0) node[diamond, anchor = west, fill = red!70, minimum size = 5mm] (76) {} (06.north) -- ++(0, 0.5) node[diamond, fill = red!70, minimum size = 5mm, anchor = south] (07) {} (07.east) -- ++(0.5, 0) node[diamond, anchor = west, draw = red!70, inner sep = 0pt, minimum size = 5mm] (17) {$X$} (17.east) -- ++(0.5, 0) node[diamond, anchor = west, fill = red!70, minimum size = 5mm] (27) {} (27.east) -- ++(0.5, 0) node[diamond, anchor = west, draw = red!70, inner sep = 0pt, minimum size = 5mm] (37) {$X$} (37.east) -- ++(0.5, 0) node[diamond, anchor = west, fill = red!70, minimum size = 5mm] (47) {} (47.east) -- ++(0.5, 0) node[diamond, anchor = west, draw = red!70, inner sep = 0pt, minimum size = 5mm] (57) {$X$} (57.east) -- ++(0.5, 0) node[diamond, anchor = west, fill = red!70, minimum size = 5mm] (67) {} (67.east) -- ++(0.5, 0) node[diamond, anchor = west, draw = red!70, inner sep = 0pt, minimum size = 5mm] (77) {$X$};
\foreach \x in {0,...,7}
\foreach \y [count=\yi] in {0,...,6}  
\draw[red] (\x\y) -- (\x\yi) (\y\x) -- (\yi\x);
\foreach \y in {1,...,6}
\pgfmathtruncatemacro{\yi}{\y+1}
\draw[very thick] (7\y) -- (7\yi);
\foreach \x in {1, 4}
\foreach \y [count=\yi] in {0,...,6}  
\draw[very thick] (\y\x) -- (\yi\x);
\foreach \y [count=\yi] in {0,...,6}  
\draw[very thick] (\y7) -- (\yi7);
\foreach \x in {0, 2, 4}
\draw[very thick] (\x0) -- (\x1) (\x1) -- (\x2);
\foreach \x in {1, 3, 5}
\draw[very thick] (\x3) -- (\x4) (\x4) -- (\x5);
\foreach \x in {0, 2, 4}
\draw[very thick] (\x6) -- (\x7);
\draw[very thick] (61) -- (60);
\node[below = 0.5cm of 30] {\textbf{(b)}};
\end{tikzpicture}
\caption{\label{fig:grid_graph} Extracting GHZ state from grid network according to Algorithm~\ref{alg:const}. The nodes labeled as $Z$ and $X$ will undergo $Z$ and $X$ measurements respectively. The remaining highlighted nodes will be part of the final GHZ state. The bold black edges indicate the repeater tree. (a) $n=7$ shows an example of odd size, and (b) $n=8$ shows an example of even size.}
\end{figure*}

A grid network is an important network architecture that connects nearest neighbors via physical links. Therefore, they are required to share entanglement within the shortest distance to produce a graph state making it relevant for quantum communication. Briegel and Raussendorf~\cite{PhysRevLett.86.910} and Mannalath and Pathak~\cite{PhysRevA.108.062614} considered $n\times n$ grid networks, and conjectured bounds for largest GHZ state as $L_B=\lceil n/2\rceil^2$ and $L_M=\lfloor\frac{n+1}{4}\rfloor(3\lfloor\frac{n-1}{2}\rfloor+4)-2$ respectively. As mentioned in Section~\ref{sec:prev}, the later conjecture is incompatible with their construction. In that section, we mentioned the corrected bound as $L_M^c=3\lfloor\frac{n+1}{4}\rfloor\lfloor\frac{n+1}{2}\rfloor$.

In this section, we propose constructions of the repeater tree in $n\times n$ grid network improving both of the above bounds $L_B$ and $L_M^c$. Although for $n=4k+3$ with $k\geq0$, our construction provides the same size as the above corrected bound $l^c_M$, for other values of $n$ our construction improves both the bounds $L_B$ and $L_M^c$. The construction is given as Algorithm~\ref{alg:const}.

In constructing the repeater tree, we consider five sets for the grid network depending on its size $n$. For odd values of $n$, two sets are given by two values of $r=n\bmod4$, and for even values of $n$, three sets are given by three values of $r=n\bmod6$. One can easily see that these five sets are pairwise disjoint and exhaust the set of all possible square grids. Algorithm~\ref{alg:const} constructs a repeater tree, say $T$, out of a $n\times n$ grid network. For $n=4k+3$, this construction would give the same graph as the construction given in Ref.~\cite{PhysRevA.108.062614}. The nodes labeled as $X$ would undergo an $X$ measurement and the nodes labeled as $G$, which have been highlighted in the figures by color-filing, would be in the final GHZ state after separating from the remaining network by performing $Z$ measurements as required. Note that the blocks in Fig.~\ref{fig:blocks}(a) and Fig.~\ref{fig:blocks}(b) contain $3$ and $2$ nodes labeled as $G$, respectively. The sizes of the GHZ states for each of the five cases are given in Table~\ref{tab:GHZ_size}. The first column shows the values of $n$. The second column lists the steps in the algorithm that produce nodes labeled as $G$. The third column indicates the number of times each step in the second column is repeated. The fourth column presents the total count of nodes labeled as $G$ at each step, including repetitions. Finally, the last column provides the size of the final GHZ state for each corresponding $n$. In all five cases, the size exceeds the previously conjectured values $L_B$ and $L_M^c$. 

It's straightforward to demonstrate through simple calculations that repeating the loops~\ref{S:l1} and~\ref{S:l2} in multiples of 6 and 4 respectively does not increase the size of GHZ states. Also, increasing the size of the repeater tree produced through the Algorithm~\ref{alg:const} would create cycles in it. Therefore, recalling Theorem~\ref{th:opt}, we can say that our algorithm would produce a GHZ state of maximum size from a given $n\times n$ grid.

Two repeater trees Fig.~\ref{fig:grid_graph}(a) and Fig.~\ref{fig:grid_graph}(b) have been produced using the Algorithm~\ref{alg:const} for $n=7$ and $8$ respectively. As mentioned in Table~\ref{tab:GHZ_size}, the size of the GHZ states would be $\frac{3}{8}(7+1)^2=24$ and $\frac{3\times4^2-2\times4+10}{6}=31$ for $n=7$ and $8$ respectively. A step-by-step construction for $n=8$ is provided in Appendix~\ref{sec:StS}.

\begin{table*}
\centering
\resizebox{\textwidth}{!}{\begin{tabular}{|c|c|c|c|c|c|c|}
\hline
\multirow{2}{*}{\textbf{Cases}}&\textbf{Conjecture in Ref.~\cite{PhysRevLett.86.910}}&\textbf{Conjecture in Ref.~\cite{PhysRevA.108.062614}}&\textbf{Correction over }$\mathbf{L_M}$&\textbf{Our Result}&\multicolumn{2}{c|}{\textbf{Our Improvement}}\\
\cline{6-7}
&$\mathbf{L_B}$&$\mathbf{L_M}$&$\mathbf{L_M^c}$&$\mathbf{L_0}$&$\mathbf{L_0-L_B}$&$\mathbf{L_0-L_M^c}$\\
\hline
$n=4k+3$&\multirow{10}{*}{$\lceil\frac{n}{2}\rceil^2$}&\multirow{10}{*}{$\lfloor\frac{n+1}{4}\rfloor\left(3\lfloor\frac{n-1}{2}\rfloor+4\right)-2$}&\multirow{10}{*}{$3\lfloor\frac{n+1}{4}\rfloor\lfloor\frac{n+1}{2}\rfloor$}&\multirow{2}{*}{$\frac{3}{8}(n+1)^2$}&\multirow{2}{*}{$\mathbf{\frac{1}{8}(n+1)^2}$}&\multirow{2}{*}{$0$}\\
$k\geq0$&&&&&&\\
\cline{1-1}
\cline{5-7}
$n=4k+1$&&&&\multirow{2}{*}{$\frac{3}{8}(n^2-1)+n$}&\multirow{2}{*}{$\mathbf{\frac{n^2+4n-5}{8}}$}&\multirow{2}{*}{$\mathbf{n}$}\\
$k\geq0$&&&&&&\\
\cline{1-1}
\cline{5-7}
$n=6k$&&&&\multirow{2}{*}{$\frac{3n^2-2n+12}{6}$}&\multirow{2}{*}{$\mathbf{\frac{3n^2-4n+24}{12}}$}&\multirow{2}{*}{$\mathbf{\frac{3n^2-8n+48}{24}\left(+\frac{3n}{4}\right)}$}\\
$k\geq1$&&&&&&\\
\cline{1-1}
\cline{5-7}
$n=6k+2$&&&&\multirow{2}{*}{$\frac{3n^2-2n+10}{6}$}&\multirow{2}{*}{$\mathbf{\frac{3n^2-4n+20}{12}}$}&\multirow{2}{*}{$\mathbf{\frac{3n^2-8n+40}{24}\left(+\frac{3n}{4}\right)}$}\\
$k\geq0$&&&&&&\\
\cline{1-1}
\cline{5-7}
$n=6k+4$&&&&\multirow{2}{*}{$\frac{3n^2-5n+20}{6}$}&\multirow{2}{*}{$\mathbf{\frac{3n^2-10n+40}{12}}$}&\multirow{2}{*}{$\mathbf{\frac{3n^2-20n+80}{24}\left(+\frac{3n}{4}\right)}$}\\
$k\geq0$&&&&&&\\
\hline
\end{tabular}}
\caption{Comparing our results with existing ones for $n\times n$ grid network. Our results improve the conjecture of Ref.~\cite{PhysRevLett.86.910} for all cases, and also improve the corrected conjecture of Ref.~\cite{PhysRevA.108.062614} for all but one case, namely $n=4k+3,\ k\geq0$. For $n=4k+3$, our result is identical to the corrected conjecture.}
\label{tab:comp}
\end{table*}

\begin{figure}
\centering
\includegraphics[width = \columnwidth]{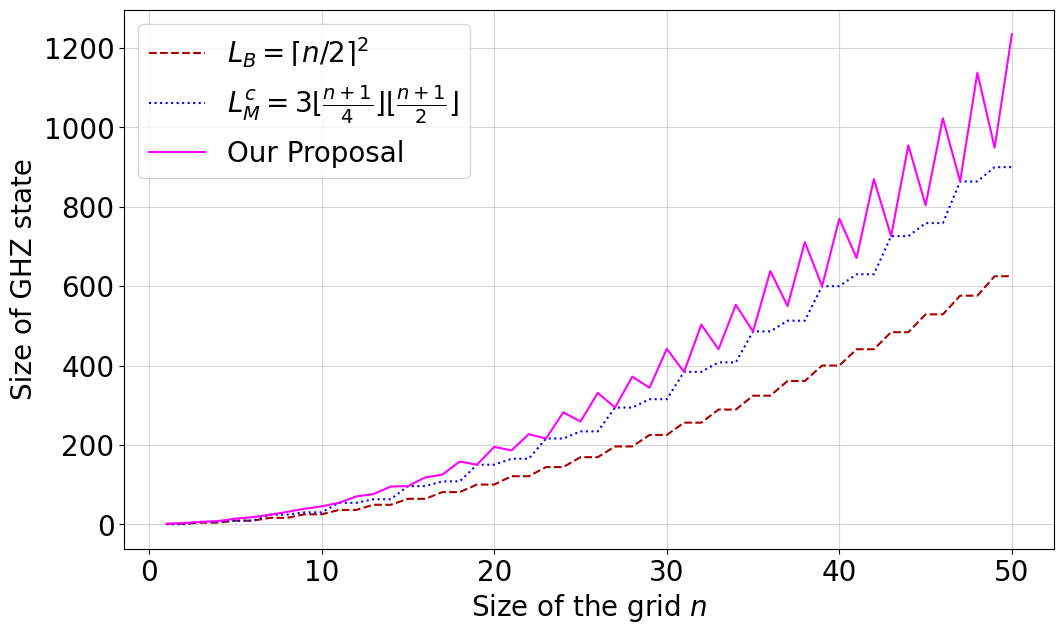}
\caption{Size of GHZ states can be extracted from a given $n\times n$ grid. The dashed line was conjectured by Briegel
and Raussendor~\cite{PhysRevLett.86.910}, the dotted line is the conjecture by Mannalath and Pathak~\cite{PhysRevA.108.062614} corrected in Section~\ref{sec:prev} and the solid line indicates the result according to our proposal given in Algorithm~\ref{alg:const}.}
\label{fig:GHZ_size_grid}
\end{figure}

A comparison of this proposal with the previous two conjectures in Ref.~\cite{PhysRevLett.86.910,PhysRevA.108.062614} is given in Table~\ref{tab:comp}. In one case, when the grid size $n$ is of the form $4k+3$ for $k\geq0$, our construction gives GHZ state of size same as in Ref.~\cite{PhysRevA.108.062614} which is larger than the size conjectured in Ref.~\cite{PhysRevLett.86.910}. In all other cases, our constructions produce larger GHZ states than both of the existing results. A graphical visualization is given in Fig.~\ref{fig:GHZ_size_grid}.
% Our claim is that for a given graph state corresponding to an $n\times n$ grid, our construction would generate the largest possible GHZ state out of the graph state using only local Pauli $X$ and Pauli $Z$ operation. This claim can be easily seen as follows. 

\subsection{GHZ State routing between Specific Users}

Until now our discussion was towards the maximum size of GHZ state that can be extracted from a given network. Although it is useful for broadcasting where we are required to reach a maximum number of people, it may not be suitable for other cases when communication is restricted between some specific users. One can think of a solution as preparing the maximum possible complete graph from a repeater tree using the proposal given in Section~\ref{sec:GHZ}, and then removing those users who are not participating in the communication from that complete graph by performing $Z$ measurements. But in this process, the whole network will be destroyed except the people who are participating in the communication. However, if we can avoid the destruction of the remaining network, we may utilize that for routing of entanglement over that part of the network. This can be done by proper choice of the repeater tree. A minimum spanning tree connecting the participating users could reduce the number of users to perform $X$ measurements. However, a spanning tree may not always satisfy the conditions for a repeater tree. Therefore, we have to find a minimum spanning tree satisfying the following conditions:
\begin{enumerate}
\item the number of consecutive $X$ nodes should be odd,
\item no consecutive $G$ node except the leaf and their parents,
\item exactly one child for each $X$ node.
\end{enumerate}
These three conditions restrict the spanning tree to be a repeater tree. If there exist multiple such spanning trees, one can use the \emph{majorization codition} from Ref.~\cite[Theorem IV.1]{PhysRevA.108.062614} to choose the best one. This will leave some part of the initial network which can be reused for routing of entanglement.

\section{\label{sec:concl}Discussion}

In this article, we significantly advance the field of GHZ state distribution by refining and extending current protocols to accommodate a greater number of users within a given network. Our primary innovation involves employing a tree structure to link the users participating in the final GHZ state. Through our analysis, we have determined that utilizing a balanced tree structure allows for a substantially higher number of users compared to other types of tree configurations.

Moreover, we explore the more practical form of network, which is a grid network. We have found a flaw in the previous work and corrected that flaw. We propose specific and detailed constructions for embedding the balanced tree into the grid network, and our results demonstrate that this approach yields markedly superior outcomes. Not only does our construction improve upon existing methods, but it also supports a larger user base, enhancing the efficiency and scalability of GHZ state-routing protocols. Also, our method for a grid network provides an optimal GHZ state. However, it is an open question whether our findings are also optimal for general graph networks.

\appendix

\section{\label{sec:StS}Construction of Repeater Tree using Algorithm~\ref{alg:const}}

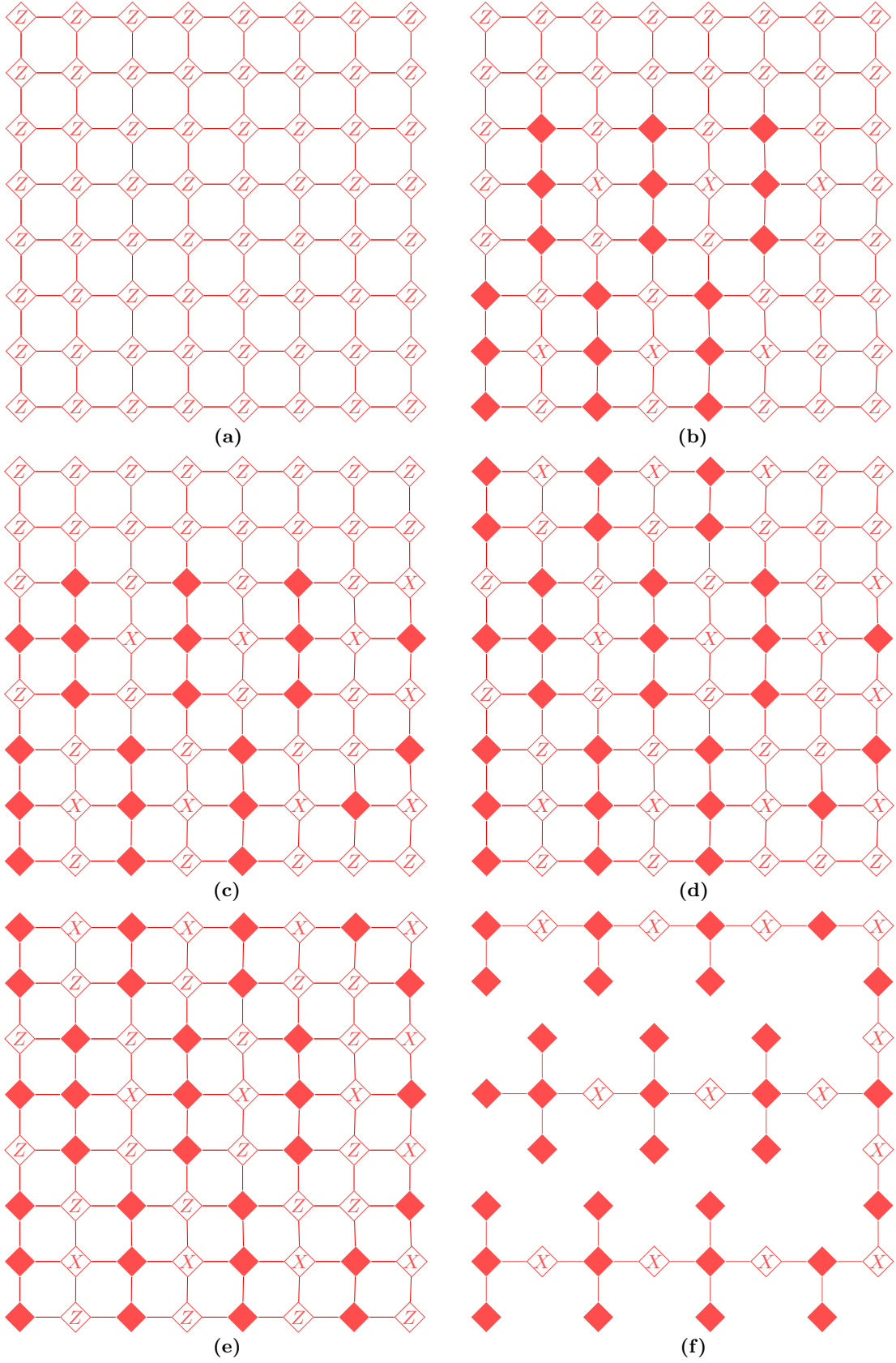
\begin{figure*}
\begin{tikzpicture}
\node[red!70, diamond, draw = red!70, inner sep = 0pt, minimum size = 5mm] at (0, 0) (00) {$Z$};
\draw[red!70] (00.east) -- ++(0.4, 0) node[diamond, anchor = west, draw = red!70, inner sep = 0pt, minimum size = 5mm] (10) {$Z$} (10.east) -- ++(0.4, 0) node[diamond, anchor = west, draw = red!70, inner sep = 0pt, minimum size = 5mm] (20) {$Z$} (20.east) -- ++(0.4, 0) node[diamond, anchor = west, draw = red!70, inner sep = 0pt, minimum size = 5mm] (30) {$Z$} (30.east) -- ++(0.4, 0) node[diamond, anchor = west, draw = red!70, inner sep = 0pt, minimum size = 5mm] (40) {$Z$} (40.east) -- ++(0.4, 0) node[diamond, anchor = west, draw = red!70, inner sep = 0pt, minimum size = 5mm] (50) {$Z$} (50.east) -- ++(0.4, 0) node[diamond, anchor = west, draw = red!70, inner sep = 0pt, minimum size = 5mm] (60) {$Z$} (60.east) -- ++(0.4, 0) node[diamond, anchor = west, draw = red!70, inner sep = 0pt, minimum size = 5mm] (70) {$Z$} (00.north) -- ++(0, 0.4) node[diamond, anchor = south, draw = red!70, inner sep = 0pt, minimum size = 5mm] (01) {$Z$} (01.east) -- ++(0.4, 0) node[diamond, anchor = west, draw = red!70, inner sep = 0pt, minimum size = 5mm] (11) {$Z$} (11.east) -- ++(0.4, 0) node[diamond, anchor = west, draw = red!70, inner sep = 0pt, minimum size = 5mm] (21) {$Z$} (21.east) -- ++(0.4, 0) node[diamond, anchor = west, draw = red!70, inner sep = 0pt, minimum size = 5mm] (31) {$Z$} (31.east) -- ++(0.4, 0) node[diamond, anchor = west, draw = red!70, inner sep = 0pt, minimum size = 5mm] (41) {$Z$} (41.east) -- ++(0.4, 0) node[diamond, anchor = west, draw = red!70, inner sep = 0pt, minimum size = 5mm] (51) {$Z$} (51.east) -- ++(0.4, 0) node[diamond, anchor = west, draw = red!70, inner sep = 0pt, inner sep=0pt, minimum size = 5mm] (61) {$Z$} (61.east) -- ++(0.4, 0) node[diamond, anchor = west, draw = red!70, inner sep=0pt, minimum size = 5mm] (71) {$Z$} (01.north) -- ++(0, 0.4) node[diamond, anchor = south, draw = red!70, inner sep = 0pt, inner sep = 0pt, minimum size = 5mm] (02) {$Z$} (02.east) -- ++(0.4, 0) node[diamond, anchor = west, draw = red!70, inner sep = 0pt, minimum size = 5mm] (12) {$Z$} (12.east) -- ++(0.4, 0) node[diamond, anchor = west, draw = red!70, inner sep = 0pt, minimum size = 5mm] (22) {$Z$} (22.east) -- ++(0.4, 0) node[diamond, anchor = west, draw = red!70, inner sep = 0pt, minimum size = 5mm] (32) {$Z$} (32.east) -- ++(0.4, 0) node[diamond, anchor = west, draw = red!70, inner sep = 0pt, minimum size = 5mm] (42) {$Z$} (42.east) -- ++(0.4, 0) node[diamond, anchor = west, draw = red!70, inner sep = 0pt, minimum size = 5mm] (52) {$Z$} (52.east) -- ++(0.4, 0) node[diamond, anchor = west, draw = red!70, inner sep = 0pt, minimum size = 5mm] (62) {$Z$} (62.east) -- ++(0.4, 0) node[diamond, anchor = west, draw = red!70, inner sep = 0pt, inner sep = 0pt, minimum size = 5mm] (72) {$Z$} (02.north) -- ++(0, 0.4) node[diamond, anchor = south, draw = red!70, inner sep = 0pt, minimum size = 5mm] (03) {$Z$} (03.east) -- ++(0.4, 0) node[diamond, anchor = west, draw = red!70, inner sep = 0pt, inner sep = 0pt, minimum size = 5mm] (13) {$Z$} (13.east) -- ++(0.4, 0) node[diamond, anchor = west, draw = red!70, inner sep = 0pt, minimum size = 5mm] (23) {$Z$} (23.east) -- ++(0.4, 0) node[diamond, anchor = west, draw = red!70, inner sep = 0pt, inner sep = 0pt, minimum size = 5mm] (33) {$Z$} (33.east) -- ++(0.4, 0) node[diamond, anchor = west, draw = red!70, inner sep = 0pt, minimum size = 5mm] (43) {$Z$} (43.east) -- ++(0.4, 0) node[diamond, anchor = west, draw = red!70, inner sep = 0pt, inner sep = 0pt, minimum size = 5mm] (53) {$Z$} (53.east) -- ++(0.4, 0) node[diamond, anchor = west, draw = red!70, inner sep = 0pt, minimum size = 5mm] (63) {$Z$} (63.east) -- ++(0.4, 0) node[diamond, anchor = west, draw = red!70, inner sep = 0pt, minimum size = 5mm] (73) {$Z$} (03.north) -- ++(0, 0.4) node[diamond, anchor = south, draw = red!70, inner sep = 0pt, inner sep = 0pt, minimum size = 5mm] (04) {$Z$} (04.east) -- ++(0.4, 0) node[diamond, anchor = west, draw = red!70, inner sep = 0pt, inner sep = 0pt, minimum size = 5mm] (14) {$Z$} (14.east) -- ++(0.4, 0) node[diamond, anchor = west, draw = red!70, inner sep = 0pt, minimum size = 5mm] (24) {$Z$} (24.east) -- ++(0.4, 0) node[diamond, anchor = west, draw = red!70, inner sep = 0pt, inner sep = 0pt, minimum size = 5mm] (34) {$Z$} (34.east) -- ++(0.4, 0) node[diamond, anchor = west, draw = red!70, inner sep = 0pt, minimum size = 5mm] (44) {$Z$} (44.east) -- ++(0.4, 0) node[diamond, anchor = west, draw = red!70, inner sep = 0pt, inner sep = 0pt, minimum size = 5mm] (54) {$Z$} (54.east) -- ++(0.4, 0) node[diamond, anchor = west, draw = red!70, inner sep = 0pt, minimum size = 5mm] (64) {$Z$} (64.east) -- ++(0.4, 0) node[diamond, anchor = west, draw = red!70, inner sep = 0pt, inner sep = 0pt, minimum size = 5mm] (74) {$Z$} (04.north) -- ++(0, 0.4) node[diamond, anchor = south, draw = red!70, inner sep = 0pt, minimum size = 5mm] (05) {$Z$} (05.east) -- ++(0.4, 0) node[diamond, anchor = west, draw = red!70, inner sep = 0pt, inner sep = 0pt, minimum size = 5mm] (15) {$Z$} (15.east) -- ++(0.4, 0) node[diamond, anchor = west, draw = red!70, inner sep = 0pt, minimum size = 5mm] (25) {$Z$} (25.east) -- ++(0.4, 0) node[diamond, anchor = west, draw = red!70, inner sep = 0pt, inner sep = 0pt, minimum size = 5mm] (35) {$Z$} (35.east) -- ++(0.4, 0) node[diamond, anchor = west, draw = red!70, inner sep = 0pt, minimum size = 5mm] (45) {$Z$} (45.east) -- ++(0.4, 0) node[diamond, anchor = west, draw = red!70, inner sep = 0pt, inner sep = 0pt, minimum size = 5mm] (55) {$Z$} (55.east) -- ++(0.4, 0) node[diamond, anchor = west, draw = red!70, inner sep=0pt, minimum size = 5mm] (65) {$Z$} (65.east) -- ++(0.4, 0) node[diamond, anchor = west, draw = red!70, inner sep = 0pt, minimum size = 5mm] (75) {$Z$} (05.north) -- ++(0, 0.4) node[diamond, draw = red!70, inner sep = 0pt, minimum size = 5mm, anchor = south] (06) {$Z$} (06.east) -- ++(0.4, 0) node[diamond, anchor = west, draw = red!70, inner sep = 0pt, minimum size = 5mm] (16) {$Z$} (16.east) -- ++(0.4, 0) node[diamond, anchor = west, draw = red!70, inner sep = 0pt, minimum size = 5mm] (26) {$Z$} (26.east) -- ++(0.4, 0) node[diamond, anchor = west, draw = red!70, inner sep = 0pt, minimum size = 5mm] (36) {$Z$} (36.east) -- ++(0.4, 0) node[diamond, anchor = west, draw = red!70, inner sep = 0pt, minimum size = 5mm] (46) {$Z$} (46.east) -- ++(0.4, 0) node[diamond, anchor = west, draw = red!70, inner sep = 0pt, minimum size = 5mm] (56) {$Z$} (56.east) -- ++(0.4, 0) node[diamond, anchor = west, draw = red!70, inner sep = 0pt, minimum size = 5mm] (66) {$Z$} (66.east) -- ++(0.4, 0) node[diamond, anchor = west, draw = red!70, inner sep = 0pt, minimum size = 5mm] (76) {$Z$} (06.north) -- ++(0, 0.4) node[diamond, draw = red!70, inner sep = 0pt, minimum size = 5mm, anchor = south] (07) {$Z$} (07.east) -- ++(0.4, 0) node[diamond, anchor = west, draw = red!70, inner sep = 0pt, minimum size = 5mm] (17) {$Z$} (17.east) -- ++(0.4, 0) node[diamond, anchor = west, draw = red!70, inner sep = 0pt, minimum size = 5mm] (27) {$Z$} (27.east) -- ++(0.4, 0) node[diamond, anchor = west, draw = red!70, inner sep = 0pt, minimum size = 5mm] (37) {$Z$} (37.east) -- ++(0.4, 0) node[diamond, anchor = west, draw = red!70, inner sep = 0pt, minimum size = 5mm] (47) {$Z$} (47.east) -- ++(0.4, 0) node[diamond, anchor = west, draw = red!70, inner sep = 0pt, minimum size = 5mm] (57) {$Z$} (57.east) -- ++(0.4, 0) node[diamond, anchor = west, draw = red!70, inner sep = 0pt, minimum size = 5mm] (67) {$Z$} (67.east) -- ++(0.4, 0) node[diamond, anchor = west, draw = red!70, inner sep = 0pt, minimum size = 5mm] (77) {$Z$};
\foreach \x in {0,...,7}
\foreach \y [count=\yi] in {0,...,6}  
\draw[red] (\x\y) -- (\x\yi) (\y\x) -- (\yi\x);
\node at (3.4, -0.5) {\textbf{(a)}};
\end{tikzpicture}
\hspace{.5cm}
\begin{tikzpicture}
\node[red!70, diamond, fill = red!70, minimum size = 5mm] at (0, 0) (00) {};
\draw[red!70] (00.east) -- ++(0.4, 0) node[diamond, anchor = west, draw = red!70, inner sep = 0pt, minimum size = 5mm] (10) {$Z$} (10.east) -- ++(0.4, 0) node[diamond, anchor = west, fill = red!70, minimum size = 5mm] (20) {} (20.east) -- ++(0.4, 0) node[diamond, anchor = west, draw = red!70, inner sep = 0pt, minimum size = 5mm] (30) {$Z$} (30.east) -- ++(0.4, 0) node[diamond, anchor = west, fill = red!70, minimum size = 5mm] (40) {} (40.east) -- ++(0.4, 0) node[diamond, anchor = west, draw = red!70, inner sep = 0pt, minimum size = 5mm] (50) {$Z$} (50.east) -- ++(0.4, 0) node[diamond, anchor = west, draw = red!70, inner sep = 0pt, minimum size = 5mm] (60) {$Z$} (60.east) -- ++(0.4, 0) node[diamond, anchor = west, draw = red!70, inner sep = 0pt, minimum size = 5mm] (70) {$Z$} (00.north) -- ++(0, 0.4) node[diamond, anchor = south, fill = red!70, minimum size = 5mm] (01) {} (01.east) -- ++(0.4, 0) node[diamond, anchor = west, draw = red!70, inner sep = 0pt, minimum size = 5mm] (11) {$X$} (11.east) -- ++(0.4, 0) node[diamond, anchor = west, fill = red!70, minimum size = 5mm] (21) {} (21.east) -- ++(0.4, 0) node[diamond, anchor = west, draw = red!70, inner sep = 0pt, minimum size = 5mm] (31) {$X$} (31.east) -- ++(0.4, 0) node[diamond, anchor = west, fill = red!70, minimum size = 5mm] (41) {} (41.east) -- ++(0.4, 0) node[diamond, anchor = west, draw = red!70, inner sep = 0pt, minimum size = 5mm] (51) {$X$} (51.east) -- ++(0.4, 0) node[diamond, anchor = west, draw = red!70, inner sep = 0pt, inner sep=0pt, minimum size = 5mm] (61) {$Z$} (61.east) -- ++(0.4, 0) node[diamond, anchor = west, draw = red!70, inner sep=0pt, minimum size = 5mm] (71) {$Z$} (01.north) -- ++(0, 0.4) node[diamond, anchor = south, fill = red!70, inner sep = 0pt, minimum size = 5mm] (02) {$Z$} (02.east) -- ++(0.4, 0) node[diamond, anchor = west, draw = red!70, inner sep = 0pt, minimum size = 5mm] (12) {$Z$} (12.east) -- ++(0.4, 0) node[diamond, anchor = west, fill = red!70, minimum size = 5mm] (22) {} (22.east) -- ++(0.4, 0) node[diamond, anchor = west, draw = red!70, inner sep = 0pt, minimum size = 5mm] (32) {$Z$} (32.east) -- ++(0.4, 0) node[diamond, anchor = west, fill = red!70, minimum size = 5mm] (42) {} (42.east) -- ++(0.4, 0) node[diamond, anchor = west, draw = red!70, inner sep = 0pt, minimum size = 5mm] (52) {$Z$} (52.east) -- ++(0.4, 0) node[diamond, anchor = west, draw = red!70, inner sep = 0pt, minimum size = 5mm] (62) {$Z$} (62.east) -- ++(0.4, 0) node[diamond, anchor = west, draw = red!70, inner sep = 0pt, inner sep = 0pt, minimum size = 5mm] (72) {$Z$} (02.north) -- ++(0, 0.4) node[diamond, anchor = south, draw = red!70, inner sep = 0pt, minimum size = 5mm] (03) {$Z$} (03.east) -- ++(0.4, 0) node[diamond, anchor = west, fill = red!70, inner sep = 0pt, minimum size = 5mm] (13) {} (13.east) -- ++(0.4, 0) node[diamond, anchor = west, draw = red!70, inner sep = 0pt, minimum size = 5mm] (23) {$Z$} (23.east) -- ++(0.4, 0) node[diamond, anchor = west, fill = red!70, inner sep = 0pt, minimum size = 5mm] (33) {} (33.east) -- ++(0.4, 0) node[diamond, anchor = west, draw = red!70, inner sep = 0pt, minimum size = 5mm] (43) {$Z$} (43.east) -- ++(0.4, 0) node[diamond, anchor = west, fill = red!70, inner sep = 0pt, minimum size = 5mm] (53) {} (53.east) -- ++(0.4, 0) node[diamond, anchor = west, draw = red!70, inner sep = 0pt, minimum size = 5mm] (63) {$Z$} (63.east) -- ++(0.4, 0) node[diamond, anchor = west, draw = red!70, inner sep = 0pt, minimum size = 5mm] (73) {$Z$} (03.north) -- ++(0, 0.4) node[diamond, anchor = south, draw = red!70, inner sep = 0pt, inner sep = 0pt, minimum size = 5mm] (04) {$Z$} (04.east) -- ++(0.4, 0) node[diamond, anchor = west, fill = red!70, inner sep = 0pt, minimum size = 5mm] (14) {} (14.east) -- ++(0.4, 0) node[diamond, anchor = west, draw = red!70, inner sep = 0pt, minimum size = 5mm] (24) {$X$} (24.east) -- ++(0.4, 0) node[diamond, anchor = west, fill = red!70, inner sep = 0pt, minimum size = 5mm] (34) {} (34.east) -- ++(0.4, 0) node[diamond, anchor = west, draw = red!70, inner sep = 0pt, minimum size = 5mm] (44) {$X$} (44.east) -- ++(0.4, 0) node[diamond, anchor = west, fill = red!70, inner sep = 0pt, minimum size = 5mm] (54) {} (54.east) -- ++(0.4, 0) node[diamond, anchor = west, draw = red!70, inner sep = 0pt, minimum size = 5mm] (64) {$X$} (64.east) -- ++(0.4, 0) node[diamond, anchor = west, draw = red!70, inner sep = 0pt, inner sep = 0pt, minimum size = 5mm] (74) {$Z$} (04.north) -- ++(0, 0.4) node[diamond, anchor = south, draw = red!70, inner sep = 0pt, minimum size = 5mm] (05) {$Z$} (05.east) -- ++(0.4, 0) node[diamond, anchor = west, fill = red!70, inner sep = 0pt, minimum size = 5mm] (15) {} (15.east) -- ++(0.4, 0) node[diamond, anchor = west, draw = red!70, inner sep = 0pt, minimum size = 5mm] (25) {$Z$} (25.east) -- ++(0.4, 0) node[diamond, anchor = west, fill = red!70, inner sep = 0pt, minimum size = 5mm] (35) {} (35.east) -- ++(0.4, 0) node[diamond, anchor = west, draw = red!70, inner sep = 0pt, minimum size = 5mm] (45) {$Z$} (45.east) -- ++(0.4, 0) node[diamond, anchor = west, fill = red!70, inner sep = 0pt, minimum size = 5mm] (55) {} (55.east) -- ++(0.4, 0) node[diamond, anchor = west, draw = red!70, inner sep = 0pt, minimum size = 5mm] (65) {$Z$} (65.east) -- ++(0.4, 0) node[diamond, anchor = west, draw = red!70, inner sep = 0pt, minimum size = 5mm] (75) {$Z$} (05.north) -- ++(0, 0.4) node[diamond, draw = red!70, inner sep = 0pt, minimum size = 5mm, anchor = south] (06) {$Z$} (06.east) -- ++(0.4, 0) node[diamond, anchor = west, draw = red!70, inner sep = 0pt, minimum size = 5mm] (16) {$Z$} (16.east) -- ++(0.4, 0) node[diamond, anchor = west, draw = red!70, inner sep = 0pt, minimum size = 5mm] (26) {$Z$} (26.east) -- ++(0.4, 0) node[diamond, anchor = west, draw = red!70, inner sep = 0pt, minimum size = 5mm] (36) {$Z$} (36.east) -- ++(0.4, 0) node[diamond, anchor = west, draw = red!70, inner sep = 0pt, minimum size = 5mm] (46) {$Z$} (46.east) -- ++(0.4, 0) node[diamond, anchor = west, draw = red!70, inner sep = 0pt, minimum size = 5mm] (56) {$Z$} (56.east) -- ++(0.4, 0) node[diamond, anchor = west, draw = red!70, inner sep = 0pt, minimum size = 5mm] (66) {$Z$} (66.east) -- ++(0.4, 0) node[diamond, anchor = west, draw = red!70, inner sep = 0pt, minimum size = 5mm] (76) {$Z$} (06.north) -- ++(0, 0.4) node[diamond, draw = red!70, inner sep = 0pt, minimum size = 5mm, anchor = south] (07) {$Z$} (07.east) -- ++(0.4, 0) node[diamond, anchor = west, draw = red!70, inner sep = 0pt, minimum size = 5mm] (17) {$Z$} (17.east) -- ++(0.4, 0) node[diamond, anchor = west, draw = red!70, inner sep = 0pt, minimum size = 5mm] (27) {$Z$} (27.east) -- ++(0.4, 0) node[diamond, anchor = west, draw = red!70, inner sep = 0pt, minimum size = 5mm] (37) {$Z$} (37.east) -- ++(0.4, 0) node[diamond, anchor = west, draw = red!70, inner sep = 0pt, minimum size = 5mm] (47) {$Z$} (47.east) -- ++(0.4, 0) node[diamond, anchor = west, draw = red!70, inner sep = 0pt, minimum size = 5mm] (57) {$Z$} (57.east) -- ++(0.4, 0) node[diamond, anchor = west, draw = red!70, inner sep = 0pt, minimum size = 5mm] (67) {$Z$} (67.east) -- ++(0.4, 0) node[diamond, anchor = west, draw = red!70, inner sep = 0pt, minimum size = 5mm] (77) {$Z$};
\foreach \x in {0,...,7}
\foreach \y [count=\yi] in {0,...,6}  
\draw[red] (\x\y) -- (\x\yi) (\y\x) -- (\yi\x);
\node at (3.4, -0.5) {\textbf{(b)}};
\end{tikzpicture}
\begin{tikzpicture}
\node[red!70, diamond, fill = red!70, minimum size = 5mm] at (0, 0) (00) {};
\draw[red!70] (00.east) -- ++(0.4, 0) node[diamond, anchor = west, draw = red!70, inner sep = 0pt, minimum size = 5mm] (10) {$Z$} (10.east) -- ++(0.4, 0) node[diamond, anchor = west, fill = red!70, minimum size = 5mm] (20) {} (20.east) -- ++(0.4, 0) node[diamond, anchor = west, draw = red!70, inner sep = 0pt, minimum size = 5mm] (30) {$Z$} (30.east) -- ++(0.4, 0) node[diamond, anchor = west, fill = red!70, minimum size = 5mm] (40) {} (40.east) -- ++(0.4, 0) node[diamond, anchor = west, draw = red!70, inner sep = 0pt, minimum size = 5mm] (50) {$Z$} (50.east) -- ++(0.4, 0) node[diamond, anchor = west, draw = red!70, inner sep = 0pt, minimum size = 5mm] (60) {$Z$} (60.east) -- ++(0.4, 0) node[diamond, anchor = west, draw = red!70, inner sep = 0pt, minimum size = 5mm] (70) {$Z$} (00.north) -- ++(0, 0.4) node[diamond, anchor = south, fill = red!70, minimum size = 5mm] (01) {} (01.east) -- ++(0.4, 0) node[diamond, anchor = west, draw = red!70, inner sep = 0pt, minimum size = 5mm] (11) {$X$} (11.east) -- ++(0.4, 0) node[diamond, anchor = west, fill = red!70, minimum size = 5mm] (21) {} (21.east) -- ++(0.4, 0) node[diamond, anchor = west, draw = red!70, inner sep = 0pt, minimum size = 5mm] (31) {$X$} (31.east) -- ++(0.4, 0) node[diamond, anchor = west, fill = red!70, minimum size = 5mm] (41) {} (41.east) -- ++(0.4, 0) node[diamond, anchor = west, draw = red!70, inner sep = 0pt, minimum size = 5mm] (51) {$X$} (51.east) -- ++(0.4, 0) node[diamond, anchor = west, fill = red!70, inner sep=0pt, minimum size = 5mm] (61) {} (61.east) -- ++(0.4, 0) node[diamond, anchor = west, draw = red!70, inner sep=0pt, minimum size = 5mm] (71) {$X$} (01.north) -- ++(0, 0.4) node[diamond, anchor = south, fill = red!70, inner sep = 0pt, minimum size = 5mm] (02) {$Z$} (02.east) -- ++(0.4, 0) node[diamond, anchor = west, draw = red!70, inner sep = 0pt, minimum size = 5mm] (12) {$Z$} (12.east) -- ++(0.4, 0) node[diamond, anchor = west, fill = red!70, minimum size = 5mm] (22) {} (22.east) -- ++(0.4, 0) node[diamond, anchor = west, draw = red!70, inner sep = 0pt, minimum size = 5mm] (32) {$Z$} (32.east) -- ++(0.4, 0) node[diamond, anchor = west, fill = red!70, minimum size = 5mm] (42) {} (42.east) -- ++(0.4, 0) node[diamond, anchor = west, draw = red!70, inner sep = 0pt, minimum size = 5mm] (52) {$Z$} (52.east) -- ++(0.4, 0) node[diamond, anchor = west, draw = red!70, inner sep = 0pt, minimum size = 5mm] (62) {$Z$} (62.east) -- ++(0.4, 0) node[diamond, anchor = west, fill = red!70, inner sep = 0pt, minimum size = 5mm] (72) {} (02.north) -- ++(0, 0.4) node[diamond, anchor = south, draw = red!70, inner sep = 0pt, minimum size = 5mm] (03) {$Z$} (03.east) -- ++(0.4, 0) node[diamond, anchor = west, fill = red!70, inner sep = 0pt, minimum size = 5mm] (13) {} (13.east) -- ++(0.4, 0) node[diamond, anchor = west, draw = red!70, inner sep = 0pt, minimum size = 5mm] (23) {$Z$} (23.east) -- ++(0.4, 0) node[diamond, anchor = west, fill = red!70, inner sep = 0pt, minimum size = 5mm] (33) {} (33.east) -- ++(0.4, 0) node[diamond, anchor = west, draw = red!70, inner sep = 0pt, minimum size = 5mm] (43) {$Z$} (43.east) -- ++(0.4, 0) node[diamond, anchor = west, fill = red!70, inner sep = 0pt, minimum size = 5mm] (53) {} (53.east) -- ++(0.4, 0) node[diamond, anchor = west, draw = red!70, inner sep = 0pt, minimum size = 5mm] (63) {$Z$} (63.east) -- ++(0.4, 0) node[diamond, anchor = west, draw = red!70, inner sep = 0pt, minimum size = 5mm] (73) {$X$} (03.north) -- ++(0, 0.4) node[diamond, anchor = south, fill = red!70, inner sep = 0pt, minimum size = 5mm] (04) {} (04.east) -- ++(0.4, 0) node[diamond, anchor = west, fill = red!70, inner sep = 0pt, minimum size = 5mm] (14) {} (14.east) -- ++(0.4, 0) node[diamond, anchor = west, draw = red!70, inner sep = 0pt, minimum size = 5mm] (24) {$X$} (24.east) -- ++(0.4, 0) node[diamond, anchor = west, fill = red!70, inner sep = 0pt, minimum size = 5mm] (34) {} (34.east) -- ++(0.4, 0) node[diamond, anchor = west, draw = red!70, inner sep = 0pt, minimum size = 5mm] (44) {$X$} (44.east) -- ++(0.4, 0) node[diamond, anchor = west, fill = red!70, inner sep = 0pt, minimum size = 5mm] (54) {} (54.east) -- ++(0.4, 0) node[diamond, anchor = west, draw = red!70, inner sep = 0pt, minimum size = 5mm] (64) {$X$} (64.east) -- ++(0.4, 0) node[diamond, anchor = west, fill = red!70, inner sep = 0pt, minimum size = 5mm] (74) {} (04.north) -- ++(0, 0.4) node[diamond, anchor = south, draw = red!70, inner sep = 0pt, minimum size = 5mm] (05) {$Z$} (05.east) -- ++(0.4, 0) node[diamond, anchor = west, fill = red!70, inner sep = 0pt, minimum size = 5mm] (15) {} (15.east) -- ++(0.4, 0) node[diamond, anchor = west, draw = red!70, inner sep = 0pt, minimum size = 5mm] (25) {$Z$} (25.east) -- ++(0.4, 0) node[diamond, anchor = west, fill = red!70, inner sep = 0pt, minimum size = 5mm] (35) {} (35.east) -- ++(0.4, 0) node[diamond, anchor = west, draw = red!70, inner sep = 0pt, minimum size = 5mm] (45) {$Z$} (45.east) -- ++(0.4, 0) node[diamond, anchor = west, fill = red!70, inner sep = 0pt, minimum size = 5mm] (55) {} (55.east) -- ++(0.4, 0) node[diamond, anchor = west, draw = red!70, inner sep = 0pt, minimum size = 5mm] (65) {$Z$} (65.east) -- ++(0.4, 0) node[diamond, anchor = west, draw = red!70, inner sep = 0pt, minimum size = 5mm] (75) {$X$} (05.north) -- ++(0, 0.4) node[diamond, draw = red!70, inner sep = 0pt, minimum size = 5mm, anchor = south] (06) {$Z$} (06.east) -- ++(0.4, 0) node[diamond, anchor = west, draw = red!70, inner sep = 0pt, minimum size = 5mm] (16) {$Z$} (16.east) -- ++(0.4, 0) node[diamond, anchor = west, draw = red!70, inner sep = 0pt, minimum size = 5mm] (26) {$Z$} (26.east) -- ++(0.4, 0) node[diamond, anchor = west, draw = red!70, inner sep = 0pt, minimum size = 5mm] (36) {$Z$} (36.east) -- ++(0.4, 0) node[diamond, anchor = west, draw = red!70, inner sep = 0pt, minimum size = 5mm] (46) {$Z$} (46.east) -- ++(0.4, 0) node[diamond, anchor = west, draw = red!70, inner sep = 0pt, minimum size = 5mm] (56) {$Z$} (56.east) -- ++(0.4, 0) node[diamond, anchor = west, draw = red!70, inner sep = 0pt, minimum size = 5mm] (66) {$Z$} (66.east) -- ++(0.4, 0) node[diamond, anchor = west, draw = red!70, inner sep = 0pt, minimum size = 5mm] (76) {$Z$} (06.north) -- ++(0, 0.4) node[diamond, draw = red!70, inner sep = 0pt, minimum size = 5mm, anchor = south] (07) {$Z$} (07.east) -- ++(0.4, 0) node[diamond, anchor = west, draw = red!70, inner sep = 0pt, minimum size = 5mm] (17) {$Z$} (17.east) -- ++(0.4, 0) node[diamond, anchor = west, draw = red!70, inner sep = 0pt, minimum size = 5mm] (27) {$Z$} (27.east) -- ++(0.4, 0) node[diamond, anchor = west, draw = red!70, inner sep = 0pt, minimum size = 5mm] (37) {$Z$} (37.east) -- ++(0.4, 0) node[diamond, anchor = west, draw = red!70, inner sep = 0pt, minimum size = 5mm] (47) {$Z$} (47.east) -- ++(0.4, 0) node[diamond, anchor = west, draw = red!70, inner sep = 0pt, minimum size = 5mm] (57) {$Z$} (57.east) -- ++(0.4, 0) node[diamond, anchor = west, draw = red!70, inner sep = 0pt, minimum size = 5mm] (67) {$Z$} (67.east) -- ++(0.4, 0) node[diamond, anchor = west, draw = red!70, inner sep = 0pt, minimum size = 5mm] (77) {$Z$};
\foreach \x in {0,...,7}
\foreach \y [count=\yi] in {0,...,6}  
\draw[red] (\x\y) -- (\x\yi) (\y\x) -- (\yi\x);
\node at (3.4, -0.5) {\textbf{(c)}};
\end{tikzpicture}
\hspace{.5cm}
\begin{tikzpicture}
\node[red!70, diamond, fill = red!70, minimum size = 5mm] at (0, 0) (00) {};
\draw[red!70] (00.east) -- ++(0.4, 0) node[diamond, anchor = west, draw = red!70, inner sep = 0pt, minimum size = 5mm] (10) {$Z$} (10.east) -- ++(0.4, 0) node[diamond, anchor = west, fill = red!70, minimum size = 5mm] (20) {} (20.east) -- ++(0.4, 0) node[diamond, anchor = west, draw = red!70, inner sep = 0pt, minimum size = 5mm] (30) {$Z$} (30.east) -- ++(0.4, 0) node[diamond, anchor = west, fill = red!70, minimum size = 5mm] (40) {} (40.east) -- ++(0.4, 0) node[diamond, anchor = west, draw = red!70, inner sep = 0pt, minimum size = 5mm] (50) {$Z$} (50.east) -- ++(0.4, 0) node[diamond, anchor = west, draw = red!70, inner sep = 0pt, minimum size = 5mm] (60) {$Z$} (60.east) -- ++(0.4, 0) node[diamond, anchor = west, draw = red!70, inner sep = 0pt, minimum size = 5mm] (70) {$Z$} (00.north) -- ++(0, 0.4) node[diamond, anchor = south, fill = red!70, minimum size = 5mm] (01) {} (01.east) -- ++(0.4, 0) node[diamond, anchor = west, draw = red!70, inner sep = 0pt, minimum size = 5mm] (11) {$X$} (11.east) -- ++(0.4, 0) node[diamond, anchor = west, fill = red!70, minimum size = 5mm] (21) {} (21.east) -- ++(0.4, 0) node[diamond, anchor = west, draw = red!70, inner sep = 0pt, minimum size = 5mm] (31) {$X$} (31.east) -- ++(0.4, 0) node[diamond, anchor = west, fill = red!70, minimum size = 5mm] (41) {} (41.east) -- ++(0.4, 0) node[diamond, anchor = west, draw = red!70, inner sep = 0pt, minimum size = 5mm] (51) {$X$} (51.east) -- ++(0.4, 0) node[diamond, anchor = west, fill = red!70, inner sep=0pt, minimum size = 5mm] (61) {} (61.east) -- ++(0.4, 0) node[diamond, anchor = west, draw = red!70, inner sep=0pt, minimum size = 5mm] (71) {$X$} (01.north) -- ++(0, 0.4) node[diamond, anchor = south, fill = red!70, inner sep = 0pt, minimum size = 5mm] (02) {$Z$} (02.east) -- ++(0.4, 0) node[diamond, anchor = west, draw = red!70, inner sep = 0pt, minimum size = 5mm] (12) {$Z$} (12.east) -- ++(0.4, 0) node[diamond, anchor = west, fill = red!70, minimum size = 5mm] (22) {} (22.east) -- ++(0.4, 0) node[diamond, anchor = west, draw = red!70, inner sep = 0pt, minimum size = 5mm] (32) {$Z$} (32.east) -- ++(0.4, 0) node[diamond, anchor = west, fill = red!70, minimum size = 5mm] (42) {} (42.east) -- ++(0.4, 0) node[diamond, anchor = west, draw = red!70, inner sep = 0pt, minimum size = 5mm] (52) {$Z$} (52.east) -- ++(0.4, 0) node[diamond, anchor = west, draw = red!70, inner sep = 0pt, minimum size = 5mm] (62) {$Z$} (62.east) -- ++(0.4, 0) node[diamond, anchor = west, fill = red!70, inner sep = 0pt, minimum size = 5mm] (72) {} (02.north) -- ++(0, 0.4) node[diamond, anchor = south, draw = red!70, inner sep = 0pt, minimum size = 5mm] (03) {$Z$} (03.east) -- ++(0.4, 0) node[diamond, anchor = west, fill = red!70, inner sep = 0pt, minimum size = 5mm] (13) {} (13.east) -- ++(0.4, 0) node[diamond, anchor = west, draw = red!70, inner sep = 0pt, minimum size = 5mm] (23) {$Z$} (23.east) -- ++(0.4, 0) node[diamond, anchor = west, fill = red!70, inner sep = 0pt, minimum size = 5mm] (33) {} (33.east) -- ++(0.4, 0) node[diamond, anchor = west, draw = red!70, inner sep = 0pt, minimum size = 5mm] (43) {$Z$} (43.east) -- ++(0.4, 0) node[diamond, anchor = west, fill = red!70, inner sep = 0pt, minimum size = 5mm] (53) {} (53.east) -- ++(0.4, 0) node[diamond, anchor = west, draw = red!70, inner sep = 0pt, minimum size = 5mm] (63) {$Z$} (63.east) -- ++(0.4, 0) node[diamond, anchor = west, draw = red!70, inner sep = 0pt, minimum size = 5mm] (73) {$X$} (03.north) -- ++(0, 0.4) node[diamond, anchor = south, fill = red!70, inner sep = 0pt, minimum size = 5mm] (04) {} (04.east) -- ++(0.4, 0) node[diamond, anchor = west, fill = red!70, inner sep = 0pt, minimum size = 5mm] (14) {} (14.east) -- ++(0.4, 0) node[diamond, anchor = west, draw = red!70, inner sep = 0pt, minimum size = 5mm] (24) {$X$} (24.east) -- ++(0.4, 0) node[diamond, anchor = west, fill = red!70, inner sep = 0pt, minimum size = 5mm] (34) {} (34.east) -- ++(0.4, 0) node[diamond, anchor = west, draw = red!70, inner sep = 0pt, minimum size = 5mm] (44) {$X$} (44.east) -- ++(0.4, 0) node[diamond, anchor = west, fill = red!70, inner sep = 0pt, minimum size = 5mm] (54) {} (54.east) -- ++(0.4, 0) node[diamond, anchor = west, draw = red!70, inner sep = 0pt, minimum size = 5mm] (64) {$X$} (64.east) -- ++(0.4, 0) node[diamond, anchor = west, fill = red!70, inner sep = 0pt, minimum size = 5mm] (74) {} (04.north) -- ++(0, 0.4) node[diamond, anchor = south, draw = red!70, inner sep = 0pt, minimum size = 5mm] (05) {$Z$} (05.east) -- ++(0.4, 0) node[diamond, anchor = west, fill = red!70, inner sep = 0pt, minimum size = 5mm] (15) {} (15.east) -- ++(0.4, 0) node[diamond, anchor = west, draw = red!70, inner sep = 0pt, minimum size = 5mm] (25) {$Z$} (25.east) -- ++(0.4, 0) node[diamond, anchor = west, fill = red!70, inner sep = 0pt, minimum size = 5mm] (35) {} (35.east) -- ++(0.4, 0) node[diamond, anchor = west, draw = red!70, inner sep = 0pt, minimum size = 5mm] (45) {$Z$} (45.east) -- ++(0.4, 0) node[diamond, anchor = west, fill = red!70, inner sep = 0pt, minimum size = 5mm] (55) {} (55.east) -- ++(0.4, 0) node[diamond, anchor = west, draw = red!70, inner sep = 0pt, minimum size = 5mm] (65) {$Z$} (65.east) -- ++(0.4, 0) node[diamond, anchor = west, draw = red!70, inner sep = 0pt, minimum size = 5mm] (75) {$X$} (05.north) -- ++(0, 0.4) node[diamond, fill = red!70, minimum size = 5mm, anchor = south] (06) {} (06.east) -- ++(0.4, 0) node[diamond, anchor = west, draw = red!70, inner sep = 0pt, minimum size = 5mm] (16) {$Z$} (16.east) -- ++(0.4, 0) node[diamond, anchor = west, fill = red!70, minimum size = 5mm] (26) {} (26.east) -- ++(0.4, 0) node[diamond, anchor = west, draw = red!70, inner sep = 0pt, minimum size = 5mm] (36) {$Z$} (36.east) -- ++(0.4, 0) node[diamond, anchor = west, fill = red!70, minimum size = 5mm] (46) {} (46.east) -- ++(0.4, 0) node[diamond, anchor = west, draw = red!70, inner sep = 0pt, minimum size = 5mm] (56) {$Z$} (56.east) -- ++(0.4, 0) node[diamond, anchor = west, draw = red!70, inner sep = 0pt, minimum size = 5mm] (66) {$Z$} (66.east) -- ++(0.4, 0) node[diamond, anchor = west, draw = red!70, inner sep = 0pt, minimum size = 5mm] (76) {$Z$} (06.north) -- ++(0, 0.4) node[diamond, fill = red!70, minimum size = 5mm, anchor = south] (07) {} (07.east) -- ++(0.4, 0) node[diamond, anchor = west, draw = red!70, inner sep = 0pt, minimum size = 5mm] (17) {$X$} (17.east) -- ++(0.4, 0) node[diamond, anchor = west, fill = red!70, minimum size = 5mm] (27) {} (27.east) -- ++(0.4, 0) node[diamond, anchor = west, draw = red!70, inner sep = 0pt, minimum size = 5mm] (37) {$X$} (37.east) -- ++(0.4, 0) node[diamond, anchor = west, fill = red!70, minimum size = 5mm] (47) {} (47.east) -- ++(0.4, 0) node[diamond, anchor = west, draw = red!70, inner sep = 0pt, minimum size = 5mm] (57) {$X$} (57.east) -- ++(0.4, 0) node[diamond, anchor = west, draw = red!70, inner sep = 0pt, minimum size = 5mm] (67) {$Z$} (67.east) -- ++(0.4, 0) node[diamond, anchor = west, draw = red!70, inner sep = 0pt, minimum size = 5mm] (77) {$Z$};
\foreach \x in {0,...,7}
\foreach \y [count=\yi] in {0,...,6}  
\draw[red] (\x\y) -- (\x\yi) (\y\x) -- (\yi\x);
\node at (3.4, -0.5) {\textbf{(d)}};
\end{tikzpicture}
\begin{tikzpicture}
\node[red!70, diamond, fill = red!70, minimum size = 5mm] at (0, 0) (00) {};
\draw[red!70] (00.east) -- ++(0.4, 0) node[diamond, anchor = west, draw = red!70, inner sep = 0pt, minimum size = 5mm] (10) {$Z$} (10.east) -- ++(0.4, 0) node[diamond, anchor = west, fill = red!70, minimum size = 5mm] (20) {} (20.east) -- ++(0.4, 0) node[diamond, anchor = west, draw = red!70, inner sep = 0pt, minimum size = 5mm] (30) {$Z$} (30.east) -- ++(0.4, 0) node[diamond, anchor = west, fill = red!70, minimum size = 5mm] (40) {} (40.east) -- ++(0.4, 0) node[diamond, anchor = west, draw = red!70, inner sep = 0pt, minimum size = 5mm] (50) {$Z$} (50.east) -- ++(0.4, 0) node[diamond, anchor = west, fill = red!70, minimum size = 5mm] (60) {} (60.east) -- ++(0.4, 0) node[diamond, anchor = west, draw = red!70, inner sep = 0pt, minimum size = 5mm] (70) {$Z$} (00.north) -- ++(0, 0.4) node[diamond, anchor = south, fill = red!70, minimum size = 5mm] (01) {} (01.east) -- ++(0.4, 0) node[diamond, anchor = west, draw = red!70, inner sep = 0pt, minimum size = 5mm] (11) {$X$} (11.east) -- ++(0.4, 0) node[diamond, anchor = west, fill = red!70, minimum size = 5mm] (21) {} (21.east) -- ++(0.4, 0) node[diamond, anchor = west, draw = red!70, inner sep = 0pt, minimum size = 5mm] (31) {$X$} (31.east) -- ++(0.4, 0) node[diamond, anchor = west, fill = red!70, minimum size = 5mm] (41) {} (41.east) -- ++(0.4, 0) node[diamond, anchor = west, draw = red!70, inner sep = 0pt, minimum size = 5mm] (51) {$X$} (51.east) -- ++(0.4, 0) node[diamond, anchor = west, fill = red!70, inner sep=0pt, minimum size = 5mm] (61) {} (61.east) -- ++(0.4, 0) node[diamond, anchor = west, draw = red!70, inner sep=0pt, minimum size = 5mm] (71) {$X$} (01.north) -- ++(0, 0.4) node[diamond, anchor = south, fill = red!70, inner sep = 0pt, minimum size = 5mm] (02) {$Z$} (02.east) -- ++(0.4, 0) node[diamond, anchor = west, draw = red!70, inner sep = 0pt, minimum size = 5mm] (12) {$Z$} (12.east) -- ++(0.4, 0) node[diamond, anchor = west, fill = red!70, minimum size = 5mm] (22) {} (22.east) -- ++(0.4, 0) node[diamond, anchor = west, draw = red!70, inner sep = 0pt, minimum size = 5mm] (32) {$Z$} (32.east) -- ++(0.4, 0) node[diamond, anchor = west, fill = red!70, minimum size = 5mm] (42) {} (42.east) -- ++(0.4, 0) node[diamond, anchor = west, draw = red!70, inner sep = 0pt, minimum size = 5mm] (52) {$Z$} (52.east) -- ++(0.4, 0) node[diamond, anchor = west, draw = red!70, inner sep = 0pt, minimum size = 5mm] (62) {$Z$} (62.east) -- ++(0.4, 0) node[diamond, anchor = west, fill = red!70, inner sep = 0pt, minimum size = 5mm] (72) {} (02.north) -- ++(0, 0.4) node[diamond, anchor = south, draw = red!70, inner sep = 0pt, minimum size = 5mm] (03) {$Z$} (03.east) -- ++(0.4, 0) node[diamond, anchor = west, fill = red!70, inner sep = 0pt, minimum size = 5mm] (13) {} (13.east) -- ++(0.4, 0) node[diamond, anchor = west, draw = red!70, inner sep = 0pt, minimum size = 5mm] (23) {$Z$} (23.east) -- ++(0.4, 0) node[diamond, anchor = west, fill = red!70, inner sep = 0pt, minimum size = 5mm] (33) {} (33.east) -- ++(0.4, 0) node[diamond, anchor = west, draw = red!70, inner sep = 0pt, minimum size = 5mm] (43) {$Z$} (43.east) -- ++(0.4, 0) node[diamond, anchor = west, fill = red!70, inner sep = 0pt, minimum size = 5mm] (53) {} (53.east) -- ++(0.4, 0) node[diamond, anchor = west, draw = red!70, inner sep = 0pt, minimum size = 5mm] (63) {$Z$} (63.east) -- ++(0.4, 0) node[diamond, anchor = west, draw = red!70, inner sep = 0pt, minimum size = 5mm] (73) {$X$} (03.north) -- ++(0, 0.4) node[diamond, anchor = south, fill = red!70, inner sep = 0pt, minimum size = 5mm] (04) {} (04.east) -- ++(0.4, 0) node[diamond, anchor = west, fill = red!70, inner sep = 0pt, minimum size = 5mm] (14) {} (14.east) -- ++(0.4, 0) node[diamond, anchor = west, draw = red!70, inner sep = 0pt, minimum size = 5mm] (24) {$X$} (24.east) -- ++(0.4, 0) node[diamond, anchor = west, fill = red!70, inner sep = 0pt, minimum size = 5mm] (34) {} (34.east) -- ++(0.4, 0) node[diamond, anchor = west, draw = red!70, inner sep = 0pt, minimum size = 5mm] (44) {$X$} (44.east) -- ++(0.4, 0) node[diamond, anchor = west, fill = red!70, inner sep = 0pt, minimum size = 5mm] (54) {} (54.east) -- ++(0.4, 0) node[diamond, anchor = west, draw = red!70, inner sep = 0pt, minimum size = 5mm] (64) {$X$} (64.east) -- ++(0.4, 0) node[diamond, anchor = west, fill = red!70, inner sep = 0pt, minimum size = 5mm] (74) {} (04.north) -- ++(0, 0.4) node[diamond, anchor = south, draw = red!70, inner sep = 0pt, minimum size = 5mm] (05) {$Z$} (05.east) -- ++(0.4, 0) node[diamond, anchor = west, fill = red!70, inner sep = 0pt, minimum size = 5mm] (15) {} (15.east) -- ++(0.4, 0) node[diamond, anchor = west, draw = red!70, inner sep = 0pt, minimum size = 5mm] (25) {$Z$} (25.east) -- ++(0.4, 0) node[diamond, anchor = west, fill = red!70, inner sep = 0pt, minimum size = 5mm] (35) {} (35.east) -- ++(0.4, 0) node[diamond, anchor = west, draw = red!70, inner sep = 0pt, minimum size = 5mm] (45) {$Z$} (45.east) -- ++(0.4, 0) node[diamond, anchor = west, fill = red!70, inner sep = 0pt, minimum size = 5mm] (55) {} (55.east) -- ++(0.4, 0) node[diamond, anchor = west, draw = red!70, inner sep = 0pt, minimum size = 5mm] (65) {$Z$} (65.east) -- ++(0.4, 0) node[diamond, anchor = west, draw = red!70, inner sep = 0pt, minimum size = 5mm] (75) {$X$} (05.north) -- ++(0, 0.4) node[diamond, fill = red!70, minimum size = 5mm, anchor = south] (06) {} (06.east) -- ++(0.4, 0) node[diamond, anchor = west, draw = red!70, inner sep = 0pt, minimum size = 5mm] (16) {$Z$} (16.east) -- ++(0.4, 0) node[diamond, anchor = west, fill = red!70, minimum size = 5mm] (26) {} (26.east) -- ++(0.4, 0) node[diamond, anchor = west, draw = red!70, inner sep = 0pt, minimum size = 5mm] (36) {$Z$} (36.east) -- ++(0.4, 0) node[diamond, anchor = west, fill = red!70, minimum size = 5mm] (46) {} (46.east) -- ++(0.4, 0) node[diamond, anchor = west, draw = red!70, inner sep = 0pt, minimum size = 5mm] (56) {$Z$} (56.east) -- ++(0.4, 0) node[diamond, anchor = west, draw = red!70, inner sep = 0pt, minimum size = 5mm] (66) {$Z$} (66.east) -- ++(0.4, 0) node[diamond, anchor = west, fill = red!70, minimum size = 5mm] (76) {} (06.north) -- ++(0, 0.4) node[diamond, fill = red!70, minimum size = 5mm, anchor = south] (07) {} (07.east) -- ++(0.4, 0) node[diamond, anchor = west, draw = red!70, inner sep = 0pt, minimum size = 5mm] (17) {$X$} (17.east) -- ++(0.4, 0) node[diamond, anchor = west, fill = red!70, minimum size = 5mm] (27) {} (27.east) -- ++(0.4, 0) node[diamond, anchor = west, draw = red!70, inner sep = 0pt, minimum size = 5mm] (37) {$X$} (37.east) -- ++(0.4, 0) node[diamond, anchor = west, fill = red!70, minimum size = 5mm] (47) {} (47.east) -- ++(0.4, 0) node[diamond, anchor = west, draw = red!70, inner sep = 0pt, minimum size = 5mm] (57) {$X$} (57.east) -- ++(0.4, 0) node[diamond, anchor = west, fill = red!70, minimum size = 5mm] (67) {} (67.east) -- ++(0.4, 0) node[diamond, anchor = west, draw = red!70, inner sep = 0pt, minimum size = 5mm] (77) {$X$};
\foreach \x in {0,...,7}
\foreach \y [count=\yi] in {0,...,6}  
\draw[red] (\x\y) -- (\x\yi) (\y\x) -- (\yi\x);
\node at (3.4, -0.5) {\textbf{(e)}};
\end{tikzpicture}
\hspace{.5cm}
\begin{tikzpicture}
\node[red!70, diamond, fill = red!70, minimum size = 5mm] at (0, 0) (00) {};
\draw[red!70] (00.north) -- ++(0, 0.4) node[diamond, anchor = south, fill = red!70, minimum size = 5mm] (01) {} (01.east) -- ++(0.4, 0) node[diamond, anchor = west, draw = red!70, inner sep = 0pt, minimum size = 5mm] (11) {$X$} (11.east) -- ++(0.4, 0) node[diamond, anchor = west, fill = red!70, minimum size = 5mm] (21) {} (21.south) -- ++(0, -0.4) node[diamond, anchor = north, fill = red!70, minimum size = 5mm] (20) {} (21.east) -- ++(0.4, 0) node[diamond, anchor = west, draw = red!70, inner sep = 0pt, minimum size = 5mm] (31) {$X$} (31.east) -- ++(0.4, 0) node[diamond, anchor = west, fill = red!70, minimum size = 5mm] (41) {} (41.south) -- ++(0, -0.4) node[diamond, anchor = north, fill = red!70, minimum size = 5mm] (40) {} (41.east) -- ++(0.4, 0) node[diamond, anchor = west, draw = red!70, inner sep = 0pt, minimum size = 5mm] (51) {$X$} (51.east) -- ++(0.4, 0) node[diamond, anchor = west, fill = red!70, inner sep=0pt, minimum size = 5mm] (61) {} (61.south) -- ++(0, -0.4) node[diamond, anchor = north, fill = red!70, minimum size = 5mm] (60) {} (61.east) -- ++(0.4, 0) node[diamond, anchor = west, draw = red!70, inner sep=0pt, minimum size = 5mm] (71) {$X$} (01.north) -- ++(0, 0.4) node[diamond, anchor = south, fill = red!70, inner sep = 0pt, minimum size = 5mm] (02) {} (21.north) -- ++(0, 0.4) node[diamond, anchor = south, fill = red!70, minimum size = 5mm] (22) {} (41.north) -- ++(0, 0.4) node[diamond, anchor = south, fill = red!70, minimum size = 5mm] (42) {} (71.north) -- ++(0, 0.4) node[diamond, anchor = south, fill = red!70, inner sep = 0pt, minimum size = 5mm] (72) {} (72.north) -- ++(0, 0.4) node[diamond, anchor = south, draw = red!70, inner sep = 0pt, minimum size = 5mm] (73) {$X$} (73.north) -- ++(0, 0.4) node[diamond, anchor = south, fill = red!70, inner sep = 0pt, minimum size = 5mm] (74) {} (74.west) -- ++(-0.4, 0) node[diamond, anchor = east, draw = red!70, inner sep = 0pt, minimum size = 5mm] (64) {$X$} (64.west) -- ++(-0.4, 0) node[diamond, anchor = east, fill = red!70, inner sep = 0pt, minimum size = 5mm] (54) {} (54.west) -- ++(-0.4, 0) node[diamond, anchor = east, draw = red!70, inner sep = 0pt, minimum size = 5mm] (44) {$X$} (54.south) -- ++(0, -0.4) node[diamond, anchor = north, fill = red!70, inner sep = 0pt, minimum size = 5mm] (53) {} (54.north) -- ++(0, 0.4) node[diamond, anchor = south, fill = red!70, minimum size = 5mm] (63) {} (44.west) -- ++(-0.4, 0) node[diamond, anchor = east, fill = red!70, minimum size = 5mm] (34) {} (34.north) -- ++(0, 0.4) node[diamond, anchor = south, fill = red!70, minimum size = 5mm] (35) {} (34.south) -- ++(0, -0.4) node[diamond, anchor = north, fill = red!70, minimum size = 5mm] (33) {} (34.west) -- ++(-0.4, 0) node[diamond, anchor = east, draw = red!70, inner sep = 0pt, minimum size = 5mm] (24) {$X$} (24.west) -- ++(-0.4, 0) node[diamond, anchor = east, fill = red!70, minimum size = 5mm] (14) {} (14.west) -- ++(-0.4, 0) node[diamond, anchor = east, fill = red!70, minimum size = 5mm] (04) {} (14.south) -- ++(0, -0.4) node[diamond, anchor = north, fill = red!70, minimum size = 5mm] (13) {} (14.north) -- ++(0, 0.4) node[diamond, anchor = south, fill = red!70, minimum size = 5mm] (15) {} (74.north) -- ++(0, 0.4) node[diamond, anchor = south, draw = red!70, inner sep = 0pt, minimum size = 5mm] (75) {$X$} (75.north) -- ++(0, 0.4) node[diamond, anchor = south, fill = red!70, inner sep = 0pt, minimum size = 5mm] (76) {} (76.north) -- ++(0, 0.4) node[diamond, anchor = south, draw = red!70, inner sep = 0pt, minimum size = 5mm] (77) {$X$} (77.west) -- ++(-0.4, 0) node[diamond, anchor = east, fill = red!70, minimum size = 5mm] (67) {} (67.west) -- ++(-0.4, 0) node[diamond, anchor = east, draw = red!70, inner sep = 0pt, minimum size = 5mm] (57) {$X$} (57.west) -- ++(-0.4, 0) node[diamond, anchor = east, fill = red!70, minimum size = 5mm] (47) {} (47.west) -- ++(-0.4, 0) node[diamond, anchor = east, draw = red!70, inner sep = 0pt, minimum size = 5mm] (37) {$X$} (47.south) -- ++(0, -0.4) node[diamond, anchor = north, fill = red!70, minimum size = 5mm] (46) {} (37.west) -- ++(-0.4, 0) node[diamond, fill = red!70, minimum size = 5mm, anchor = east] (27) {} (27.west) -- ++(-0.4, 0) node[diamond, anchor = east, draw = red!70, inner sep = 0pt, minimum size = 5mm] (17) {$X$} (17.west) -- ++(-0.4, 0) node[diamond, anchor = east, fill = red!70, inner sep = 0pt, minimum size = 5mm] (07) {} (07.south) -- ++(0, -0.4) node[diamond, anchor = north, fill = red!70, minimum size = 5mm] (06) {} (27.south) -- ++(0, -0.4) node[diamond, anchor = north, fill = red!70, minimum size = 5mm] (26) {};
\node at (3.4, -0.5) {\textbf{(f)}};
\end{tikzpicture}
\caption{A step-by-step construction of repeater tree from an $8\times 8$ grid using Algorithm~\ref{alg:const}. (a) Step~\ref{S:init}, (b) Step~\ref{S:eij1} and~\ref{S:eij2}, (c) Step~\ref{S:eix} and~\ref{S:ei}, (d) Step~\ref{S:e4j}, (e) Step~\ref{S:e4},~\ref{S:e4x} and~\ref{S:e} and (f) the repeater tree have been shown here.}
\end{figure*}

% \nocite{*}

\bibliography{apssamp}% Produces the bibliography via BibTeX.

\begin{thebibliography}{50}%
\makeatletter
\providecommand \@ifxundefined [1]{%
 \@ifx{#1\undefined}
}%
\providecommand \@ifnum [1]{%
 \ifnum #1\expandafter \@firstoftwo
 \else \expandafter \@secondoftwo
 \fi
}%
\providecommand \@ifx [1]{%
 \ifx #1\expandafter \@firstoftwo
 \else \expandafter \@secondoftwo
 \fi
}%
\providecommand \natexlab [1]{#1}%
\providecommand \enquote  [1]{``#1''}%
\providecommand \bibnamefont  [1]{#1}%
\providecommand \bibfnamefont [1]{#1}%
\providecommand \citenamefont [1]{#1}%
\providecommand \href@noop [0]{\@secondoftwo}%
\providecommand \href [0]{\begingroup \@sanitize@url \@href}%
\providecommand \@href[1]{\@@startlink{#1}\@@href}%
\providecommand \@@href[1]{\endgroup#1\@@endlink}%
\providecommand \@sanitize@url [0]{\catcode `\\12\catcode `\$12\catcode `\&12\catcode `\#12\catcode `\^12\catcode `\_12\catcode `\%12\relax}%
\providecommand \@@startlink[1]{}%
\providecommand \@@endlink[0]{}%
\providecommand \url  [0]{\begingroup\@sanitize@url \@url }%
\providecommand \@url [1]{\endgroup\@href {#1}{\urlprefix }}%
\providecommand \urlprefix  [0]{URL }%
\providecommand \Eprint [0]{\href }%
\providecommand \doibase [0]{http://dx.doi.org/}%
\providecommand \selectlanguage [0]{\@gobble}%
\providecommand \bibinfo  [0]{\@secondoftwo}%
\providecommand \bibfield  [0]{\@secondoftwo}%
\providecommand \translation [1]{[#1]}%
\providecommand \BibitemOpen [0]{}%
\providecommand \bibitemStop [0]{}%
\providecommand \bibitemNoStop [0]{.\EOS\space}%
\providecommand \EOS [0]{\spacefactor3000\relax}%
\providecommand \BibitemShut  [1]{\csname bibitem#1\endcsname}%
\let\auto@bib@innerbib\@empty
%</preamble>
\bibitem [{\citenamefont {Sharma}\ \emph {et~al.}(2023)\citenamefont {Sharma}, \citenamefont {Ramkumar}, \citenamefont {Kaur}, \citenamefont {Hasija}, \citenamefont {Mittal},\ and\ \citenamefont {Singh}}]{sharma2023post}%
  \BibitemOpen
  \bibfield  {author} {\bibinfo {author} {\bibfnamefont {S.}~\bibnamefont {Sharma}}, \bibinfo {author} {\bibfnamefont {K.~R.}\ \bibnamefont {Ramkumar}}, \bibinfo {author} {\bibfnamefont {A.}~\bibnamefont {Kaur}}, \bibinfo {author} {\bibfnamefont {T.}~\bibnamefont {Hasija}}, \bibinfo {author} {\bibfnamefont {S.}~\bibnamefont {Mittal}}, \ and\ \bibinfo {author} {\bibfnamefont {B.}~\bibnamefont {Singh}},\ }in\ \href {https://doi.org/10.1007/978-981-19-6383-4_3} {\emph {\bibinfo {booktitle} {Modern Electronics Devices and Communication Systems}}},\ \bibinfo {editor} {edited by\ \bibinfo {editor} {\bibfnamefont {R.}~\bibnamefont {Agrawal}}, \bibinfo {editor} {\bibfnamefont {C.}~\bibnamefont {Kishore~Singh}}, \bibinfo {editor} {\bibfnamefont {A.}~\bibnamefont {Goyal}}, \ and\ \bibinfo {editor} {\bibfnamefont {D.~K.}\ \bibnamefont {Singh}}}\ (\bibinfo  {publisher} {Springer Nature Singapore},\ \bibinfo {address} {Singapore},\ \bibinfo {year} {2023})\ pp.\ \bibinfo {pages} {23--38}\BibitemShut {NoStop}%
\bibitem [{\citenamefont {Shor}(1994)}]{365700}%
  \BibitemOpen
  \bibfield  {author} {\bibinfo {author} {\bibfnamefont {P.}~\bibnamefont {Shor}},\ }in\ \href {\doibase 10.1109/SFCS.1994.365700} {\emph {\bibinfo {booktitle} {Proceedings 35th Annual Symposium on Foundations of Computer Science}}}\ (\bibinfo {year} {1994})\ pp.\ \bibinfo {pages} {124--134}\BibitemShut {NoStop}%
\bibitem [{\citenamefont {Grover}(1996)}]{10.1145/237814.237866}%
  \BibitemOpen
  \bibfield  {author} {\bibinfo {author} {\bibfnamefont {L.~K.}\ \bibnamefont {Grover}},\ }in\ \href {\doibase 10.1145/237814.237866} {\emph {\bibinfo {booktitle} {Proceedings of the Twenty-Eighth Annual ACM Symposium on Theory of Computing}}},\ \bibinfo {series and number} {STOC '96}\ (\bibinfo  {publisher} {Association for Computing Machinery},\ \bibinfo {address} {New York, NY, USA},\ \bibinfo {year} {1996})\ p.\ \bibinfo {pages} {212–219}\BibitemShut {NoStop}%
\bibitem [{\citenamefont {Bennett}\ and\ \citenamefont {Brassard}(2014)}]{BENNETT20147}%
  \BibitemOpen
  \bibfield  {author} {\bibinfo {author} {\bibfnamefont {C.~H.}\ \bibnamefont {Bennett}}\ and\ \bibinfo {author} {\bibfnamefont {G.}~\bibnamefont {Brassard}},\ }\href {\doibase https://doi.org/10.1016/j.tcs.2014.05.025} {\bibfield  {journal} {\bibinfo  {journal} {Theoretical Computer Science}\ }\textbf {\bibinfo {volume} {560}},\ \bibinfo {pages} {7} (\bibinfo {year} {2014})},\ \bibinfo {note} {theoretical Aspects of Quantum Cryptography – celebrating 30 years of BB84}\BibitemShut {NoStop}%
\bibitem [{\citenamefont {Ekert}(1991)}]{PhysRevLett.67.661}%
  \BibitemOpen
  \bibfield  {author} {\bibinfo {author} {\bibfnamefont {A.~K.}\ \bibnamefont {Ekert}},\ }\href {\doibase 10.1103/PhysRevLett.67.661} {\bibfield  {journal} {\bibinfo  {journal} {Phys. Rev. Lett.}\ }\textbf {\bibinfo {volume} {67}},\ \bibinfo {pages} {661} (\bibinfo {year} {1991})}\BibitemShut {NoStop}%
\bibitem [{\citenamefont {Lo}\ \emph {et~al.}(2005)\citenamefont {Lo}, \citenamefont {Ma},\ and\ \citenamefont {Chen}}]{PhysRevLett.94.230504}%
  \BibitemOpen
  \bibfield  {author} {\bibinfo {author} {\bibfnamefont {H.-K.}\ \bibnamefont {Lo}}, \bibinfo {author} {\bibfnamefont {X.}~\bibnamefont {Ma}}, \ and\ \bibinfo {author} {\bibfnamefont {K.}~\bibnamefont {Chen}},\ }\href {\doibase 10.1103/PhysRevLett.94.230504} {\bibfield  {journal} {\bibinfo  {journal} {Phys. Rev. Lett.}\ }\textbf {\bibinfo {volume} {94}},\ \bibinfo {pages} {230504} (\bibinfo {year} {2005})}\BibitemShut {NoStop}%
\bibitem [{\citenamefont {Liao}\ \emph {et~al.}(2017)\citenamefont {Liao}, \citenamefont {Cai}, \citenamefont {Liu}, \citenamefont {Zhang}, \citenamefont {Li}, \citenamefont {Ren}, \citenamefont {Yin}, \citenamefont {Shen}, \citenamefont {Cao}, \citenamefont {Li} \emph {et~al.}}]{liao2017satellite}%
  \BibitemOpen
  \bibfield  {author} {\bibinfo {author} {\bibfnamefont {S.-K.}\ \bibnamefont {Liao}}, \bibinfo {author} {\bibfnamefont {W.-Q.}\ \bibnamefont {Cai}}, \bibinfo {author} {\bibfnamefont {W.-Y.}\ \bibnamefont {Liu}}, \bibinfo {author} {\bibfnamefont {L.}~\bibnamefont {Zhang}}, \bibinfo {author} {\bibfnamefont {Y.}~\bibnamefont {Li}}, \bibinfo {author} {\bibfnamefont {J.-G.}\ \bibnamefont {Ren}}, \bibinfo {author} {\bibfnamefont {J.}~\bibnamefont {Yin}}, \bibinfo {author} {\bibfnamefont {Q.}~\bibnamefont {Shen}}, \bibinfo {author} {\bibfnamefont {Y.}~\bibnamefont {Cao}}, \bibinfo {author} {\bibfnamefont {Z.-P.}\ \bibnamefont {Li}},  \emph {et~al.},\ }\href {https://doi.org/10.1038/nature23655} {\bibfield  {journal} {\bibinfo  {journal} {Nature}\ }\textbf {\bibinfo {volume} {549}},\ \bibinfo {pages} {43} (\bibinfo {year} {2017})}\BibitemShut {NoStop}%
\bibitem [{\citenamefont {Xu}\ \emph {et~al.}(2020)\citenamefont {Xu}, \citenamefont {Ma}, \citenamefont {Zhang}, \citenamefont {Lo},\ and\ \citenamefont {Pan}}]{RevModPhys.92.025002}%
  \BibitemOpen
  \bibfield  {author} {\bibinfo {author} {\bibfnamefont {F.}~\bibnamefont {Xu}}, \bibinfo {author} {\bibfnamefont {X.}~\bibnamefont {Ma}}, \bibinfo {author} {\bibfnamefont {Q.}~\bibnamefont {Zhang}}, \bibinfo {author} {\bibfnamefont {H.-K.}\ \bibnamefont {Lo}}, \ and\ \bibinfo {author} {\bibfnamefont {J.-W.}\ \bibnamefont {Pan}},\ }\href {\doibase 10.1103/RevModPhys.92.025002} {\bibfield  {journal} {\bibinfo  {journal} {Rev. Mod. Phys.}\ }\textbf {\bibinfo {volume} {92}},\ \bibinfo {pages} {025002} (\bibinfo {year} {2020})}\BibitemShut {NoStop}%
\bibitem [{\citenamefont {Gr{\"u}nenfelder}\ \emph {et~al.}(2023)\citenamefont {Gr{\"u}nenfelder}, \citenamefont {Boaron}, \citenamefont {Resta}, \citenamefont {Perrenoud}, \citenamefont {Rusca}, \citenamefont {Barreiro}, \citenamefont {Houlmann}, \citenamefont {Sax}, \citenamefont {Stasi}, \citenamefont {El-Khoury} \emph {et~al.}}]{grunenfelder2023fast}%
  \BibitemOpen
  \bibfield  {author} {\bibinfo {author} {\bibfnamefont {F.}~\bibnamefont {Gr{\"u}nenfelder}}, \bibinfo {author} {\bibfnamefont {A.}~\bibnamefont {Boaron}}, \bibinfo {author} {\bibfnamefont {G.~V.}\ \bibnamefont {Resta}}, \bibinfo {author} {\bibfnamefont {M.}~\bibnamefont {Perrenoud}}, \bibinfo {author} {\bibfnamefont {D.}~\bibnamefont {Rusca}}, \bibinfo {author} {\bibfnamefont {C.}~\bibnamefont {Barreiro}}, \bibinfo {author} {\bibfnamefont {R.}~\bibnamefont {Houlmann}}, \bibinfo {author} {\bibfnamefont {R.}~\bibnamefont {Sax}}, \bibinfo {author} {\bibfnamefont {L.}~\bibnamefont {Stasi}}, \bibinfo {author} {\bibfnamefont {S.}~\bibnamefont {El-Khoury}},  \emph {et~al.},\ }\href {https://doi.org/10.1038/s41566-023-01168-2} {\bibfield  {journal} {\bibinfo  {journal} {Nature Photonics}\ }\textbf {\bibinfo {volume} {17}},\ \bibinfo {pages} {422} (\bibinfo {year} {2023})}\BibitemShut {NoStop}%
\bibitem [{\citenamefont {Das}\ and\ \citenamefont {Paul}(2020)}]{doi:10.1142/S0219749920500380}%
  \BibitemOpen
  \bibfield  {author} {\bibinfo {author} {\bibfnamefont {N.}~\bibnamefont {Das}}\ and\ \bibinfo {author} {\bibfnamefont {G.}~\bibnamefont {Paul}},\ }\href {\doibase 10.1142/S0219749920500380} {\bibfield  {journal} {\bibinfo  {journal} {International Journal of Quantum Information}\ }\textbf {\bibinfo {volume} {18}},\ \bibinfo {pages} {2050038} (\bibinfo {year} {2020})}\BibitemShut {NoStop}%
\bibitem [{\citenamefont {Deng}\ \emph {et~al.}(2003)\citenamefont {Deng}, \citenamefont {Long},\ and\ \citenamefont {Liu}}]{PhysRevA.68.042317}%
  \BibitemOpen
  \bibfield  {author} {\bibinfo {author} {\bibfnamefont {F.-G.}\ \bibnamefont {Deng}}, \bibinfo {author} {\bibfnamefont {G.~L.}\ \bibnamefont {Long}}, \ and\ \bibinfo {author} {\bibfnamefont {X.-S.}\ \bibnamefont {Liu}},\ }\href {\doibase 10.1103/PhysRevA.68.042317} {\bibfield  {journal} {\bibinfo  {journal} {Phys. Rev. A}\ }\textbf {\bibinfo {volume} {68}},\ \bibinfo {pages} {042317} (\bibinfo {year} {2003})}\BibitemShut {NoStop}%
\bibitem [{\citenamefont {Das}\ \emph {et~al.}(2021)\citenamefont {Das}, \citenamefont {Paul},\ and\ \citenamefont {Majumdar}}]{das2021quantum}%
  \BibitemOpen
  \bibfield  {author} {\bibinfo {author} {\bibfnamefont {N.}~\bibnamefont {Das}}, \bibinfo {author} {\bibfnamefont {G.}~\bibnamefont {Paul}}, \ and\ \bibinfo {author} {\bibfnamefont {R.}~\bibnamefont {Majumdar}},\ }\href {https://doi.org/10.1007/s10773-021-04952-4} {\bibfield  {journal} {\bibinfo  {journal} {International Journal of Theoretical Physics}\ }\textbf {\bibinfo {volume} {60}},\ \bibinfo {pages} {4044} (\bibinfo {year} {2021})}\BibitemShut {NoStop}%
\bibitem [{\citenamefont {Zhou}\ \emph {et~al.}(2023)\citenamefont {Zhou}, \citenamefont {Xu}, \citenamefont {Zhong},\ and\ \citenamefont {Sheng}}]{PhysRevApplied.19.014036}%
  \BibitemOpen
  \bibfield  {author} {\bibinfo {author} {\bibfnamefont {L.}~\bibnamefont {Zhou}}, \bibinfo {author} {\bibfnamefont {B.-W.}\ \bibnamefont {Xu}}, \bibinfo {author} {\bibfnamefont {W.}~\bibnamefont {Zhong}}, \ and\ \bibinfo {author} {\bibfnamefont {Y.-B.}\ \bibnamefont {Sheng}},\ }\href {\doibase 10.1103/PhysRevApplied.19.014036} {\bibfield  {journal} {\bibinfo  {journal} {Phys. Rev. Appl.}\ }\textbf {\bibinfo {volume} {19}},\ \bibinfo {pages} {014036} (\bibinfo {year} {2023})}\BibitemShut {NoStop}%
\bibitem [{\citenamefont {Cao}\ \emph {et~al.}(2021)\citenamefont {Cao}, \citenamefont {Wang}, \citenamefont {Liang}, \citenamefont {Chai},\ and\ \citenamefont {Peng}}]{PhysRevApplied.16.024012}%
  \BibitemOpen
  \bibfield  {author} {\bibinfo {author} {\bibfnamefont {Z.}~\bibnamefont {Cao}}, \bibinfo {author} {\bibfnamefont {L.}~\bibnamefont {Wang}}, \bibinfo {author} {\bibfnamefont {K.}~\bibnamefont {Liang}}, \bibinfo {author} {\bibfnamefont {G.}~\bibnamefont {Chai}}, \ and\ \bibinfo {author} {\bibfnamefont {J.}~\bibnamefont {Peng}},\ }\href {\doibase 10.1103/PhysRevApplied.16.024012} {\bibfield  {journal} {\bibinfo  {journal} {Phys. Rev. Appl.}\ }\textbf {\bibinfo {volume} {16}},\ \bibinfo {pages} {024012} (\bibinfo {year} {2021})}\BibitemShut {NoStop}%
\bibitem [{\citenamefont {Hillery}\ \emph {et~al.}(1999)\citenamefont {Hillery}, \citenamefont {Bu\ifmmode~\check{z}\else \v{z}\fi{}ek},\ and\ \citenamefont {Berthiaume}}]{PhysRevA.59.1829}%
  \BibitemOpen
  \bibfield  {author} {\bibinfo {author} {\bibfnamefont {M.}~\bibnamefont {Hillery}}, \bibinfo {author} {\bibfnamefont {V.}~\bibnamefont {Bu\ifmmode~\check{z}\else \v{z}\fi{}ek}}, \ and\ \bibinfo {author} {\bibfnamefont {A.}~\bibnamefont {Berthiaume}},\ }\href {\doibase 10.1103/PhysRevA.59.1829} {\bibfield  {journal} {\bibinfo  {journal} {Phys. Rev. A}\ }\textbf {\bibinfo {volume} {59}},\ \bibinfo {pages} {1829} (\bibinfo {year} {1999})}\BibitemShut {NoStop}%
\bibitem [{\citenamefont {Xiao}\ \emph {et~al.}(2004)\citenamefont {Xiao}, \citenamefont {Lu~Long}, \citenamefont {Deng},\ and\ \citenamefont {Pan}}]{PhysRevA.69.052307}%
  \BibitemOpen
  \bibfield  {author} {\bibinfo {author} {\bibfnamefont {L.}~\bibnamefont {Xiao}}, \bibinfo {author} {\bibfnamefont {G.}~\bibnamefont {Lu~Long}}, \bibinfo {author} {\bibfnamefont {F.-G.}\ \bibnamefont {Deng}}, \ and\ \bibinfo {author} {\bibfnamefont {J.-W.}\ \bibnamefont {Pan}},\ }\href {\doibase 10.1103/PhysRevA.69.052307} {\bibfield  {journal} {\bibinfo  {journal} {Phys. Rev. A}\ }\textbf {\bibinfo {volume} {69}},\ \bibinfo {pages} {052307} (\bibinfo {year} {2004})}\BibitemShut {NoStop}%
\bibitem [{\citenamefont {Zhang}\ \emph {et~al.}(2005)\citenamefont {Zhang}, \citenamefont {Li},\ and\ \citenamefont {Man}}]{PhysRevA.71.044301}%
  \BibitemOpen
  \bibfield  {author} {\bibinfo {author} {\bibfnamefont {Z.-j.}\ \bibnamefont {Zhang}}, \bibinfo {author} {\bibfnamefont {Y.}~\bibnamefont {Li}}, \ and\ \bibinfo {author} {\bibfnamefont {Z.-x.}\ \bibnamefont {Man}},\ }\href {\doibase 10.1103/PhysRevA.71.044301} {\bibfield  {journal} {\bibinfo  {journal} {Phys. Rev. A}\ }\textbf {\bibinfo {volume} {71}},\ \bibinfo {pages} {044301} (\bibinfo {year} {2005})}\BibitemShut {NoStop}%
\bibitem [{\citenamefont {Guo}\ and\ \citenamefont {Guo}(2003)}]{GUO2003247}%
  \BibitemOpen
  \bibfield  {author} {\bibinfo {author} {\bibfnamefont {G.-P.}\ \bibnamefont {Guo}}\ and\ \bibinfo {author} {\bibfnamefont {G.-C.}\ \bibnamefont {Guo}},\ }\href {\doibase https://doi.org/10.1016/S0375-9601(03)00074-4} {\bibfield  {journal} {\bibinfo  {journal} {Physics Letters A}\ }\textbf {\bibinfo {volume} {310}},\ \bibinfo {pages} {247} (\bibinfo {year} {2003})}\BibitemShut {NoStop}%
\bibitem [{\citenamefont {Gottesman}(2000)}]{PhysRevA.61.042311}%
  \BibitemOpen
  \bibfield  {author} {\bibinfo {author} {\bibfnamefont {D.}~\bibnamefont {Gottesman}},\ }\href {\doibase 10.1103/PhysRevA.61.042311} {\bibfield  {journal} {\bibinfo  {journal} {Phys. Rev. A}\ }\textbf {\bibinfo {volume} {61}},\ \bibinfo {pages} {042311} (\bibinfo {year} {2000})}\BibitemShut {NoStop}%
\bibitem [{\citenamefont {Tittel}\ \emph {et~al.}(2001)\citenamefont {Tittel}, \citenamefont {Zbinden},\ and\ \citenamefont {Gisin}}]{PhysRevA.63.042301}%
  \BibitemOpen
  \bibfield  {author} {\bibinfo {author} {\bibfnamefont {W.}~\bibnamefont {Tittel}}, \bibinfo {author} {\bibfnamefont {H.}~\bibnamefont {Zbinden}}, \ and\ \bibinfo {author} {\bibfnamefont {N.}~\bibnamefont {Gisin}},\ }\href {\doibase 10.1103/PhysRevA.63.042301} {\bibfield  {journal} {\bibinfo  {journal} {Phys. Rev. A}\ }\textbf {\bibinfo {volume} {63}},\ \bibinfo {pages} {042301} (\bibinfo {year} {2001})}\BibitemShut {NoStop}%
\bibitem [{\citenamefont {Zhou}\ \emph {et~al.}(2018)\citenamefont {Zhou}, \citenamefont {Yu}, \citenamefont {Yan}, \citenamefont {Jia}, \citenamefont {Zhang}, \citenamefont {Xie},\ and\ \citenamefont {Peng}}]{PhysRevLett.121.150502}%
  \BibitemOpen
  \bibfield  {author} {\bibinfo {author} {\bibfnamefont {Y.}~\bibnamefont {Zhou}}, \bibinfo {author} {\bibfnamefont {J.}~\bibnamefont {Yu}}, \bibinfo {author} {\bibfnamefont {Z.}~\bibnamefont {Yan}}, \bibinfo {author} {\bibfnamefont {X.}~\bibnamefont {Jia}}, \bibinfo {author} {\bibfnamefont {J.}~\bibnamefont {Zhang}}, \bibinfo {author} {\bibfnamefont {C.}~\bibnamefont {Xie}}, \ and\ \bibinfo {author} {\bibfnamefont {K.}~\bibnamefont {Peng}},\ }\href {\doibase 10.1103/PhysRevLett.121.150502} {\bibfield  {journal} {\bibinfo  {journal} {Phys. Rev. Lett.}\ }\textbf {\bibinfo {volume} {121}},\ \bibinfo {pages} {150502} (\bibinfo {year} {2018})}\BibitemShut {NoStop}%
\bibitem [{\citenamefont {Liao}\ \emph {et~al.}(2021)\citenamefont {Liao}, \citenamefont {Liu}, \citenamefont {Zhu},\ and\ \citenamefont {Guo}}]{PhysRevA.103.032410}%
  \BibitemOpen
  \bibfield  {author} {\bibinfo {author} {\bibfnamefont {Q.}~\bibnamefont {Liao}}, \bibinfo {author} {\bibfnamefont {H.}~\bibnamefont {Liu}}, \bibinfo {author} {\bibfnamefont {L.}~\bibnamefont {Zhu}}, \ and\ \bibinfo {author} {\bibfnamefont {Y.}~\bibnamefont {Guo}},\ }\href {\doibase 10.1103/PhysRevA.103.032410} {\bibfield  {journal} {\bibinfo  {journal} {Phys. Rev. A}\ }\textbf {\bibinfo {volume} {103}},\ \bibinfo {pages} {032410} (\bibinfo {year} {2021})}\BibitemShut {NoStop}%
\bibitem [{\citenamefont {Chong}\ and\ \citenamefont {Hwang}(2010)}]{CHONG20101192}%
  \BibitemOpen
  \bibfield  {author} {\bibinfo {author} {\bibfnamefont {S.-K.}\ \bibnamefont {Chong}}\ and\ \bibinfo {author} {\bibfnamefont {T.}~\bibnamefont {Hwang}},\ }\href {\doibase https://doi.org/10.1016/j.optcom.2009.11.007} {\bibfield  {journal} {\bibinfo  {journal} {Optics Communications}\ }\textbf {\bibinfo {volume} {283}},\ \bibinfo {pages} {1192} (\bibinfo {year} {2010})}\BibitemShut {NoStop}%
\bibitem [{\citenamefont {Liu}\ \emph {et~al.}(2013)\citenamefont {Liu}, \citenamefont {Gao}, \citenamefont {Huang},\ and\ \citenamefont {Wen}}]{liu2013multiparty}%
  \BibitemOpen
  \bibfield  {author} {\bibinfo {author} {\bibfnamefont {B.}~\bibnamefont {Liu}}, \bibinfo {author} {\bibfnamefont {F.}~\bibnamefont {Gao}}, \bibinfo {author} {\bibfnamefont {W.}~\bibnamefont {Huang}}, \ and\ \bibinfo {author} {\bibfnamefont {Q.-y.}\ \bibnamefont {Wen}},\ }\href {https://doi.org/10.1007/s11128-012-0492-6} {\bibfield  {journal} {\bibinfo  {journal} {Quantum information processing}\ }\textbf {\bibinfo {volume} {12}},\ \bibinfo {pages} {1797} (\bibinfo {year} {2013})}\BibitemShut {NoStop}%
\bibitem [{\citenamefont {Das}\ and\ \citenamefont {Majumdar}(2020)}]{das2020comment}%
  \BibitemOpen
  \bibfield  {author} {\bibinfo {author} {\bibfnamefont {N.}~\bibnamefont {Das}}\ and\ \bibinfo {author} {\bibfnamefont {R.}~\bibnamefont {Majumdar}},\ }\href {https://doi.org/10.1142/S0219749920500392} {\bibfield  {journal} {\bibinfo  {journal} {International Journal of Quantum Information}\ }\textbf {\bibinfo {volume} {18}},\ \bibinfo {pages} {2050039} (\bibinfo {year} {2020})}\BibitemShut {NoStop}%
\bibitem [{\citenamefont {Yang}\ \emph {et~al.}(2022)\citenamefont {Yang}, \citenamefont {Lv}, \citenamefont {Gao}, \citenamefont {Zhou},\ and\ \citenamefont {Shi}}]{yang2022detector}%
  \BibitemOpen
  \bibfield  {author} {\bibinfo {author} {\bibfnamefont {Y.-G.}\ \bibnamefont {Yang}}, \bibinfo {author} {\bibfnamefont {X.-L.}\ \bibnamefont {Lv}}, \bibinfo {author} {\bibfnamefont {S.}~\bibnamefont {Gao}}, \bibinfo {author} {\bibfnamefont {Y.-H.}\ \bibnamefont {Zhou}}, \ and\ \bibinfo {author} {\bibfnamefont {W.-M.}\ \bibnamefont {Shi}},\ }\href {https://doi.org/10.1007/s10773-022-05052-7} {\bibfield  {journal} {\bibinfo  {journal} {International Journal of Theoretical Physics}\ }\textbf {\bibinfo {volume} {61}},\ \bibinfo {pages} {50} (\bibinfo {year} {2022})}\BibitemShut {NoStop}%
\bibitem [{\citenamefont {Chen}\ \emph {et~al.}(2021)\citenamefont {Chen}, \citenamefont {Zhang}, \citenamefont {Liu}, \citenamefont {Jiang}, \citenamefont {Zhang}, \citenamefont {Han}, \citenamefont {Ma}, \citenamefont {Hu}, \citenamefont {Li}, \citenamefont {Liu} \emph {et~al.}}]{chen2021twin}%
  \BibitemOpen
  \bibfield  {author} {\bibinfo {author} {\bibfnamefont {J.-P.}\ \bibnamefont {Chen}}, \bibinfo {author} {\bibfnamefont {C.}~\bibnamefont {Zhang}}, \bibinfo {author} {\bibfnamefont {Y.}~\bibnamefont {Liu}}, \bibinfo {author} {\bibfnamefont {C.}~\bibnamefont {Jiang}}, \bibinfo {author} {\bibfnamefont {W.-J.}\ \bibnamefont {Zhang}}, \bibinfo {author} {\bibfnamefont {Z.-Y.}\ \bibnamefont {Han}}, \bibinfo {author} {\bibfnamefont {S.-Z.}\ \bibnamefont {Ma}}, \bibinfo {author} {\bibfnamefont {X.-L.}\ \bibnamefont {Hu}}, \bibinfo {author} {\bibfnamefont {Y.-H.}\ \bibnamefont {Li}}, \bibinfo {author} {\bibfnamefont {H.}~\bibnamefont {Liu}},  \emph {et~al.},\ }\href {https://doi.org/10.1038/s41566-021-00828-5} {\bibfield  {journal} {\bibinfo  {journal} {Nature Photonics}\ }\textbf {\bibinfo {volume} {15}},\ \bibinfo {pages} {570} (\bibinfo {year} {2021})}\BibitemShut {NoStop}%
\bibitem [{\citenamefont {Avesani}\ \emph {et~al.}(2021)\citenamefont {Avesani}, \citenamefont {Calderaro}, \citenamefont {Schiavon}, \citenamefont {Stanco}, \citenamefont {Agnesi}, \citenamefont {Santamato}, \citenamefont {Zahidy}, \citenamefont {Scriminich}, \citenamefont {Foletto}, \citenamefont {Contestabile} \emph {et~al.}}]{avesani2021full}%
  \BibitemOpen
  \bibfield  {author} {\bibinfo {author} {\bibfnamefont {M.}~\bibnamefont {Avesani}}, \bibinfo {author} {\bibfnamefont {L.}~\bibnamefont {Calderaro}}, \bibinfo {author} {\bibfnamefont {M.}~\bibnamefont {Schiavon}}, \bibinfo {author} {\bibfnamefont {A.}~\bibnamefont {Stanco}}, \bibinfo {author} {\bibfnamefont {C.}~\bibnamefont {Agnesi}}, \bibinfo {author} {\bibfnamefont {A.}~\bibnamefont {Santamato}}, \bibinfo {author} {\bibfnamefont {M.}~\bibnamefont {Zahidy}}, \bibinfo {author} {\bibfnamefont {A.}~\bibnamefont {Scriminich}}, \bibinfo {author} {\bibfnamefont {G.}~\bibnamefont {Foletto}}, \bibinfo {author} {\bibfnamefont {G.}~\bibnamefont {Contestabile}},  \emph {et~al.},\ }\href {https://doi.org/10.1038/s41534-021-00421-2} {\bibfield  {journal} {\bibinfo  {journal} {npj Quantum Information}\ }\textbf {\bibinfo {volume} {7}},\ \bibinfo {pages} {93} (\bibinfo {year} {2021})}\BibitemShut {NoStop}%
\bibitem [{\citenamefont {Pirandola}\ \emph {et~al.}(2017)\citenamefont {Pirandola}, \citenamefont {Laurenza}, \citenamefont {Ottaviani},\ and\ \citenamefont {Banchi}}]{pirandola2017fundamental}%
  \BibitemOpen
  \bibfield  {author} {\bibinfo {author} {\bibfnamefont {S.}~\bibnamefont {Pirandola}}, \bibinfo {author} {\bibfnamefont {R.}~\bibnamefont {Laurenza}}, \bibinfo {author} {\bibfnamefont {C.}~\bibnamefont {Ottaviani}}, \ and\ \bibinfo {author} {\bibfnamefont {L.}~\bibnamefont {Banchi}},\ }\href {https://doi.org/10.1038/ncomms15043} {\bibfield  {journal} {\bibinfo  {journal} {Nature communications}\ }\textbf {\bibinfo {volume} {8}},\ \bibinfo {pages} {1} (\bibinfo {year} {2017})}\BibitemShut {NoStop}%
\bibitem [{\citenamefont {Dahlberg}\ \emph {et~al.}(2020)\citenamefont {Dahlberg}, \citenamefont {Helsen},\ and\ \citenamefont {Wehner}}]{dahlberg2020transforming}%
  \BibitemOpen
  \bibfield  {author} {\bibinfo {author} {\bibfnamefont {A.}~\bibnamefont {Dahlberg}}, \bibinfo {author} {\bibfnamefont {J.}~\bibnamefont {Helsen}}, \ and\ \bibinfo {author} {\bibfnamefont {S.}~\bibnamefont {Wehner}},\ }\href {https://doi.org/10.22331/q-2020-10-22-348} {\bibfield  {journal} {\bibinfo  {journal} {Quantum}\ }\textbf {\bibinfo {volume} {4}},\ \bibinfo {pages} {348} (\bibinfo {year} {2020})}\BibitemShut {NoStop}%
\bibitem [{\citenamefont {Liao}\ \emph {et~al.}(2018)\citenamefont {Liao}, \citenamefont {Cai}, \citenamefont {Handsteiner}, \citenamefont {Liu}, \citenamefont {Yin}, \citenamefont {Zhang}, \citenamefont {Rauch}, \citenamefont {Fink}, \citenamefont {Ren}, \citenamefont {Liu}, \citenamefont {Li}, \citenamefont {Shen}, \citenamefont {Cao}, \citenamefont {Li}, \citenamefont {Wang}, \citenamefont {Huang}, \citenamefont {Deng}, \citenamefont {Xi}, \citenamefont {Ma}, \citenamefont {Hu}, \citenamefont {Li}, \citenamefont {Liu}, \citenamefont {Koidl}, \citenamefont {Wang}, \citenamefont {Chen}, \citenamefont {Wang}, \citenamefont {Steindorfer}, \citenamefont {Kirchner}, \citenamefont {Lu}, \citenamefont {Shu}, \citenamefont {Ursin}, \citenamefont {Scheidl}, \citenamefont {Peng}, \citenamefont {Wang}, \citenamefont {Zeilinger},\ and\ \citenamefont {Pan}}]{PhysRevLett.120.030501}%
  \BibitemOpen
  \bibfield  {author} {\bibinfo {author} {\bibfnamefont {S.-K.}\ \bibnamefont {Liao}}, \bibinfo {author} {\bibfnamefont {W.-Q.}\ \bibnamefont {Cai}}, \bibinfo {author} {\bibfnamefont {J.}~\bibnamefont {Handsteiner}}, \bibinfo {author} {\bibfnamefont {B.}~\bibnamefont {Liu}}, \bibinfo {author} {\bibfnamefont {J.}~\bibnamefont {Yin}}, \bibinfo {author} {\bibfnamefont {L.}~\bibnamefont {Zhang}}, \bibinfo {author} {\bibfnamefont {D.}~\bibnamefont {Rauch}}, \bibinfo {author} {\bibfnamefont {M.}~\bibnamefont {Fink}}, \bibinfo {author} {\bibfnamefont {J.-G.}\ \bibnamefont {Ren}}, \bibinfo {author} {\bibfnamefont {W.-Y.}\ \bibnamefont {Liu}}, \bibinfo {author} {\bibfnamefont {Y.}~\bibnamefont {Li}}, \bibinfo {author} {\bibfnamefont {Q.}~\bibnamefont {Shen}}, \bibinfo {author} {\bibfnamefont {Y.}~\bibnamefont {Cao}}, \bibinfo {author} {\bibfnamefont {F.-Z.}\ \bibnamefont {Li}}, \bibinfo {author} {\bibfnamefont {J.-F.}\ \bibnamefont {Wang}}, \bibinfo {author} {\bibfnamefont {Y.-M.}\ \bibnamefont {Huang}}, \bibinfo
  {author} {\bibfnamefont {L.}~\bibnamefont {Deng}}, \bibinfo {author} {\bibfnamefont {T.}~\bibnamefont {Xi}}, \bibinfo {author} {\bibfnamefont {L.}~\bibnamefont {Ma}}, \bibinfo {author} {\bibfnamefont {T.}~\bibnamefont {Hu}}, \bibinfo {author} {\bibfnamefont {L.}~\bibnamefont {Li}}, \bibinfo {author} {\bibfnamefont {N.-L.}\ \bibnamefont {Liu}}, \bibinfo {author} {\bibfnamefont {F.}~\bibnamefont {Koidl}}, \bibinfo {author} {\bibfnamefont {P.}~\bibnamefont {Wang}}, \bibinfo {author} {\bibfnamefont {Y.-A.}\ \bibnamefont {Chen}}, \bibinfo {author} {\bibfnamefont {X.-B.}\ \bibnamefont {Wang}}, \bibinfo {author} {\bibfnamefont {M.}~\bibnamefont {Steindorfer}}, \bibinfo {author} {\bibfnamefont {G.}~\bibnamefont {Kirchner}}, \bibinfo {author} {\bibfnamefont {C.-Y.}\ \bibnamefont {Lu}}, \bibinfo {author} {\bibfnamefont {R.}~\bibnamefont {Shu}}, \bibinfo {author} {\bibfnamefont {R.}~\bibnamefont {Ursin}}, \bibinfo {author} {\bibfnamefont {T.}~\bibnamefont {Scheidl}}, \bibinfo {author} {\bibfnamefont {C.-Z.}\
  \bibnamefont {Peng}}, \bibinfo {author} {\bibfnamefont {J.-Y.}\ \bibnamefont {Wang}}, \bibinfo {author} {\bibfnamefont {A.}~\bibnamefont {Zeilinger}}, \ and\ \bibinfo {author} {\bibfnamefont {J.-W.}\ \bibnamefont {Pan}},\ }\href {\doibase 10.1103/PhysRevLett.120.030501} {\bibfield  {journal} {\bibinfo  {journal} {Phys. Rev. Lett.}\ }\textbf {\bibinfo {volume} {120}},\ \bibinfo {pages} {030501} (\bibinfo {year} {2018})}\BibitemShut {NoStop}%
\bibitem [{\citenamefont {Simon}(2017)}]{simon2017towards}%
  \BibitemOpen
  \bibfield  {author} {\bibinfo {author} {\bibfnamefont {C.}~\bibnamefont {Simon}},\ }\href {https://doi.org/10.1038/s41566-017-0032-0} {\bibfield  {journal} {\bibinfo  {journal} {Nature Photonics}\ }\textbf {\bibinfo {volume} {11}},\ \bibinfo {pages} {678} (\bibinfo {year} {2017})}\BibitemShut {NoStop}%
\bibitem [{\citenamefont {Hillery}\ \emph {et~al.}(2006)\citenamefont {Hillery}, \citenamefont {Ziman}, \citenamefont {Bužek},\ and\ \citenamefont {Bieliková}}]{HILLERY200675}%
  \BibitemOpen
  \bibfield  {author} {\bibinfo {author} {\bibfnamefont {M.}~\bibnamefont {Hillery}}, \bibinfo {author} {\bibfnamefont {M.}~\bibnamefont {Ziman}}, \bibinfo {author} {\bibfnamefont {V.}~\bibnamefont {Bužek}}, \ and\ \bibinfo {author} {\bibfnamefont {M.}~\bibnamefont {Bieliková}},\ }\href {\doibase https://doi.org/10.1016/j.physleta.2005.09.010} {\bibfield  {journal} {\bibinfo  {journal} {Physics Letters A}\ }\textbf {\bibinfo {volume} {349}},\ \bibinfo {pages} {75} (\bibinfo {year} {2006})}\BibitemShut {NoStop}%
\bibitem [{\citenamefont {Thapliyal}\ \emph {et~al.}(2017)\citenamefont {Thapliyal}, \citenamefont {Sharma},\ and\ \citenamefont {Pathak}}]{doi:10.1142/S0219749917500071}%
  \BibitemOpen
  \bibfield  {author} {\bibinfo {author} {\bibfnamefont {K.}~\bibnamefont {Thapliyal}}, \bibinfo {author} {\bibfnamefont {R.~D.}\ \bibnamefont {Sharma}}, \ and\ \bibinfo {author} {\bibfnamefont {A.}~\bibnamefont {Pathak}},\ }\href {\doibase 10.1142/S0219749917500071} {\bibfield  {journal} {\bibinfo  {journal} {International Journal of Quantum Information}\ }\textbf {\bibinfo {volume} {15}},\ \bibinfo {pages} {1750007} (\bibinfo {year} {2017})}\BibitemShut {NoStop}%
\bibitem [{\citenamefont {Mishra}\ \emph {et~al.}(2022)\citenamefont {Mishra}, \citenamefont {Thapliyal}, \citenamefont {Parakh},\ and\ \citenamefont {Pathak}}]{mishra2022quantum}%
  \BibitemOpen
  \bibfield  {author} {\bibinfo {author} {\bibfnamefont {S.}~\bibnamefont {Mishra}}, \bibinfo {author} {\bibfnamefont {K.}~\bibnamefont {Thapliyal}}, \bibinfo {author} {\bibfnamefont {A.}~\bibnamefont {Parakh}}, \ and\ \bibinfo {author} {\bibfnamefont {A.}~\bibnamefont {Pathak}},\ }\href {https://doi.org/10.1140/epjqt/s40507-022-00133-2} {\bibfield  {journal} {\bibinfo  {journal} {EPJ Quantum Technology}\ }\textbf {\bibinfo {volume} {9}},\ \bibinfo {pages} {14} (\bibinfo {year} {2022})}\BibitemShut {NoStop}%
\bibitem [{\citenamefont {Greenberger}\ \emph {et~al.}(1990)\citenamefont {Greenberger}, \citenamefont {Horne}, \citenamefont {Shimony},\ and\ \citenamefont {Zeilinger}}]{10.1119/1.16243}%
  \BibitemOpen
  \bibfield  {author} {\bibinfo {author} {\bibfnamefont {D.~M.}\ \bibnamefont {Greenberger}}, \bibinfo {author} {\bibfnamefont {M.~A.}\ \bibnamefont {Horne}}, \bibinfo {author} {\bibfnamefont {A.}~\bibnamefont {Shimony}}, \ and\ \bibinfo {author} {\bibfnamefont {A.}~\bibnamefont {Zeilinger}},\ }\href {\doibase 10.1119/1.16243} {\bibfield  {journal} {\bibinfo  {journal} {American Journal of Physics}\ }\textbf {\bibinfo {volume} {58}},\ \bibinfo {pages} {1131} (\bibinfo {year} {1990})}\BibitemShut {NoStop}%
\bibitem [{\citenamefont {Schlingemann}\ and\ \citenamefont {Werner}(2001)}]{PhysRevA.65.012308}%
  \BibitemOpen
  \bibfield  {author} {\bibinfo {author} {\bibfnamefont {D.}~\bibnamefont {Schlingemann}}\ and\ \bibinfo {author} {\bibfnamefont {R.~F.}\ \bibnamefont {Werner}},\ }\href {\doibase 10.1103/PhysRevA.65.012308} {\bibfield  {journal} {\bibinfo  {journal} {Phys. Rev. A}\ }\textbf {\bibinfo {volume} {65}},\ \bibinfo {pages} {012308} (\bibinfo {year} {2001})}\BibitemShut {NoStop}%
\bibitem [{\citenamefont {Christandl}\ and\ \citenamefont {Wehner}(2005)}]{christandl2005quantum}%
  \BibitemOpen
  \bibfield  {author} {\bibinfo {author} {\bibfnamefont {M.}~\bibnamefont {Christandl}}\ and\ \bibinfo {author} {\bibfnamefont {S.}~\bibnamefont {Wehner}},\ }in\ \href {https://doi.org/10.1007/11593447_12} {\emph {\bibinfo {booktitle} {International conference on the theory and application of cryptology and information security}}}\ (\bibinfo {organization} {Springer},\ \bibinfo {year} {2005})\ pp.\ \bibinfo {pages} {217--235}\BibitemShut {NoStop}%
\bibitem [{\citenamefont {D\"ur}\ \emph {et~al.}(2014)\citenamefont {D\"ur}, \citenamefont {Skotiniotis}, \citenamefont {Fr\"owis},\ and\ \citenamefont {Kraus}}]{PhysRevLett.112.080801}%
  \BibitemOpen
  \bibfield  {author} {\bibinfo {author} {\bibfnamefont {W.}~\bibnamefont {D\"ur}}, \bibinfo {author} {\bibfnamefont {M.}~\bibnamefont {Skotiniotis}}, \bibinfo {author} {\bibfnamefont {F.}~\bibnamefont {Fr\"owis}}, \ and\ \bibinfo {author} {\bibfnamefont {B.}~\bibnamefont {Kraus}},\ }\href {\doibase 10.1103/PhysRevLett.112.080801} {\bibfield  {journal} {\bibinfo  {journal} {Phys. Rev. Lett.}\ }\textbf {\bibinfo {volume} {112}},\ \bibinfo {pages} {080801} (\bibinfo {year} {2014})}\BibitemShut {NoStop}%
\bibitem [{\citenamefont {Komar}\ \emph {et~al.}(2014)\citenamefont {Komar}, \citenamefont {Kessler}, \citenamefont {Bishof}, \citenamefont {Jiang}, \citenamefont {S{\o}rensen}, \citenamefont {Ye},\ and\ \citenamefont {Lukin}}]{komar2014quantum}%
  \BibitemOpen
  \bibfield  {author} {\bibinfo {author} {\bibfnamefont {P.}~\bibnamefont {Komar}}, \bibinfo {author} {\bibfnamefont {E.~M.}\ \bibnamefont {Kessler}}, \bibinfo {author} {\bibfnamefont {M.}~\bibnamefont {Bishof}}, \bibinfo {author} {\bibfnamefont {L.}~\bibnamefont {Jiang}}, \bibinfo {author} {\bibfnamefont {A.~S.}\ \bibnamefont {S{\o}rensen}}, \bibinfo {author} {\bibfnamefont {J.}~\bibnamefont {Ye}}, \ and\ \bibinfo {author} {\bibfnamefont {M.~D.}\ \bibnamefont {Lukin}},\ }\href {https://doi.org/10.1038/nphys3000} {\bibfield  {journal} {\bibinfo  {journal} {Nature Physics}\ }\textbf {\bibinfo {volume} {10}},\ \bibinfo {pages} {582} (\bibinfo {year} {2014})}\BibitemShut {NoStop}%
\bibitem [{\citenamefont {Khabiboulline}\ \emph {et~al.}(2019)\citenamefont {Khabiboulline}, \citenamefont {Borregaard}, \citenamefont {De~Greve},\ and\ \citenamefont {Lukin}}]{PhysRevLett.123.070504}%
  \BibitemOpen
  \bibfield  {author} {\bibinfo {author} {\bibfnamefont {E.~T.}\ \bibnamefont {Khabiboulline}}, \bibinfo {author} {\bibfnamefont {J.}~\bibnamefont {Borregaard}}, \bibinfo {author} {\bibfnamefont {K.}~\bibnamefont {De~Greve}}, \ and\ \bibinfo {author} {\bibfnamefont {M.~D.}\ \bibnamefont {Lukin}},\ }\href {\doibase 10.1103/PhysRevLett.123.070504} {\bibfield  {journal} {\bibinfo  {journal} {Phys. Rev. Lett.}\ }\textbf {\bibinfo {volume} {123}},\ \bibinfo {pages} {070504} (\bibinfo {year} {2019})}\BibitemShut {NoStop}%
\bibitem [{\citenamefont {Hahn}\ \emph {et~al.}(2019)\citenamefont {Hahn}, \citenamefont {Pappa},\ and\ \citenamefont {Eisert}}]{hahn2019quantum}%
  \BibitemOpen
  \bibfield  {author} {\bibinfo {author} {\bibfnamefont {F.}~\bibnamefont {Hahn}}, \bibinfo {author} {\bibfnamefont {A.}~\bibnamefont {Pappa}}, \ and\ \bibinfo {author} {\bibfnamefont {J.}~\bibnamefont {Eisert}},\ }\href {https://doi.org/10.1038/s41534-019-0191-6} {\bibfield  {journal} {\bibinfo  {journal} {npj Quantum Information}\ }\textbf {\bibinfo {volume} {5}},\ \bibinfo {pages} {76} (\bibinfo {year} {2019})}\BibitemShut {NoStop}%
\bibitem [{\citenamefont {Hein}\ \emph {et~al.}(2004)\citenamefont {Hein}, \citenamefont {Eisert},\ and\ \citenamefont {Briegel}}]{PhysRevA.69.062311}%
  \BibitemOpen
  \bibfield  {author} {\bibinfo {author} {\bibfnamefont {M.}~\bibnamefont {Hein}}, \bibinfo {author} {\bibfnamefont {J.}~\bibnamefont {Eisert}}, \ and\ \bibinfo {author} {\bibfnamefont {H.~J.}\ \bibnamefont {Briegel}},\ }\href {https://link.aps.org/doi/10.1103/PhysRevA.69.062311} {\bibfield  {journal} {\bibinfo  {journal} {Phys. Rev. A}\ }\textbf {\bibinfo {volume} {69}},\ \bibinfo {pages} {062311} (\bibinfo {year} {2004})}\BibitemShut {NoStop}%
\bibitem [{\citenamefont {Mannalath}\ and\ \citenamefont {Pathak}(2023)}]{PhysRevA.108.062614}%
  \BibitemOpen
  \bibfield  {author} {\bibinfo {author} {\bibfnamefont {V.}~\bibnamefont {Mannalath}}\ and\ \bibinfo {author} {\bibfnamefont {A.}~\bibnamefont {Pathak}},\ }\href {https://link.aps.org/doi/10.1103/PhysRevA.108.062614} {\bibfield  {journal} {\bibinfo  {journal} {Phys. Rev. A}\ }\textbf {\bibinfo {volume} {108}},\ \bibinfo {pages} {062614} (\bibinfo {year} {2023})}\BibitemShut {NoStop}%
\bibitem [{\citenamefont {\ifmmode~\dot{Z}\else \.{Z}\fi{}ukowski}\ \emph {et~al.}(1993)\citenamefont {\ifmmode~\dot{Z}\else \.{Z}\fi{}ukowski}, \citenamefont {Zeilinger}, \citenamefont {Horne},\ and\ \citenamefont {Ekert}}]{PhysRevLett.71.4287}%
  \BibitemOpen
  \bibfield  {author} {\bibinfo {author} {\bibfnamefont {M.}~\bibnamefont {\ifmmode~\dot{Z}\else \.{Z}\fi{}ukowski}}, \bibinfo {author} {\bibfnamefont {A.}~\bibnamefont {Zeilinger}}, \bibinfo {author} {\bibfnamefont {M.~A.}\ \bibnamefont {Horne}}, \ and\ \bibinfo {author} {\bibfnamefont {A.~K.}\ \bibnamefont {Ekert}},\ }\href {\doibase 10.1103/PhysRevLett.71.4287} {\bibfield  {journal} {\bibinfo  {journal} {Phys. Rev. Lett.}\ }\textbf {\bibinfo {volume} {71}},\ \bibinfo {pages} {4287} (\bibinfo {year} {1993})}\BibitemShut {NoStop}%
\bibitem [{\citenamefont {Bennett}\ \emph {et~al.}(1993)\citenamefont {Bennett}, \citenamefont {Brassard}, \citenamefont {Cr\'epeau}, \citenamefont {Jozsa}, \citenamefont {Peres},\ and\ \citenamefont {Wootters}}]{PhysRevLett.70.1895}%
  \BibitemOpen
  \bibfield  {author} {\bibinfo {author} {\bibfnamefont {C.~H.}\ \bibnamefont {Bennett}}, \bibinfo {author} {\bibfnamefont {G.}~\bibnamefont {Brassard}}, \bibinfo {author} {\bibfnamefont {C.}~\bibnamefont {Cr\'epeau}}, \bibinfo {author} {\bibfnamefont {R.}~\bibnamefont {Jozsa}}, \bibinfo {author} {\bibfnamefont {A.}~\bibnamefont {Peres}}, \ and\ \bibinfo {author} {\bibfnamefont {W.~K.}\ \bibnamefont {Wootters}},\ }\href {\doibase 10.1103/PhysRevLett.70.1895} {\bibfield  {journal} {\bibinfo  {journal} {Phys. Rev. Lett.}\ }\textbf {\bibinfo {volume} {70}},\ \bibinfo {pages} {1895} (\bibinfo {year} {1993})}\BibitemShut {NoStop}%
\bibitem [{\citenamefont {Van~den Nest}\ \emph {et~al.}(2004)\citenamefont {Van~den Nest}, \citenamefont {Dehaene},\ and\ \citenamefont {De~Moor}}]{PhysRevA.69.022316}%
  \BibitemOpen
  \bibfield  {author} {\bibinfo {author} {\bibfnamefont {M.}~\bibnamefont {Van~den Nest}}, \bibinfo {author} {\bibfnamefont {J.}~\bibnamefont {Dehaene}}, \ and\ \bibinfo {author} {\bibfnamefont {B.}~\bibnamefont {De~Moor}},\ }\href {https://link.aps.org/doi/10.1103/PhysRevA.69.022316} {\bibfield  {journal} {\bibinfo  {journal} {Phys. Rev. A}\ }\textbf {\bibinfo {volume} {69}},\ \bibinfo {pages} {022316} (\bibinfo {year} {2004})}\BibitemShut {NoStop}%
\bibitem [{\citenamefont {Meignant}\ \emph {et~al.}(2019)\citenamefont {Meignant}, \citenamefont {Markham},\ and\ \citenamefont {Grosshans}}]{PhysRevA.100.052333}%
  \BibitemOpen
  \bibfield  {author} {\bibinfo {author} {\bibfnamefont {C.}~\bibnamefont {Meignant}}, \bibinfo {author} {\bibfnamefont {D.}~\bibnamefont {Markham}}, \ and\ \bibinfo {author} {\bibfnamefont {F.}~\bibnamefont {Grosshans}},\ }\href {https://link.aps.org/doi/10.1103/PhysRevA.100.052333} {\bibfield  {journal} {\bibinfo  {journal} {Phys. Rev. A}\ }\textbf {\bibinfo {volume} {100}},\ \bibinfo {pages} {052333} (\bibinfo {year} {2019})}\BibitemShut {NoStop}%
\bibitem [{\citenamefont {Briegel}\ and\ \citenamefont {Raussendorf}(2001)}]{PhysRevLett.86.910}%
  \BibitemOpen
  \bibfield  {author} {\bibinfo {author} {\bibfnamefont {H.~J.}\ \bibnamefont {Briegel}}\ and\ \bibinfo {author} {\bibfnamefont {R.}~\bibnamefont {Raussendorf}},\ }\href {https://link.aps.org/doi/10.1103/PhysRevLett.86.910} {\bibfield  {journal} {\bibinfo  {journal} {Phys. Rev. Lett.}\ }\textbf {\bibinfo {volume} {86}},\ \bibinfo {pages} {910} (\bibinfo {year} {2001})}\BibitemShut {NoStop}%
\bibitem [{\citenamefont {de~Jong}\ \emph {et~al.}(2024)\citenamefont {de~Jong}, \citenamefont {Hahn}, \citenamefont {Tcholtchev}, \citenamefont {Hauswirth},\ and\ \citenamefont {Pappa}}]{PhysRevResearch.6.013330}%
  \BibitemOpen
  \bibfield  {author} {\bibinfo {author} {\bibfnamefont {J.}~\bibnamefont {de~Jong}}, \bibinfo {author} {\bibfnamefont {F.}~\bibnamefont {Hahn}}, \bibinfo {author} {\bibfnamefont {N.}~\bibnamefont {Tcholtchev}}, \bibinfo {author} {\bibfnamefont {M.}~\bibnamefont {Hauswirth}}, \ and\ \bibinfo {author} {\bibfnamefont {A.}~\bibnamefont {Pappa}},\ }\href {\doibase 10.1103/PhysRevResearch.6.013330} {\bibfield  {journal} {\bibinfo  {journal} {Phys. Rev. Res.}\ }\textbf {\bibinfo {volume} {6}},\ \bibinfo {pages} {013330} (\bibinfo {year} {2024})}\BibitemShut {NoStop}%
\end{thebibliography}%

\end{document}